\documentclass[12pt]{article}\usepackage[cmtip,arrow]{xy}\usepackage{pb-diagram,pb-xy}%
\input xy
\xyoption{all}
\usepackage{heck}
\usepackage{cite}
\usepackage{graphicx}
\usepackage{makeidx}
\usepackage{multicol}
\usepackage{geometry}
\usepackage{amsfonts}
\usepackage{mathrsfs}
\usepackage{amssymb}
\usepackage{amsmath}%

\providecommand{\U}[1]{\protect\rule{.1in}{.1in}}
\numberwithin{equation}{section}

\newcommand{\bea}{\begin{eqnarray}}
\newcommand{\eea}{\end{eqnarray}}
\newcommand{\be}{\begin{equation}}
\newcommand{\ee}{\end{equation}}

\newcommand{\bem}{\begin{pmatrix}}
\newcommand{\eem}{\end{pmatrix}}

\def\Tr{{\rm Tr}}
\def\U{\Upsilon}

\def\ca{{\cal A}}

\def\cc{{\cal C}}

\def\ch{{\cal H}}

\def\cm{{\cal M}}
\def\cn{{\cal N}}
\def\co{{\cal O}}

\def\cq{{\cal Q}}

\def\cu{{\cal U}}
\def\cv{{\cal V}}

\def\cx{{\cal X}}

\def \N {{\mathcal N}}
\def \Z {{\mathbb Z}}
\def \C {{\mathbb C}}
\def \R {{\mathbb R}}
\def\PP {{\mathbb P}}

\DeclareMathOperator{\Pexp}{Pexp}

\begin{document}

\bibliographystyle{utphys}

\date{June, 2010}


\institution{SISSA}{\centerline{${}^{1}$Scuola Internazionale Superiore di Studi Avanzati, via Bonomea 265, I-34100 Trieste, ITALY}}

\institution{UT}{\centerline{${}^2$Department of Mathematics, University of Texas at Austin, Austin, TX 78712, USA}}

\institution{HarvardU}{\centerline{${}^{3}$Jefferson Physical Laboratory, Harvard University, Cambridge, MA 02138, USA}}%

\title{R-Twisting and 4d/2d Correspondences}
%

\authors{Sergio Cecotti\worksat{\SISSA}\footnote{e-mail: {\tt cecotti@sissa.it}}, Andrew Neitzke\worksat{\UT}\footnote{e-mail: {\tt neitzke@math.utexas.edu}} and Cumrun Vafa\worksat{\HarvardU}\footnote{e-mail: {\tt vafa@physics.harvard.edu}}}%

\abstract{We show how aspects of the R-charge of $\cn=2$ CFTs in four dimensions
are encoded in the q-deformed Kontsevich-Soibelman monodromy operator, built from their dyon spectra.
In particular, the monodromy operator should have finite order if the R-charges
are rational.  We verify this for a number of examples including those arising
from pairs of ADE singularities on a Calabi-Yau threefold (some of which are dual to 6d $(2,0)$ ADE theories
suitably fibered over the plane).  In these cases we find
that our monodromy maps to that of the Y-systems, studied by Zamolodchikov
in the context of TBA.
 Moreover we find that the trace of the (fractional) q-deformed KS monodromy is given by the characters
of 2d conformal field theories associated to the corresponding TBA (i.e. integrable
deformations of the generalized parafermionic systems).   The Verlinde algebra gets realized through evaluation of  line operators at the loci of the associated hyperK\"ahler manifold
fixed under R-symmetry action.  Moreover, we propose how the TBA system
arises as part of the $\cn=2$ theory in 4 dimensions.
Finally, we initiate a classification of $\cn=2$ superconformal theories in 4 dimensions based on their quiver data
and find that this classification problem is mapped to the classification of $\cn=2$ theories in 2 dimensions,
and use this to classify all the 4d, $\cn=2$ theories with up to 3 generators for BPS
states.}

\maketitle

\tableofcontents

\pagebreak

\section{Introduction}

Supersymmetric gauge theories have very special properties which are ``protected'' from quantum corrections 
by the supersymmetry.  There is an interesting spectrum of possibilities --- at one
end we have the case of maximal supersymmetry where the theory is very rigid and almost all quantities are protected,
while with lower supersymmetry quantum corrections affect even the low energy amplitudes.  ${\cal N}=2$ supersymmetric
gauge theories in four dimensions offer a middle ground where many quantities are protected, but
there is sufficient flexibility to include a wide range of interesting physical effects.
In particular, these theories have a low energy Lagrangian which is completely specified by 
holomorphic data, often computable in terms of a Seiberg-Witten curve \cite{Seiberg:1994rs}.

This is also the case where there is a stable class of states, the BPS particles, whose masses are protected
by the supersymmetry algebra. These particles are generically protected from decay due to the combination of
charge conservation and conservation of energy.
However, as one changes the parameters in the theory, BPS particles can in principle decay,
when the phases of the central charges of at least two of them align in the complex plane.  The loci in parameter space 
where such a decay occurs are called ``walls of marginal stability'', and the problem of determining the spectrum
as one crosses them is the problem of ``wall-crossing.''
In fact the possibility of such wall-crossing was essential for a consistent picture of the low energy dynamics 
of ${\cal N}=2$ gauge theory \cite{Seiberg:1994rs}.

This situation parallels the simpler case of ${\cal N}=2$ theories in 2 dimensions.  For massive
${\cal N}=2$ theories one also has BPS particles, in this case kinks interpolating between two vacua,
whose mass is protected by supersymmetry algebra.  The central charge is again a complex number, and again the
BPS particles can decay when two central charges become aligned as one varies parameters.
In that context it was discovered \cite{Cecotti:1993rm} that the wall-crossing phenomenon can be captured without
knowing many details of the theory:  all one needs to know to predict the spectrum of BPS states after wall-crossing
is the spectrum of the BPS states before the wall-crossing and the ordering of the phases of central charges near the wall.  
Indeed, the wall-crossing behavior is captured by the statement that certain product of matrices built from the soliton numbers
do not change as one crosses the wall.  This structure has found a close parallel in $d=4$, ${\cal N}=2$ 
gauge theories.
In these theories the jumping phenomenon can again be captured
in terms of a certain object which does {\it not} jump:  in this case rather than a 
finite-dimensional matrix it turns out to
be a symplectomorphism of a torus, constructed as a product of elementary symplectomorphisms coming from 
the relevant BPS states.   This wall-crossing formula was discovered in the context of Donaldson-Thomas
theory by Kontsevich-Soibelman in \cite{ks1}.  In \cite{Gaiotto:2008cd} this formula was 
proven to give the correct wall-crossing for $\cn=2$ gauge theories.
Furthermore, a more refined $q$-deformed version (taking into account the spin of BPS states), 
has been advanced in \cite{Dimofte:2009bv,Dimofte:2009tm}, 
and proven for $\cn=2$ theories which arise from M5-branes in \cite{Cecotti:2009uf}.

It thus becomes very natural to consider the ``BPS monodromy'' $M(q)$, a product of $q$-deformed 
symplectomorphisms corresponding to all of the BPS states of the theory (or a fraction of them
in case there are extra R-symmetries).
Up to conjugation this is a completely wall-crossing invariant object.
So we have a simple physical question:  {\it what invariant information is the BPS monodromy capturing}?

The analogous question has been answered in the case of ${\cal N}=2$ theories in $d=2$ \cite{Cecotti:1993rm}:  viewing the theory we study
as a massive perturbation of some CFT, the BPS monodromy captures the R-charges of the chiral fields of that CFT.  
This puts severe restrictions on what the BPS spectrum of 
${\cal N}=2$ theories can be.  In particular, the eigenvalues of the monodromy
should lie on the unit circle --- this condition already puts strong constraints on
the possible BPS spectra.  This led to a classification program for ${\cal N}=2$ theories in $d=2$.
In particular it was shown that conformal theories with R-charges less than 1 can be
classified using this procedure, and correspond to A-D-E Dynkin diagrams:  the nodes
correspond to vacua and the links correspond to kinks
interpolating between the vacua (in some chamber).  This was then related
to the minimal ${\cal N}=2$ CFTs in 2 dimensions.  It was also shown how
to use this procedure to classify ${\cal N}=2$ theories with up to three vacua. 

Given the parallel between the 4d and 2d cases, it is natural to ask whether this classification
program can be imported to the case of ${\cal N}=2$ in $d=4$.  In particular
it seems natural to imagine that the BPS monodromy operator is related to R-charges of some
${\cal N}=2$ CFT, and that this could be used for a classification program for ${\cal N}=2$ theories
in 4 dimensions.

This paper takes some first steps toward this program.  In particular, we formulate what the 
trace of $M(q)$, and certain fractional powers of it (such as the `half-monodromy') compute from the perspective of the 4d path integral.
We first approach this question in the context of 4d theories which arise from M5-branes, 
using the connection between the monodromy and the topological string previously observed for these cases 
in \cite{Cecotti:2009uf}.  We find that the computation of 
$\Tr M(q)^k$ is related, at the conformal point, to a path-integral computation:
\begin{equation} \label{pitrace}
Z_{KS}(q, k) := Z_{{\cal N}=2}[MC_q\times S^1_{g^k}] = \mathrm{Tr}(M(q)^k).
\end{equation}
Here $MC_q$ is the ``Melvin Cigar'', defined by
$$MC_q=(C\times S^1)_q,$$
where $C$ is the topologically twisted cigar (also used in the story of $tt^*$ geometry)
and as we go around $S^1$ we rotate $C$ by an angle $\lambda_s$, where $q=\exp(2\pi i\lambda_s)$.
Moreover, the other circle of the 4d space is also twisted:  as we go around it we twist by
$g^k$, where $g$ is an appropriate element of the $U(1)$ R-symmetry group
of the ${\cal N}=2$ theory at the conformal point.
Since $C$ is non-compact, the above path-integral requires a boundary condition,
and we show that this indeed matches a similar choice which arises in the computation of the trace of
the monodromy.  

However, there are cases of ${\cal N}=2$ gauge theories which are
not realized by M5-branes.  Thus the methods of \cite{Cecotti:2009uf} cannot be applied to them directly.
For these more general cases we offer an alternative derivation of \eqref{pitrace}.  As
a by-product, this also provides an alternative general derivation of
the wall-crossing formula for $\cn=2$ theories in 4 dimensions.

Our path-integral interpretation of the monodromy leads to several predictions.
For example, if we have a conformal $\cn=2$ theory
where the denominators of the R-charges of chiral operators all divide $r$, \eqref{pitrace} implies the 
{\it prediction} that insertion of monodromy operator is periodic:
$$M^{r}(q)=1.$$
(It is believed that some such $r$ exists for any $\cn=2$ conformal theory.)
If $|q|<1$, we additionally
argue that $\mathrm{Tr}(M(q))$ should give a quasi-modular function of $q$.

We consider a number of examples to check these predictions.
These include the case of Argyres-Douglas models, corresponding
to M5-branes with singularities of the type $y^2=x^n$.  In these cases the R-charges
have common denominator $r=n+2$, and the monodromy operator
is indeed periodic, in the sense that $M^{k+r}=M^k$ (at least up to an overall $c$-number),
as is expected from the path-integral argument.
This investigation, however, reveals a much more detailed story.
We find
that the possible choices of R-symmetry invariant boundary conditions on the cigar 
naturally correspond with the eigenvectors of a Verlinde algebra for 
an associated RCFT, and that the Verlinde algebra itself is generated 
by the reduction of certain canonical line operators of the 4d theory.  Moreover,
by considering the insertion of line operators $X_i$ along the circle in the Melvin Cigar geometry, we obtain
different characters of RCFT:
$$\mathrm{Tr}\prod(X_i^{n_i})M(q)=\chi_{n_i}(q)$$
In particular, we find that for the 4d CFT corresponding to 
an M5-brane with the singularity $y^m=x^n$, 
such computations (with $M(q)$ replaced by its $m$-th root)
give characters of the level $m$ $SU(n)$ parafermionic systems.
When $m=2$, using the full monodromy $M(q)$ instead of its square root,
we similarly find characters of
$(2,n+2)$ minimal models for the $W^{sl(n-1)}$ algebra.

We also discuss $\N=2$ CFTs which do not necessarily come from M5-branes.
For example, consider Type IIB superstring compactification on a Calabi-Yau threefold $X$.
If $X$ develops an isolated singularity we expect a corresponding ${\cal N}=2$
CFT.  The singularities of the type $f(x,y)+uv=0$ correspond to M5-brane
CFTs, but these do not exhaust the possibilities --- there are more general 
singularities which involve all variables.  For example, given a {\it pair} $(G, G')$ of A-D-E groups
we can form a singularity of the type
$$W_{G} (x,y) + W_{G'} (u,v)=0.$$
The BPS spectra of these $\cn = 2$ theories
have not yet been computed.  However, there is a rather natural conjecture.
It was observed in \cite{Gaiotto:2008cd} that the BPS spectrum is naturally
associated to a certain integral equation which has the form of a Thermodynamic
Bethe Ansatz.  From this TBA one can build a discrete dynamical system known as
the $Y$-system.\footnote{A closely related $Y$-system played an important role in \cite{Alday:2010vh}.}
Similar $Y$-systems have been studied in the theory of cluster algebras,
and in particular there is a family of cluster algebras which are labeled exactly
by pairs $(G, G')$.  These turn out to be related to Zamolodchikov's
TBA systems \cite{Zamolodchikov:1991et} describing certain massive integrable
deformations of CFT's in 2 dimensions.
If we identify the 
$Y$-system associated to these theories
with the one that would come from \cite{Gaiotto:2008cd}, 
we get a prediction for the spectrum of BPS states (in one chamber),
or equivalently for the BPS monodromy $M$.
The common denominator of the R-charges for these conformal theories
divides $r = h+h'$, where $h$, $h'$ are the
dual Coxeter numbers of $G$, $G'$.  So combining our conjectures 
we would predict that $M$ obeys $M^r = 1$.
Exactly this periodicity has very
recently been proven \cite{keller-periodicity}, 
and we take this as an evidence for our picture.  To be precise, we identify 
our operator $M$ with the $h'$-th or $h$-th powers of the operators 
introduced in \cite{keller-periodicity}.
(Incidentally, assuming this conjecture, we actually know both the BPS spectrum and the central charge functions 
for these models --- the latter being given by periods of the Calabi-Yau threefold; applying the 
methods of \cite{Gaiotto:2008cd} to these data, we would expect to get an
explicit construction of a new family of hyperk\"ahler metrics, which are not Hitchin systems
as far as we know.)

This mysterious relation with RCFT as well as its connection with
TBA can be explained, at least in some cases,
using string dualities.  By realizing
the gauge system in terms of M5 branes and applying various dualities, we map
the $(G,A_{m-1})$ geometry to $SU(m)$ gauge groups on $\mathbb{C}^2/\Gamma$ 
where $\Gamma$ is the discrete subgroup of $SU(2)$ leading to $G$-type singularities.  It
is known that the instanton partition function of this theory, according to Nakajima \cite{MR95i:53051},
form a representation of the current algebra of $G$ at level $m$.  This turns out to be
consistent with our observation that computing the trace of the fractional monodromy for 
$(G,A_{m-1})$ pairs leads to parafermions of $G$ at level $m$.
Moreover, in the case $G=A_{n-1}$, applying another
string duality as in \cite{Dijkgraaf:2007sw} maps the system
to $m$ D4 branes ending on $n$ D6 branes, and the bifundamental matter
fields gauged by the dynamical $SU(m)$ gauge symmetry leads to parafermionic
system of $SU(n)$ at level $m$, again matching what we found in computing
the trace of the monodromy.  This last duality also suggests how to identify the 
2d space where Zamolodchikov's
TBA system \cite{Zamolodchikov:1991et} lives, thus demystifying its appearance
in the context of $\cn=2$ theories in 4 dimensions.

Finally we consider the problem we started with:  the classification of ${\cal N}=2$ theories in 4 dimensions.
Surprisingly, we find that the classification problem is actually related to the exact classification of
${\cal N}=2$ theories in $d=2$ dimensions!  In particular, it seems that some of the results in that
context can be borrowed for our use in 4d.  We offer some explanation of this:
when the $\cn=2$ theory has a superstring realization, 
the quiver which captures its BPS degeneracies is the same
as a quiver governing the $\cn=2$ {\it worldsheet} theory.
For example, when the $\cn=2$ theory arises from Type IIB strings on
a singular Calabi-Yau, the 2d worldsheet theory is a certain Landau-Ginzburg model
coupled to Liouville fields.

Applying this correspondence, we use the classification program of 2d $\cn=2$ theories in \cite{Cecotti:1993rm}
to initiate classification of the quiver types of possible 4d theories.  More specifically
for BPS spectra which are generated from up to three basic objects in an appropriate sense, the
classification results of \cite{Cecotti:1993rm} leads to a complete classification for the 4d case.
In particular, we rule out 4d theories whose BPS spectrum is generated from
two basic charges $\gamma_1, \gamma_2$ for which the electromagnetic inner product
$\lvert\langle \gamma_1, \gamma_2 \rangle\rvert > 2$.  Similarly, we classify 
the allowed $\cn=2$ theories with three BPS generators.  Moreover we find that
the 4d theory is conformal if and only if the corresponding monodromy has finite order.
This mirrors the case in 2d, where the monodromy matrix has infinite order for non-conformal
theories such as $\C \PP^n$ sigma models.

So in this paper we have found two apparently different links between $\cn=2$ 4d and 2d physics, where
the 2d is either part of the target or on the worldsheet.
We should note that quite a few links between supersymmetric theories in 4 dimensions and
theories in 2 dimensions are known, and more have been appearing recently --- e.g. \cite{Harvey:1995tg,
Bershadsky:1995vm,geom-lang-n4,Nekrasov:2009rc,Nekrasov:2010ka,Alday:2009aq,Gaiotto:2009fs} are some prominent examples.
While our constructions here involve ingredients which are certainly familiar 
(reduction on two circles plays a prominent role,
as does a deformation which resembles the $\Omega$-background), we do not know a precise
connection between our story and previously known ones.

The organization of this paper is as follows:   In Section 2 we
review aspects of open A-branes for topological strings and the spacetime interpretation of
the open topological amplitudes.  In Section 3 we review aspects of the construction in
\cite{Cecotti:2009uf}.  Furthermore,
we formulate in 4-dimensional gauge theory terms what the (quantum) monodromy
computes.  In Section 4 we show how these results can be generalized to arbitrary $\cn=2$ theories
in 4 dimensions, regardless of whether they arise from M5-branes.
In Section 5 we discuss in more detail the structure of the quantum monodromy operator.
In Section 6 we consider the quantum monodromy and the associated TBA system for the ADE type singularities
and verify our conjecture for this class.  
In Section 7  we consider a class of examples where the $\cn=2$ CFT's are
obtained by compactification of type IIB on general hypersurface singular Calabi-Yau threefolds.  
 In Section 8, we consider the case of pairs of ADE singularities
and the associated quantum monodromy.
  In Sections 9,10 we study the trace of the quantum monodromy operator
for both irrational and rational $q$ respectively, for some of the examples presented.
In particular we see the appearance of RCFT characters and the Verlinde algebra.
 In Section 11 using string dualities we explain why the RCFT's, as well as the
corresponding TBA system appear
in our theories.
In Section 12 we advance a conjecture relating 
the 4d and 2d classifications of $\cn=2$ theories.   In Section 13 we use
the 2d classification to classify 4d theories with up to 3 generator for BPS lattices, and identify
the corresponding theories.   For clarity of presentation,
we postpone various extensions and elaborations of the ideas in the main body of the paper to the appendices.

\section{Open A-branes and the physical interpretation of topological amplitudes} \label{top-review}

While the main statements of this paper are formulated independently of the topological string, we will
use topological strings as a tool for finding and proving them.
In this section we therefore 
review some facts about open topological strings and their relation to physical superstrings.

\subsection{The open topological A model}

Consider a Calabi-Yau 3-fold $K$.  For most of our examples $K$ will be non-compact.
Furthermore, consider a Lagrangian submanifold $L \subset K$.  
We will study the A model topological string on $K$, with $M$ A branes supported
on $L$.  As has been shown in \cite{Witten:1992fb}, the open string sector gives rise in the target space to a 
$U(M)$ Chern-Simons theory on $L$, where $\lambda_s$ plays the role of the (quantum corrected) Chern-Simons coupling constant.  The
partition function will be expressed in terms of
$$q = \exp (2\pi i \lambda_s).$$
In the usual approach to Chern-Simons theory with compact gauge group, 
quantization of the coupling implies that $q$ is ``rational'' in the sense that
$$q^N=1$$
for some integer $N$.  In the context of the topological string we usually consider the case
where $q$ is not rational, corresponding to a more general choice for the
Chern-Simons coupling (see e.g. \cite{Klemm:2010tm}). The meaning of such irrational $q$ has recently been clarified from the
path-integral viewpoint in \cite{Witten:2010cx}.
In this paper we will be interested in both the rational and irrational cases.

Were it not for worldsheet instantons, the partition function of the theory would simply be
$$Z^{open}_{top}=Z^L_{CS}(q, M)$$
where 
$$Z^L_{CS}(q, M)=\int DA \ \exp(CS(A,\lambda_s)).$$
Worldsheet instantons give further
corrections to $Z^{open}_{top}$ \cite{Witten:1992fb}.  
For example, for an isolated worldsheet instanton given by a holomorphic
disc $\cc$, with boundary on a curve $\gamma \subset L$, the instanton correction
produces an insertion of the complexified holonomy
\begin{equation} \label{defUgamma}
\cu_\gamma = e^{- \int_\cc k} \Pexp\left(i \int_\gamma A\right)
\end{equation}
into the Chern-Simons path integral (where $k$ is the K\"ahler form on $K$).

For more general worldsheets with higher genus
and many boundaries, the precise form of the corrections to Chern-Simons theory which arise in this way appears
rather complicated.  However, it has been argued \cite{Ooguri:1999bv,Labastida:2000yw} that these contributions actually
do admit a simple structure.  The simplicity becomes most apparent once we embed the topological 
string into the physical superstring, to which we now turn.

\subsection{Embedding in the physical superstring} \label{embedding}

We consider Type IIA string theory on the background
$$K \times \R^4.$$
As before we take a Lagrangian submanifold $L \subset K$.  We now wrap
$M$ D4-branes on
$$L\times \R^2 \subset K\times \R^4.$$
The topological string we discussed above computes quantities related to this physical string theory.  
For example, topological amplitudes 
at fixed genus show up as superpotential terms or gravitational corrections.
To see the whole topological partition function appear at once we use the idea
in \cite{Dijkgraaf:2006um} which shows how the partition function of closed
topological string can be reformulated as a computation in a specific background in M-theory.
Below we are interested in the open string version of the same idea which involves a
simple extension of this setup considered, e.g. in  \cite{Aganagic:2009cg,DSV-unpub}.

For this reformulation we need to make a further modification of the target space of the physical theory.
We begin by lifting to M-theory on
$$K\times S^1\times \R^4$$
so that the D4-branes are replaced by M5-branes on
$$L\times S^1\times \R^2\subset K \times S^1\times \R^4.$$
So far this is not a modification, just another way of describing the same system.
The desired modification is to replace $\R^4$ with Taub-NUT space, which we 
denote by $TN$, and in addition make a twist:  as we go around the M-theory
circle $S^1$, we rotate $TN$ by $\lambda_s$.  (The word ``rotate'' is used slightly loosely here:
in coordinates $(z_1, z_2)$ which identify
$TN$ with $\C^2$, our action is $(z_1, z_2) \to (q z_1, q^{-1} z_2)$.)
We denote this geometry
$$K\times (S^1\times TN)_q.$$
%
The M5-branes now occupy (see \cite{Aganagic:2009cg,DSV-unpub})
$$L\times (S^1\times C)_q \subset K\times (S^1\times TN)_q$$
where $C$ is the two-dimensional subspace $\{z_2 = 0\}$ of $TN$, metrically a ``cigar'',
and the rotation acts by $z_1\rightarrow qz_1$.  From now on we call this twisted space $(S^1 \times C)_q$
a ``Melvin cigar,'' by analogy to the Melvin universe, and denote it $MC_q$.
Let $X$ denote this M-theory background, i.e. the geometry plus the fivebrane configuration:
$$X = ( L\times  MC_q) \subset K\times (S^1\times TN)_q. $$

The partition function of M-theory on this twisted background is nothing but the topological partition function:
$$Z_{M-theory}(X; q) = Z_{top}(K,L; q).$$
From now on we will focus on the contributions from the {\it open} string sector, $Z^{open}_{top}(K,L; q)$.

\subsection{Integrating out the BPS states}

As noted before, $Z^{open}_{top}(K,L; q)$
is the Chern-Simons path integral on $L$, with insertions corresponding to holomorphic curves $\cc$ ending on $L$.
We have identified this with $Z_{M-theory}(X; q)$, and in that language one
can think of the Chern-Simons path integral as the integral over some
light modes living on the M5-brane.  How do the instanton corrections appear in this language?
The answer is that the effective Lagrangian for these light modes is not {\it just} Chern-Simons:  it can receive corrections from integrating 
out massive degrees of freedom.  Since we are doing a BPS computation the relevant massive objects should be BPS; here they are
just M2-branes ending on the M5-branes.

In general, integrating out such M2-branes would involve 
a complicated combinatorial structure (characterized by Young diagrams), 
since there are $M$ different M5-branes
among which the various boundaries can be distributed, as has been studied in \cite{Ooguri:1999bv,Labastida:2000yw}.  We will
be mostly interested in the case $M=1$, where the structure simplifies dramatically.

How do these states contribute to the partition function?  We first recall how the closed M2 branes
contribute in the closed string context \cite{Dijkgraaf:2006um}:  One considers the gas of M2 branes
bound to the TN geomery, and take into account how each mode of the M2 branes constributes to the
partition function. The same idea works for the open string setup, with the only difference
that the open M2 branes are restricted to lie on a 2d cigar-like subspace of TN, where the M5 brane occupies.

So let us consider a single M2-brane, ending on some cycle $\gamma \subset L$.  Upon dimensional reduction,
this M2-brane produces a quantum field $\Phi$ in the three dimensions $(S^1 \times C)_q$, 
with (left) spin $s$.
At fixed ``time'' (coordinate along $S^1$), any configuration $\Phi(z_1) = \sum_n \Phi_n z_1^n$ 
is BPS.  Each mode $\Phi_n$ thus gives an oscillator creating states in the BPS Hilbert space.
First, recall that the partition function is a {\it twisted} trace over the BPS Hilbert space, because of the rotation by $z_1 \to q z_1$.
This rotation transforms $\Phi_n$ by a factor $q^{n + s + 1/2}$.
Moreover, the oscillator $\Phi_n$ is charged, which also affects its path-integral contribution, as follows.
The $B$ field on the M5-brane here is a 2-form on $L\times MC_q$.  Reduction to zero modes along $L$
gives $b_1(L)$ $U(1)$ gauge fields; in particular, integration over any specific 1-cycle
$\gamma \subset L$ gives a specific linear combination of gauge fields in $MC_q$, which we call
$A_\gamma$.  Denote its holonomy on $S^1$ as $e^{i \theta_\gamma}$.
Then a state containing an M2-brane ending on $\gamma$ 
is weighted by a factor $\cu_\gamma = \exp(i \theta_\gamma$).  So altogether $\Phi_n$ creates a state whose path-integral contribution is 
weighted by a factor $q^{n + s + 1/2} \cu_\gamma$.

The total contribution from our M2-brane is thus\footnote{Note that the top component
of the multiplet has spin $s+{1\over 2}$ which determines if we have the partition functions
of fermionic or bosonic type characterized by the $(-1)^{2s}$ in the above formula.}
\begin{equation} \label{bps-contribution}
O(\gamma, s) = \prod_{n=0}^{\infty} (1-q^{n+s+{1\over 2}} \cu_\gamma)^{(-1)^{2s}}.
\end{equation}
(Note that this is essentially the quantum dilogarithm.)
If we reduce on $S^1$ to go back to Type II,
the $\cu_\gamma$ appearing here is identified with the holonomy around $\gamma$ of the $U(1)$ gauge field on the D4-brane.  
Our naive action for these light modes was just Chern-Simons on $L$; 
\eqref{bps-contribution} gives an {\it operator} which gets inserted as a correction
into the Chern-Simons path integral.

Introduce an index $\alpha$ to keep track of the contributions from different M2-branes.  Each one contributes 
an operator $O^\alpha(\gamma, s)$ of the form \eqref{bps-contribution}, with $s$ and $\gamma$ depending on $\alpha$.
Then altogether what we have found is
\begin{equation} \label{cscorr}
Z^{open}_{top} = Z_{M-theory}(X) = \big\langle \prod_{\alpha} O^{\alpha}(\gamma, s) \big\rangle
\end{equation}
where $\langle \cdots \rangle$ denotes the correlation function of
operators in the Chern-Simons theory on $L$.

\subsection{The case of one M5-brane on $\Sigma \times S^1$} \label{one-M5}

The case of main interest for this paper is when we only have $M=1$ M5-brane, and our
Lagrangian submanifold has the topology 
$$L=\Sigma \times S^1,$$
where $S^1$ is non-contractible inside $K$.  In such a situation each holomorphic M2-brane $\cc_\alpha$ which ends on $L$
is bounding some cycle of $\Sigma$, and sitting at a point $t_\alpha \in S^1$ (parameterize $S^1$ by $0\leq t\leq 1$). 

Viewing $S^1$ as the ``time'' direction for the Chern-Simons
theory, we can compute \eqref{cscorr} in the Hamiltonian formulation.
If we are dealing with the rational case $q^N=1$, i.e. $U(1)$ Chern-Simons theory at level $N$, 
then the Hilbert space $\ch(\Sigma)$ has dimension $N^g$, where $g$ is the genus of $\Sigma$.
Let $\gamma_i \in H_1(\Sigma, \Z)$:  then the standard quantization of Chern-Simons \cite{Witten:1989hf} gives
\begin{equation} \label{op-algebra}
\cu_{\gamma_1} \cu_{\gamma_2}=q^{\langle \gamma_1,\gamma_2\rangle} \cu_{\gamma_2} \cu_{\gamma_1}.
\end{equation}
So $\ch(\Sigma)$ is a representation of this algebra, sometimes referred to as the `quantum torus algebra'.

Then \eqref{cscorr} becomes
\begin{equation} \label{top-trace}
Z_{M-theory}(X) = \Tr \, T \left( \prod_{\alpha} O^{\alpha}(\gamma,s)(t_{\alpha}) \right)
\end{equation}
where $T$ denotes the time-ordered product.
This trace does not depend on the precise values of the $t_\alpha$, but it does generally depend on their ordering:
the reason is that if $\langle \gamma_1,\gamma_2 \rangle \not=0$ then
$$[O(\gamma_1, s_1), O(\gamma_2, s_2)] \neq 0.$$

\section{BPS states and R-twisting} \label{rtwist}

In this section we review and extend the construction of \cite{Cecotti:2009uf}.
We will also get our first payoff:  a prediction relating the spectrum of BPS states
in certain $\cn=2$ theories to the spectrum of R-charges of relevant operators 
at the conformal point.

\subsection{Our setup}

Consider M-theory on flat space
$$
\C^2_{x,y} \times (\C_z \times \R_p) \times \R^4
$$
(subscripts give the coordinates we will use on the space),
with an M5-brane wrapped on the locus
$$
\Sigma \times \{ z = 0, p = 0 \} \times \R^4
$$
where $\Sigma$ is a (non-compact) Riemann surface
$$\Sigma = \{f(x,y) = 0\} \subset \C^2_{x,y}.$$
This gives an $\cn=2$ theory in the last $\R^4$ \cite{Klemm:1996bj,Witten:1997sc}, where
$\Sigma$ is the Seiberg-Witten curve, and $\lambda = y dx$ is the Seiberg-Witten differential.
In what follows we are going to use the topological string as a way of getting information about this $\cn=2$
theory.

First, consider compactifying on two circles, 
thus replacing $\R^4$ by
$$\R^4 \rightsquigarrow S^1\times S^1 \times \R^2.$$
Then further modify the geometry as follows.  Let $g$ be some symmetry
of $\C^2_{x,y}$ preserving $\Sigma$ (which hence also gives a symmetry of the $\cn=2$ theory
in $\R^4$.)  As we go around the first circle, we make a twist of
$\C^2_{x,y}$ by $g$.  We write the resulting space as 
$$
\C^2_{x,y} \times (\C_z \times \R_p) \times S^1_g \times S^1 \times \R^2
$$
(a slight abuse of notation since strictly speaking it is not a product.)\footnote{Since 
we will use this language frequently, let us spell it out a
bit more:  by $\C^2_{x,y} \times S^1_g$ we mean $\C^2_{x,y} \times [0,1]$ modulo the
identification $((x,y), 1) \sim (g(x,y), 0)$.  (The rest of the factors are just bystanders.)}
We still have an M5-brane on 
$$\Sigma \times \{ z = 0, p = 0 \} \times S^1_g \times S^1 \times \R^2.$$

Let us view the last $S^1$ as the small ``M-theory circle.''  Then reducing to Type IIA we get
$\C^2_{x,y} \times (\C_z \times \R_p) \times S^1_g  \times \R^2$, 
with a D4-brane wrapping $\Sigma \times \{ z = 0, p = 0 \} \times S^1_g \times \R^2$.
Now comes the surprising move:  we divide our space up into $6+4$ in an unusual way.
Write 
$$K = \C^2_{x,y} \times \R_p \times S^1_g,$$ leaving
$\C_z \times \R^2$ as the remaining $4$ dimensions.  Our D4-brane now wraps
the product of $L = \Sigma\times S^1_g \subset K$ and $\{z = 0\} \times \R^2$.
We do not yet specify $g$ or the Calabi-Yau structure on $K$.

\subsection{Topological A model}

So far we have arrived at a threefold $K = \C^2_{x,y} \times \R_p \times S^1_g$
with the subspace $L = \Sigma\times S^1_g$.  We now consider the topological A model on $K$, with a brane
on $L$.

As we reviewed in Section \ref{one-M5}, the topological partition function should have an expression
of the form
\begin{equation} \label{ztrm}
Z_{top}^{open}(K, L) = \Tr\,M
\end{equation}
where 
\begin{equation}
 M = T(\prod_\alpha O^\alpha)
\end{equation}
and the $O^\alpha$ are the instanton corrections.

If we choose $g = 1$, then the story would be particularly trivial:
the instanton corrections would actually vanish (one way to understand this
is that in the physical setup we would get higher supersymmetry here), so we would
have $M = 1$.

Now suppose that $g$ is nontrivial but has finite order,
$$g^r=1.$$
In this case the instanton corrections are nontrivial, so $M \neq 1$.  Twisting by $g^k$
($k \in \Z$) instead of $g$ similarly defines an operator $M_k$.  Since $g^r = 1$,
we have $M_r = 1$.
On the other hand, we can view the $g^k$-twisted geometry as obtained by gluing together $k$ copies of the
$g$-twisted geometry, and our computation of the instanton corrections was purely local in the time direction,
so
$$M_k = M^k.$$
In particular, it follows that
$$M^r = 1.$$

\subsection{SCFT}

So far we have not chosen a specific $g$.  Now let us specialize
to the case where $\Sigma$ is singular and our theory in $\R^4$ is actually
an $\cn = 2$ SCFT.  We will consider two particular cases of interest:  $g$ 
may be an appropriate element of the R-symmetry group of the SCFT, or a certain `square root' thereof --- 
or more generally a fractional power, when there are extra symmetries.

For concreteness, consider the case
$$f(x,y)=y^m - x^n.$$
It is generally believed that these examples give rise to 4d SCFTs.  
The case $(m,n) = (2, 3)$ and its generalization to $(2, n)$ are the original
SCFTs studied by Argyres-Douglas \cite{Argyres:1995jj}.  In this case, following \cite{Argyres:1995xn} 
we should assign R-charges to the coordinates $(x, y)$, in such a way that $f$ is homogeneous 
(else we will not get a symmetry) and
$dx\,dy$ has charge $1$.  This is because $d\theta$ has R-charge $-1/2$ and so
the prepotential ${\cal F}$ has R-charge $2$ (so that $\int d^4\theta {\cal F}$ has R-charge
0), which implies that
the BPS masses  related to $a, a_D=\partial{\cal F}/\partial a$, given by integrals of $ydx$,
have R-charge 1.
These conditions fix the R-charges as
$$[x]={m\over (n+m)}, \qquad [y]={n\over m+n}.$$
So we will take $g$ to act by
$$(x, y) \to (\omega^m x, \omega^n y)$$
where $\omega^{n+m}=1$.
Then, writing 
$$
\zeta = e^{p + i \varphi}, 
$$
$K$ is a $\C^2$-bundle over $\C^\times_\zeta$, 
{\it locally} identified with $\C^\times_\zeta \times \C^2_{x,y}$, with the transition function
$$
(\zeta,x,y) \sim (e^{2\pi i} \zeta,\omega^m x, \omega^n y).
$$

Now we can specify the Calabi-Yau structure of $K$.  We choose local complex coordinates to be
$$(w_1 = x+{\overline y}, \  w_2=y-{\overline x}, \ \zeta)$$
with the holomorphic 3-form
$$ \Omega = d \zeta \, d w_1 \, d w_2 $$
and K\"ahler form 
$$ k = i \frac{d\zeta d{\overline \zeta}}{\zeta{\overline \zeta}} +i\, dw_i\, d{\overline w}_i. $$
Note that even though  $w_i$ are not global coordinates, $k$ is globally defined
(because $dw_i d{\overline w_i} = dx dy + d \bar{x} d\bar{y}$ which is invariant under $g$.)

One can check directly that our brane, given locally by
$$
L = \Sigma \times \{ \lvert{\zeta\rvert} = 1 \} \subset \C^2_{x,y} \times \C^\times_\zeta,
$$
is Lagrangian as it should be.

One can check that the dimensions and R-charges of the CFT operators are all integral multiples of $1/(n+m)$. 
This can be seen by using the assumption that adding any operator to the action will deform
the SW curve.  Consider an operator ${\cal O}_\alpha$.  Deforming the 4d action by
$$\int d^4\theta\sum a_\alpha {\cal O}_\alpha,$$
will deform the SW curve in a way depending on $a_\alpha$.  Using 
the dimensions $[x]$, $[y]$ we can read off the R-charge of $a_\alpha$,
then also determines the R-charge of ${\cal O}_\alpha$ using $[{\cal O}_\alpha] = 2 - [a_\alpha]$.
Consider for example the cases $(2,n)$.  Let us start with the $(2,3)$ case.
Then the most general deformation we will have is given by
$$y^2-x^3+g(x,y)=0.$$
Using the fact that $[dx\,dy]=1$, we can assign dimensions
$[y]=3/5, [x]=2/5$.    Note that this means the coefficient of the constant term $1$ in $g$  has dimension $6/5$,
and that of $x$ has dimension $4/5$.  These two are dual:  the addition of $1$ to $g$ corresponds to vev of a field
of dimension $6/5$ and the term linear in $x$ corresponds to the dimension of the parameter $t$ which couples
to.  Similarly monomial $a_{k,l}x^ky^l \subset g(x,y)$ will have a dimension
$6/5$ which means that $a_{k,l}$ has dimension
$$[a_{k,l}]={6\over 5} -{2k+3l\over 5}$$ 
which therefore implies that this should be a mass parameter which couples to a field ${\cal{O}}_{k,l}$
$$\int d^4\theta \sum a_{k,l} {\cal O}_{k,l}$$
of dimension
$$[{\cal O}_{k,l}]=2-\left({6\over 5}-{2k+3l\over 5}\right)={2k+3l+4\over 5}.$$ 
However, not all $l,k$ are independent.  In fact we can get rid of a lot of them by field redefinitions.
The symmetries of the theory are those which are compatible with the SW
differential, \textit{i.e.}, they should preserve $dx \wedge dy$, \textit{i.e.}\! 
they are arbitrary symplectic transformations.
Let $f(x,y)$ be any function and use it to generate symplectic transformations
on $x, y$ by the usual Poisson bracket.  Thus the most general transformation which should be viewed
as trivial is given by
$$\{f,y^2-x^3\}=2y {\partial f\over \partial x}+3x^2{\partial f\over \partial y}=0$$
If we take $f(x,y)=x^my^n$ this implies that 
$$2my^{n+1}x^{m-1}+3nx^{m+2}y^{n-1}=(2my^2+3nx^3)x^{m-1}y^{n-1}=0$$
This means that a basis for the chiral rings of this model correspond to the mass parameters in front
of the monomial given by (using the shift vectors in the monomial degrees by $(-3,2)$ and $(3,-2)$)
$$x,x^3,x^4,x^6,x^7,...,x^r,...$$
where $r=3k-1$ is eliminated.  In the above we did not start with 1 because
that is already the dual vev to $x$.\footnote{Note that this is the same as the space
of physical fields of $A_2$ minimal model coupled to topological gravity in 2d \cite{Li:1990ke}
(it would be interesting to ask if the correlations of that 2d theory have any connection
with the 4d CFT correlators).}
 Thus the dimension of the corresponding
chiral operators, as discussed above is given by 
\begin{equation}\label{adcase}
6/5,10/5,12/5,16/5,...,(2r+4)/5,...\quad r\not= -1 \quad mod \ 3
\end{equation}

For the $y^2=x^n$ models the generalization of these dimensions are (for $n$ odd):

$${1\over n+2}(n+3,n+5,..,n+2k+1,...), \ {\rm except} \ k=(n+1)/2  \mod  n$$
(where the first $(n-1)/2$ terms are the relevant ones (analog of 6/5), and the rest are descendant).
A similar expression works for $n$ even:
$${1\over n+2}(n+4,n+6,..,n+2k+2,...), \ {\rm except} \ k=n/2  \mod  n$$
Thus we see that for $n$ odd the dimension of R-charges are an integral multiple
of ${1\over n+2}$ and for $n$ even, and integral multiple of ${2\over n+2}$.
In other words we learn that the monodromy operator $M^r=1$ where
$$r=(n+2) \quad n={\rm odd}$$
$$r={(n+2)\over 2}  \quad n={\rm even}.$$
Similarly one can extend these to the more general $(n,m)$ theories.
Let $d=gcd(m,n)$.  Then  we find $r=(n+m)/d$.

A specially interesting class of theories correspond to where the M5 brane has
an $ADE$ type singularity.  For the $A_{n-1}$ type, corresponding to $(2,n)$, we have already seen that
$r=n+2$ for odd $n$ and $r=(n+2)/2$ for even $n$.  We can easily generalize the above analysis for the
dimensions $[x], [y]$ and determine the R-charges for the $D$ and $E$ series:  
$$D_n:\ x^{n-1}+xy^2=0$$
$$E_6: \ x^3+y^4=0$$
$$E_7: \ x^3+xy^3=0$$
$$E_8:\ x^3+y^5=0.$$
For the $D_n$ case, we find the common denominator is
$r=n$ for $n$ odd and $r=n/2$ for $n$ even.  Similarly, for the $E$ series we find
$$r_{E_6}=7$$
$$r_{E_7}=5$$
$$r_{E_8}=8.$$

\subsection{Deforming away from the SCFT point} \label{deforming}

We cannot compute the topological partition function directly in the above setup, 
because $\Sigma$ is singular.  We would like to deform away from the conformal point, 
replacing $\Sigma$ by
\begin{equation} \label{deformed-sw}
\tilde\Sigma = \{ y^m-x^n+\sum_{0 \le k < n,\,0 \le l < m} c_{k,l} x^k y^l = 0 \} \subset \C^2_{x,y}.
\end{equation}
In the four-dimensional language $c_{k,l}$ are parameters which move the theory away from 
the conformal point (Coulomb branch vevs and/or mass deformations).

Naively this deformation would not be allowed:  $\tilde\Sigma$ is not $g$-invariant, 
precisely because the R-symmetry is only present at the conformal point.
The construction of \cite{Cecotti:2010qn} motivates a way
around this difficulty:  replace $f$ by
$${\tilde f}=y^m-x^n+\sum_{0 \le k < n,\,0 \le l < m} \zeta^{{\frac{mn-km-ln}{m+n}}} c_{k,l} x^k y^l. $$
The brane $L = \{\tilde f = 0\}$ is nonsingular, so now 
we can evaluate the contributions from BPS states.
It is convenient to change variables to
$${\tilde x}=\zeta^{m\over n+m} x, \qquad {\tilde y}=\zeta^{n\over n+m} y.$$
The new ${\tilde x}, {\tilde y}$ are globally defined, and
$${\tilde f}({\tilde x},{\tilde y}) = \zeta^{nm}({\tilde y}^m-{\tilde x}^n+\sum_{0 \le k < n,\,0 \le l < m} c_{k,l}{\tilde x}^k{\tilde y}^l).$$
So at any fixed $\zeta$, $L$ looks complex-analytically like a copy of the deformed Seiberg-Witten curve $\tilde\Sigma$
from \eqref{deformed-sw}.  Moreover, at fixed $\zeta$ the K\"ahler form $k$ restricts to
$$-i\,k = dw_i\wedge d{\overline w}_i = \zeta \, d{\tilde x}\,d{\tilde y}.$$
The BPS states correspond to holomorphic curves ${\cal C} \subset K$ ending on $L$ --- where ``holomorphic''
refers to the complex structure on $K$, in which $w_1$, $w_2$, $\zeta$ are complex coordinates.
Such a holomorphic curve necessarily sits at some fixed $\zeta = e^{i t}$, has boundary on $\tilde\Sigma$, and has
\begin{equation} \label{phasecc}
\int_{\cal C} k = i\,\zeta \int_{\cal C} d{\tilde x}\,d{\tilde y} =i\, \zeta\  Z> 0,
\end{equation}
where $Z$ is the BPS central charge.  We thus see that the phase of the corresponding
 BPS charge correlates with the phase of $\zeta$, i.e. the choice of point $\theta_1$ on $S^1$.

As before, let us label the various holomorphic curves $\cc$ by the index $\alpha$; they sit at various 
$\zeta_\alpha = e^{i t_\alpha}$.  According to \eqref{top-trace} the topological partition function is
$$Z^{open}_{top}(K, L) = \Tr\,M$$
where
\begin{equation} \label{defM}
M = T\Big( \prod_{\alpha} O^{\alpha}(\gamma,s)(t_\alpha)\Big). 
\end{equation}
Furthermore, the computation of $M$ is topological and does not depend on the size of the coefficients
$c_{k,l}$ which deformed $f$ away from the conformal fixed point.  Taking the limit
$c_{k,l}\rightarrow 0$ we learn that
\begin{equation} \label{morder}
M^r=1.
\end{equation}

Now we come to our first payoff:  we ask what is the meaning of this result for the $\cn=2$ theory
on $\R^4$.  The answer is that the holomorphic curves $\cc_\alpha$ ending on $\tilde\Sigma$ 
give rise to BPS states in that theory, with charge $\gamma$, spin $s$, and phase of the central charge 
$t_\alpha$ (as follows from \eqref{phasecc}).  So the data that goes into $M$ in \eqref{defM} is simply the BPS spectrum
of the $\cn=2$ theory; and we have shown that $M$ so defined obeys the very nontrivial equation
\eqref{morder}.
This is a remarkable prediction:  it says that a particular product
of operators, built from quantum dilogarithms in a manner dictated by the BPS spectrum, 
is actually trivial\,!
Later in this paper we will check this prediction in various examples.

Finally we should admit to one gap in the above discussion.
When $\Sigma$ is singular our brane $L$ is Lagrangian.  Unfortunately, after the
deformation this is no longer the case.
It was argued in \cite{Cecotti:2009uf} that by taking a suitable limit this problem may 
be avoided (as the worldsheet configurations which detect it become infinite action).
In later sections of this paper we will propose a scheme for
preserving the spacetime supersymmetry even in the presence of this non-Lagrangian brane,
by introducing a kind of R-symmetry twist as we go around $S^1$, 
analogous to a construction performed in \cite{Cecotti:2010qn} in two-dimensional theories.
It is natural to expect that this mechanism for preserving spacetime supersymmetry also
has a manifestation on the worldsheet; in other words, there should be some way of deforming the
A model which makes our deformed brane admit a supersymmetric boundary condition.
It would be important to clarify this point.

In Appendix \ref{app:hk-ambient}, we generalize the above construction to the case where
the ambient $\C^2$ is replaced with a more general hyperk\"ahler manifold.

\subsubsection{The half-monodromy $Y$, and fractional monodromy $K$}\label{YK}
In the previous sections we studied the case where $g$ is given by the R-charge twisting.   More precisely we have
$$g=(-1)^F {\rm exp}( 2\pi i R)$$
where we have to insert a $(-1)^F$ in the path-integral in order to preserve the supersymmetry, as ${\rm exp} (2\pi i R)$
action on bosons and fermions differ by a sign.  This leads to the insertion of operator $M$ in the topological
theory setup.    We would be interested in taking a square root of this twisting.  In other words we wish to define
a twisting ${\tilde g}$ satisfying
$${\tilde g}^2=g.$$
We will take 
$${\tilde g}={\rm C}\ {\rm exp}(i\pi J_{12}) \ {\rm exp}(i \pi R)$$
where ${\rm C}$ is the charge conjugation operator, and ${\rm exp}(i \pi J_{12})$ is a $180^\circ$ degree rotation in the 2-plane, $w\rightarrow -w$ (which
we identify with the plane of the cigar geometry $C$).  The insertion of $\rm C$ in the above guarantees that
$\tilde g$ does not change the central charge $Z$ of the $\cn=2$ algebra.  This is because ${\rm exp}(i\pi R)$ will
change $Z\rightarrow -Z$ and $\rm C$ removes this action.

Note that at the level of the M5-brane in the Calabi-Yau, when we go around the
circle, the action of ${\tilde g}$ takes
$$dx\wedge dy\rightarrow -dx\wedge dy,$$
$$dw\rightarrow -dw.$$
This combined operation preserves $dx\wedge dy\wedge dw$, which is compatible with
preserving supersymmetry.

The corresponding operator in the topological string setup we will denote by $Y$.
It is the same as going around half the circle and inserting an operator $I$.
Let $S_{{1\over 2}-0}$ denote the contribution of the solitons as we go half the circle
around:
$$S_{{1\over 2}-0}=T\Big(\prod_{\alpha \in \text{half circle}} O^\alpha\Big)$$

Then we have the insertion of the operator $I$ which conjugates it to
$$I\  S_{{1\over 2}-0} I=S_{1-{1\over 2}}$$
 Then the half Monodromy operator is\footnote{\ Here and elsewhere $\sim$ stands for equality up to conjugacy.}
$$Y=S_{{1\over 2}-0}\, I\sim I\, S_{1-{1\over 2}}$$
and $M$, the full monodromy is represented by
$$Y^2=(I \ S_{{1\over 2}-0}) (I \ S_{{1\over2}-0})= S_{1-{1\over 2}}S_{{1\over 2}-0}=S_{1-0}=M$$
The structure of this operator and how $I$ acts on the topological string fields and its generalization
to fractional monodromies is discussed in section \ref{strucmon}.

While the half-monodromy operator will always exist for arbitrary $\cn=2$ theories
due to CPT symmetry, in some cases we can also have fractional monodromy operators.
This can happen if, as we deform the CFT, we can preserve a discrete subgroup of the
R-symmetry.  Suppose we have a $\mathbb{Z}_k$ discrete R-symmetry away
from the CFT point which acts on the central charge $Z$ by
$Z\rightarrow {\rm exp}(2\pi i/k)\ Z$.  In such a case the BPS spectrum and the $S$ operator can be similarly
decomposed in terms of the contribution of soliton in the pie wedges of size $2\pi/k$, and we would
expect that $M=K^k$, i.e. the we should be able to take a $k$-th root of the monodromy operator.
To preserve supersymmetry, this operation will be accompanied by a $-2\pi/k$ rotation of the cigar
about its tip.
The operators $Y$ and  $K$ will be discussed in more detail in \S.\ref{strucmon}.

\subsection{The four-dimensional perspective}

Finally let us reconsider our construction from the purely four-dimensional point of view.

We identified $\Tr\,M$ as the topological partition function $Z_{top}^{open}(K, L)$,
which is a generating function for certain amplitudes in M-theory on 
$$\C^2_{x,y} \times (\C_z \times \R_p) \times S^1_g \times S^1 \times \R^2.$$
However, as we explained
in Section \ref{embedding}, $Z_{top}^{open}(K, L)$ can also be understood directly as the partition function of
M-theory on a different background, where we replace the 4-dimensional space
$\C_z \times \R^2$ by Taub-NUT space $TN$, and additionally twist by a rotation of $TN$ as we go around
the M-theory circle.
We can describe this partition function in purely four-dimensional terms:  it amounts
to considering the original four-dimensional $\cn=2$ theory not on
$\R^4$ but on
$$ X = S^1_g\times (S^1\times C)_q $$
where now the twisting by $g$ is just interpreted as an internal $R$-symmetry twist,
rather than geometrically.

\section{A purely four-dimensional approach} \label{pure}

Let us briefly recapitulate what we have said so far.
Consider an $\cn=2$ field theory in four dimensions, obtained as the worldvolume theory
on an M5-brane whose internal part is a Riemann surface.  We argued that: 

\begin{itemize}
\item Attached to this theory 
there is a natural Hilbert space $\ch$, which is a representation of an algebra \eqref{op-algebra}
of operators $\cu_\gamma$ (both depending on an auxiliary parameter $q$).

\item There is a natural operator $M$ acting on $\ch$,
which is a product of elementary operators of the form 
\eqref{bps-contribution} corresponding to the various BPS states of the theory,
taken in the order of the phases of their central charges.

\item If the field theory has a conformal point, where the
dimensions of all relevant chiral operators are rational with denominators dividing $r$, then
$M^r = 1$.

\item The trace of $M$ is equal to the partition function of the theory considered 
on the background $S^1_g \times MC_q$.
\end{itemize}

Note that the above statements do not refer to the M5-brane, and
therefore it is reasonable to suspect that they hold generally for any $\cn=2$ theory in $d=4$.
In this section we sketch a
re-derivation of these statements for general $\cn=2$ theories using purely four-dimensional arguments.
As a byproduct, this gives an alternative proof of the wall-crossing formula which holds 
without assuming that the $\cn=2$ theory descends from an M5-brane.

\subsection{Hilbert space and operators}

We begin with a generic ${\cal N}=2$ supersymmetric gauge theory in $d=4$.    
Compactify this theory on $S^1$.  This yields a three-dimensional field theory,
which at low energies is a sigma model into a hyperk\"ahler manifold $\cm$.
As discussed at length in \cite{Gaiotto:2008cd,Gaiotto:2009hg}, $\cm$ admits important
canonically defined coordinate functions $\cx_\gamma$, which can be thought of as a kind of
complexification of the holonomies of the Abelian gauge fields around $S^1$, or more precisely as the 
vacuum expectation values of certain supersymmetric line operators wrapped around $S^1$.
From the perspective of the 3d theory, we can think of each $\cx_\gamma$ as a chiral point operator.

We next compactify this three-dimensional theory on 
the cigar geometry $C$ (with an appropriate topological twist, embedding the $U(1)$ holonomy
in the $SU(2)_R$ symmetry).  So altogether we have replaced
$\R^4$ by $S^1\times C \times \R_t$.
This compactification reduces the supersymmetry
by 1/2.  We end up with an effective one-dimensional theory on $\R_t$
with 4 supercharges, with chiral point operators $\cx_\gamma(z, t)$ ($z \in C$, $t \in \R_t$).
The operators $\cx_\gamma(z, t)$ are actually independent of $(z, t)$ (up to exact terms involving
$Q$-trivial contributions), 
thanks to the topological supersymmetry.  In particular, their ordering in $t$ is irrelevant, since they 
can be exchanged without ever colliding, by displacing them in $C$.

Next we pass to the quotient
$$Y := (S^1\times C) / G \times \R_t$$
where $G$ is a discrete cyclic group.  We will consider the case of $G$ being
finite and infinite.  For the finite case,  which we take it to be $\Z_N$, its generator acts by a shift
of $1/N$-th around the circle and at the same time rotating $C$ by $2\pi/N$ around
its tip $p$ at $z=0$.  In the infinite
case we replace $S^1$ by $\R$ and mod out by simultaneously translating $\R$ by
a shift and rotating $C$ by an angle $\theta$.  In either case  the geometry is the same
as having a $C$ fibered over $S^1$ which rotates by
$$z\rightarrow qz$$
$$q = e^{2 \pi i \theta}.$$  In other words,  $(S^1\times C) / G$ is just the Melvin cigar $MC_q$.

In Hamiltonian quantization of the theory, we obtain a Hilbert space of vacuum states on $MC_q$.
This is our desired $\ch$.

To get operators on $\ch$, we take supersymmetric line operators of the 4-dimensional theory, 
wrapped around loops in $MC_q$.  $\pi_1(MC_q)$ is cyclic, with a generator $\rho$.
Lifted to $S^1 \times C$, a generic loop in the class 
$\rho$ looks like a little arc which traverses around $S^1$ 
while going around $p$ by an arclength $\theta$.  In the rational case,
when $q^N=1$, note that $\rho^N$ lifts to the closed loop in $S^1 \times C$
which just runs around $S^1$, and its projection on $C$ can be a constant map to any point.

Wrapping supersymmetric line operators around loops in the class $\rho$ gives new loop operators 
$\cu_\gamma$.  In order to be supersymmetric these operators have to sit at the tip $p$ of $C$.   One
quick way to see this is to think of a stringy realization where they are F1 or D1 branes, which clearly
have to wrap geodesic cycles in order to minimize their energy:  the shortest arc going around $p$
is one which just sits at $p$.

In the case where the theory we consider comes from an M5-brane, the $\cu_\gamma$ should be identified
with the operators we called $\cu_\gamma$ in Section \ref{top-review}.  In particular,
as discussed there, they are also complexified in the context of topological
strings by including the K\"ahler class.
Moreover, the space we here called $\R_t$ is identified
with the time direction in the Hamiltonian quantization of the Chern-Simons theory.

In the Chern-Simons context we know that the ordering of the $\cu_\gamma(t)$ in $t$ matters:  
indeed they obey the noncommutative algebra
\eqref{op-algebra}.
How could such a noncommutative algebra arise for the $\cu_\gamma$ from the perspective
of the 4d theory?
The point is that unlike the $\cx_\gamma(z, t)$ which were labeled by points in 3-dimensional space,
the $\cu_\gamma(t)$ are labeled just by points on the line.
We can use supersymmetry to show that $\cu_\gamma(t)$
is independent of $t$, {\it except} that we cannot pass through singular configurations where
two of them collide, and now we have no room to move them around one another.
So $\cu_\gamma$ need not commute with $\cu_{\gamma'}$.

We would like to argue more precisely that the $\cu_\gamma$ obey the analog of \eqref{op-algebra}:
\begin{equation} \label{op-algebra-2}
\cu_{\gamma_1} \cu_{\gamma_2}=q^{\langle \gamma_1,\gamma_2\rangle}\, \cu_{\gamma_2} \cu_{\gamma_1}.
\end{equation}
We will show \eqref{op-algebra-2} directly in a moment, but let us first marshal some indirect evidence.
First, we have already
shown that this commutation relation arises in the case where our $U(1)^r$ theory that come from M5-branes,
but the commutation relations are an IR question and hence should be independent of 
the UV details.
Second, as a consistency check, note that the above commutation relations imply that if $q^N = 1$ then 
$\cu_\gamma^N$
commutes with all $\cu_{\gamma'}$.  This fits perfectly with our picture:  a loop operator in the class $\rho^N$ 
{\it can} be moved away from $p$, so $\cu_\gamma^N$ depends on $(z,t)$ rather than
just $t$, and hence we can reorder the $t$'s without $\cu_\gamma^N$ colliding with $\cu_{\gamma'}$.

Now we show \eqref{op-algebra-2} directly in four-dimensional terms.  Without loss of generality we will consider 
the case of a single $U(1)$ theory.  Let us first consider the
theory before dividing out by the $G$ action.  We consider the effective theory on
$C$ obtained by reducing from $4$ dimensions to $2$ along the internal space $\R_t \times S^1$. (So
we consider the Euclidean time $\R_t$ as a spatial direction; we 
can also consider replacing $\R_t$ by a circle, but this does not
change our argument below.)  Let $\phi_\gamma(t,z) = \log \cx_\gamma(t,z)$ be the complexified
holonomies along $S^1$, which as discussed before are naturally supersymmetric operators.  
From the two-dimensional point of view, we can think of them as an infinite collection of chiral fields
corresponding to different values of $t$.  Choose an electric-magnetic duality frame, so that we have 
basis elements $\gamma_e$, $\gamma_m$ and corresponding electric and magnetic holonomies $\phi_e$, $\phi_m$.
The 2d theory then contains a superpotential term of the form\footnote{This is an analog of the statement 
that 10-dimensional super Yang-Mills, when reduced to four dimensions and written in $\cn=1$ notation, 
contains a superpotential which has the form of a Chern-Simons
term \cite{Marcus:1983wb}; to see that this 
superpotential is indeed present, note that (labeling the $\R_t$ direction as $0$,
the $S^1$ as $1$, and the cigar as $23$) it would lead to the potential
$$V\propto \bigg|{\delta W\over \delta \phi_e}\bigg|^2+\bigg|{\delta W\over \delta \phi_e}\bigg|^2=\bigg({d\phi_e\over dt}\bigg)^2+\bigg({d\phi_m\over dt}\bigg)^2=|F^e_{01}|^2+|F^m_{01}|^2=|F^e_{01}|^2+|F^e_{23}|^2$$
which is part of the gauge theory action.}
$$\int d^2z\,d^2\theta\  [ W]\sim \int_C d^2z\, d^2\theta\ \bigg[\int dt\ \phi_e(t) {d\over dt}\phi_m(t)\bigg].$$
It is known that in the presence of such a superpotential, in order to preserve supersymmetry on a manifold
with boundary, we need to include bosonic boundary terms in the action \cite{Warner:1995ay}.  
This has been discussed in detail in \cite{Hori:2000ck} in the case of the cigar;
as shown there the desired boundary term is
\begin{equation} \label{boundary}
\delta S= \int_{\partial C}W=\int_{\partial C\times \R_t} \ \phi_e(t) {d\over dt}\phi_m(t).
\end{equation}
So we have found that in the theory on $\R_t \times S^1\times C$ the action includes the boundary term
\eqref{boundary}.  Let us return to the original context of the Hamiltonian quantization,
viewing $\R_t$ as the time.  Then the term \eqref{boundary}
implies that $\phi_e$ and $\phi_m$ are canonically conjugate, so
$[\phi_e, \phi_m]=i \ {\rm const}.$
To fix the overall constant we can carefully fix the constants in all the above
equivalences, or use the fact that in this case
$\cu_e $ and $\cu_m$ should commute, because the line operators are free to move on $C$.
The correct answer is the minimal one consistent with this commutation relation, i.e.
$$[\phi_e, \phi_m]=i.$$

Now let us pass to the quotient by $G = \Z_N$.  This leads to a boundary term which
is bigger by a factor of $N$,
$$\delta S=N\int_{\partial C\times \R_t} \ \phi_e(t) {d\over dt}\phi_m(t),$$
and thus to
$$[\phi_e,\phi_m]=i/N, \qquad \cu_e\, \cu_m = q\,\cu_m\, \cu_e.$$
Similarly, taking $q=e^{2\pi i K / N}$ gives the same story with $N$ replaced by $N/K$.
Finally, the irrational case (at least for $q$ a pure phase) 
can be obtained by successive approximations using the rational case.

\subsection{Partition function}

Our next step is to compactify time on a circle, and introduce the R-twisting:  so now we consider the
theory on
$$ MC_q \times S^1_g.$$
To define precisely what we mean by R-twisting, we apply the approach of \cite{Cecotti:2010qn}, to which we refer
for more details.
That paper discussed supersymmetric quantum mechanics with 4 supercharges, which is just what we have here
if we dimensionally reduce along $MC_q$ to 1 dimension.

The recipe of \cite{Cecotti:2010qn} for the partition function is, roughly, to compute it in the Hamiltonian formulation.
In fact, we have a time-dependent Hamiltonian, which includes
delta-function instanton contributions at special times.  In the IR limit, the computation is projected
to the ground states as usual; so at generic times 
we have the trivial evolution in the Hilbert space $\ch$ of ground states, and at special times $t_\alpha$ we
get operators $O^\alpha(t_\alpha)$ which mix the different ground states.
Setting
$$
M = T \left( \prod_\alpha O^\alpha(t_\alpha) \right)
$$
we then have
$$
Z = \Tr\,M.
$$
This is the formula we have been shooting for; it just remains to see why $O^\alpha$ have the form
\eqref{bps-contribution}.

The relevant instantons here are the Euclidean world-lines of BPS particles of the 4-d field
theory, going around the nontrivial loop $\rho \subset MC_q$, at some fixed time $t$.
In the R-twisted background, supersymmetry dictates that $t$ coincides with the phase of the 
central charge of the instanton in the sense of the supersymmetric quantum mechanics, 
which in turn coincides with the central charge of the BPS particle in the sense of the original
4-d field theory.  So the $O^\alpha$ correspond to BPS particles, and appear in the order of
the phases of their central charges, as they should.

To see the precise form of $O^\alpha$, first note that we actually get not just one instanton
for each BPS particle but an infinite tower of them, corresponding to the possible quantum states of the 
particles along $C$:  as in \cite{Dijkgraaf:2006um}, these correspond to holomorphic
Fourier modes $z^{n}$ on $C$ for all $n \ge 0$.  
 The contribution coming  from each such  particle with top component spin $s+1/2$
gets weighted in the path-integral by $q^{n+s+1/2}$ as well as  by
$\cu_\gamma$
  from the transformation of the wavefunction
as the particle goes around the loop.

Now we are essentially  ready to apply the machinery of \cite{Cecotti:2010qn} to determine
$O^\alpha$. 
The only small additional subtlety here is that we have multiple contributions arising at the 
same $t_\alpha$, corresponding to states of a ``gas'' of modes attached to the tip of the cigar.  
As we will argue in Appendix \ref{appaligned}, in such a case
we get a simple generalization of the result of \cite{Cecotti:2010qn}, where all the BPS 
states contribute independently, generating bosonic or fermionic Fock spaces 
depending on their spin:
$$O^\alpha = \prod_{n=0}^{\infty} (1-q^{n+s+1/2} \cu_\gamma)^{(-1)^{2s}}.$$
(The rational case is more subtle but can be obtained by taking a suitable
limit of the above formula, as we will discuss later in this paper.)
More precisely, for each $\cn=2$ multiplet of the form 
$$spin(j)\otimes (hypermultiplet)$$
we obtain the product of $2j+1$ such quantum dilogarithms, one for each value of $-j\leq s\leq j$.

This is the form we expected; so now we have completed our rederivation of the statements 
we listed at the beginning of this section.

\subsection{Some further predictions}

We can make a number of additional predictions/conjectures about the trace of $M$, 
based on its path-integral interpretation.

\subsubsection{Rational case}

Let us first discuss the case $q^N = 1$.  We have been studying the path integral on the geometry
$$MC_q \times S^1_g$$
which is non-compact.  Thus in order to define $\ch$ properly, and hence $\Tr\,M$,
we will have to specify the boundary conditions for the fields
on $MC_q$.  A convenient way to do this is to do it ``upstairs'', i.e. pass from $MC_q$ to the $N$-fold cover $C \times S^1$
(undoing the quotient), then reduce on $S^1$, and fix boundary conditions in the resulting
effective 3d theory.  Recalling that this effective theory is a sigma model into the
hyperk\"ahler space $\cm$, let us choose Dirichlet boundary conditions specified by a
{\it point} of $\cm$.  This amounts to fixing the values of all the coordinate functions
$\cx_\gamma$, or in the language of the quotient, to fixing the values of $\cu_\gamma^N$.

Now suppose we are at the conformal point, so that
the twisting by $g$ corresponds literally to twisting by the $R$-charge.  $g$ is realized
as a symmetry of $\cm$ (preserving the full hyperk\"ahler structure).  
If we choose our boundary condition
to correspond to a point which is not a fixed point of the $g$-action, then the path integral
should simply vanish
(by the arguments of Witten on twisting by an operator
which preserves supersymmetry \cite{Witten:1982df}).
Thus, in order to get a non-vanishing
partition function, 
we will need to choose a boundary condition corresponding to a point of $\cm$ which is {\it fixed}
under the $g$-action.  In the examples we will see how this prediction is realized.

\subsubsection{General $q$}\label{subsub:generalq}

Suppose now we let $q$ be a more general complex number
(and it turns out to be convenient to take it $|q|\leq 1$).
In this case the story becomes rather interesting.

How should the partition function $Z(q)$ look?  Let us view our setup as the compactification
of the ${\cal N}=2$ theory on $C$ down to $T^2$, where we put twists
$(q,g)$ along the two circles of $T^2$:  as we go around one circle we map
$z\rightarrow qz$ in $C$, and as we go around the other circle we do the internal R-twisting.
Since our path-integral computation is supersymmetric, it should localize to configurations which are
constant along $T^2$ and hence invariant under both twistings.  In particular, the only field configurations 
$\Phi$ on $C$ that contribute should be the ones which satisfy
$$\Phi(z)=\Phi(qz).$$
This looks a bit as if we are replacing $C$ with a torus
with modulus $q=\exp(2\pi i \tau)$.
It is then natural to ask whether the partition function
$$Z_Y(q) = \Tr\,M(q)$$ 
should be expected to be a modular function of $q$.  At least it should be
invariant under $\tau \rightarrow \tau+1$; in examples we will find that this is true, up to an overall factor.
However, symmetry under $\tau \rightarrow -1/\tau$ is not at all obvious.
The reason for this is that the field configurations we are considering are really living on 
the whole cigar $C$, 
there is  no symmetry between the A--cycle of the torus, $\gamma_A\equiv \{z\rightarrow \exp(2\pi i)z\}$, and the B--cycle,
$\gamma_B\equiv \{z\rightarrow q z\}$, because the field configurations are inherited
from the ones on cigar.
 So, in general, $\mathrm{Tr}\, M(q)$ will not be modular invariant, nevetheless it should be close
to one, because, were it not for boundary effects at the origin or the infinity of the cigar, it would
have been modular.  In the examples we will find that, up to multiplication by $q^c$ for some $c$, we
obtain objects which are modular with respect to the level $r$ subgroup of $SL(2,\mathbb{Z})$ usually denoted as $\Gamma_1(r)$, where $r$ is the order of the monodromy operator.  We do not have a general explanation
of this fact, apart for the relation with the RCFT models to be discussed in the next subsection.
 
One can also consider starting from general $q$ and taking the limit $q \to 1$.
In this case at least formally it looks like we are going back to the rational case, 
where (as we just discussed) the $\cu_\gamma$ should be localized near the fixed point locus of the R-symmetry
action.  In particular we would expect that when we compute
$${\rm Tr}\ \cu_\gamma M(q)$$
and take the $q\rightarrow 1$ limit, the path-integral should vanish unless $\langle \cu_\gamma\rangle$ is at the fixed
points of the R-symmetry action.
 We will see this happening explicitly in the examples.

\subsubsection{Connections with RCFT  Models}

In both the rational and irrational cases above we have indicated that the fixed points of 
the R-symmetry action play an important role.
In fact, looking at them more closely in the examples to follow, we will find a much 
richer story:  the algebra of functions on the set of R-symmetry invariant boundary
conditions is naturally identified with the Verlinde algebra of a 2-dimensional rational CFT!  
Evaluating these functions on the fixed points thus corresponds to {\it diagonalizing} the Verlinde algebra.
We note that the result 
is reminiscent of a construction given in \cite{Hori:2000ck} in the context of 
minimal Landau-Ginzburg models.
In the irrational case with $|q|<1$, we will find another connection to the same RCFT's:  namely, computing
$\Tr\, M(q)$, or its fractional powers,  will give their characters!

We do have a partial explanation for why and which  RCFT's appear for us, at least in some examples, 
based on various string dualities.
We will make these connections more precise in
Section \ref{speculate}, after we have presented the examples in the following sections.

\subsubsection{Action of the monodromy on line operators and quantum Frobenius property}

As we take the line operator around the time circle, it does not come back to itself.  It will
get conjugated by the evolution operator, which is represented by the monodromy:
$$\cu_\gamma \rightarrow M(q)^{-1}\,\cu_\gamma\, M(q)$$
In the limit that $q\rightarrow 1$ we call this the classical action of the monodromy on the line
operators. 

Note that the path-integral description of this computation leads to a prediction:
The operator representing $\cu_\gamma$ in the irrational case in the limit $q\rightarrow 1$
is untwisted and can be deformed away.  In particular that can be identified in the rational
case $q^N=1$ with the operator $\cu_\gamma^N$.  Therefore we predict that
{\it the action of} $M(q\rightarrow 1)$ on $\cu_\gamma$ is the same as the action of
$M(q)$ on $\cu_\gamma^N$ when $q^N=1$.  This turns out to be a highly non-trivial
fact and is known as the `quantum Frobenius property' in the context of cluster
algebras and quantum groups \cite{MR2567745}. In particular, it gives a conceptual unification of many properties observed in exactly solvable models, see \textit{e.g.}\! \cite{baxter}. In this paper we shall discuss the quantum Frobenius property in detail, in particular for its connections with the Verlinde algebra of RCFTs.

\section{Structure of the quantum monodromy}\label{strucmon}

In the previous sections we have seen that the monodromy operator $M$ has the form
\begin{equation}\label{whatM}
M= T\prod_{t_\mathrm{BPS}=0}^{2\pi} O(\gamma,s,q)(t_\mathrm{BPS})
\end{equation}
where the time--ordering corresponds to ordering in the phase $e^{i\, t_\mathrm{BPS}}$ of the central charge of the BPS states.
We write
\begin{equation}\label{whatO}
 O(\gamma,s,q)= \Psi(q^s\, \cu_\gamma; q)^{(-1)^{2s}}
\end{equation}
where the function $\Psi(x; q)$ is the \textit{quantum dilogarithm}\footnote{\ Our definition of the quantum dilogarithm differs slightly from other definitions in the literature. Explicitly, $\Psi(x;q)=\Psi(x/\sqrt{q})$, where $\Psi(\cdot)$ is the function defined in ref.\cite{Faddeev:1993rs}
and $\Psi(x;q)=\boldsymbol{\Psi}_{q^2}(-x)^{-1}$, where $\boldsymbol{\Psi}_q(\cdot)$ is the function defined in ref.\cite{MR2567745}.}.
Quantum dilogarithm is uniquely characterized by the $q$--difference equations
\begin{equation}\label{qdiff}
 \Psi(qx;q)=(1-q^{1/2}x)^{-1}\, \Psi(x;q),\qquad \Psi(q^{-1}x)=(1-q^{-1/2}x)\,\Psi(x;q),
\end{equation}
and normalization condition $\Psi(0;q)=1$. This implies the general identity
\begin{equation}\label{genidentity}
 \Psi(x,q^{-1})=\Psi(x,q)^{-1},
\end{equation}
since both sides satisfy eqn.\eqref{qdiff}. If $q=\exp(2\pi i\tau)$ with $\tau$ in the upper--half plane, we can write the solution to eqn.\eqref{qdiff} in terms of a (convergent) infinite product
\begin{equation}\label{infprodd}
 \Psi(x;q)=\prod_{k=0}^\infty\big(1-q^{k+1/2}x\big).
\end{equation}
If $q$ is a root of unity (as in the physical Chern--Simons theory) the solution to eqn.\eqref{qdiff} 
is slightly subtler and is discussed in \S.\,\ref{sec:quantFrob}.

In eqn.\eqref{whatO} $\cu_\gamma$ is an operator labelled by a point $\gamma$ in some lattice $\Gamma$ equipped with a skew--symmetric integer valued pairing $\langle\cdot,\cdot\rangle$ so that the commutation relation between the $\cu_\gamma$'s is given by eqn.\eqref{op-algebra-2}. We shall refer to the non--commutative algebra generated by the $\cu_\gamma$'s with the relations
\eqref{op-algebra-2} as the \textit{quantum torus algebra}. 

In the physical CS theory the $\cu_\gamma$'s are unitary and $q$ is a root of unity. In this case, the quantum torus algebra has an anti--automorphism given by Hermitian conjugation
\begin{equation}
 \cu_\gamma\mapsto \cu_\gamma^{-1}\equiv \cu_{-\gamma},\qquad q\mapsto q^{-1}.
\end{equation} 
This transformation is an anti--automorphism of the torus algebra even for $|q|\neq 1$, although it is not Hermitian conjugation any longer. We shall denote it by a dagger.
\medskip

The form of the operator $M$, eqn.\eqref{whatM}, is further restricted by PCT. Indeed, the BPS states with phase $t_\mathrm{BPS}+\pi$ are the PCT conjugates of those with phase $t_\mathrm{BPS}$. Write $S(t^\prime,t)$ for the time--ordered product
\begin{equation}
 S(t^\prime,t)= T \prod_{t_{\mathrm{BPS}}=t}^{t^\prime} \Psi(q^s\,\cu_\gamma;q)^{(-1)^{2s}}.
\end{equation}
Then PCT relates $S(t^\prime+\pi,t+\pi)$ with $S(t,t^\prime)$.  $S(t^\prime,t)$ is an element of the (quantum version of the) Kontsevich--Soibelman group \cite{ks1}, and, as discussed in \cite{Cecotti:2010qn} for the $2d$ case, the map $S(t^\prime,t)\rightarrow S(t^\prime+\pi,t+\pi)$ should be a group homomorphism which, in the case of physical CS, is induced by PCT and hence inverts the signs of all charges. The Hermitian conjugation $\dagger$ is an \emph{anti}--automorphism, so, just as in $2d$, we have to compose $\dagger$ with the inverse to get a true group automorphism
%
\begin{equation}
 S(t^\prime+\pi,t+\pi)= \big(S(t^\prime,t)^\dagger\big)^{-1},
\end{equation} 
and the monodromy $M$ reads
\begin{equation}\label{strucmod}
 M=S(2\pi,\pi)\,S(\pi,0)= \big(S(\pi,0)^\dagger\big)^{-1}\,S(\pi,0)
\end{equation} 
in terms of the half--circle time--ordered product $S(\pi,0)$.

One has
\begin{equation}
 (S(t^\prime,t)^\dagger)^{-1} = T\prod_{t_\mathrm{BPS}=t}^{t^\prime} \Psi(q^{-s}\cu_{-\gamma};q^{-1})^{(-1)^{2s-1}} \equiv 
 T\prod_{t_\mathrm{BPS}=t}^{t^\prime} \Psi(q^{-s}\cu_{-\gamma};q)^{(-1)^{2s}},
\end{equation} 
where we used the identity \eqref{genidentity}. Thus the net effect of PCT is just to invert the charge $\gamma\leftrightarrow-\gamma$ and the sign of spin $s\leftrightarrow -s$.\footnote{\ PCT holds in the physical theory, that is for $|q|=1$. We shall use the same expressions for $M$ (analytically continued) even if $|q|<1$, since they correspond to the expressions obtained from the topological theory in the previous sections.}

\subsection{Explicit form of the half-monodromy Y(q)}\label{explicit}
As already noted in \ref{YK} we can take a square root of $M$.  Here
we show how this is implemented.  Since all the
BPS operators come with all the allowed values of $s$ between $-j\leq s\leq j$, the net effect
of having the second half of the monodromy operator is simply to reflect $\gamma\rightarrow -\gamma$.
In particular let us define the operator $I$ acting on the quantum torus algebra by
$$I\,  \cu_\gamma =\cu_{-\gamma}I =\cu_\gamma^{-1} I.$$
Note that this is the same as the action of the charge conjugation operator $C$ on the
line operators, replacing each particle with the particle in the conjugate representation.  Then
the half-monodromy operator $Y(q)$ can be defined by
 \begin{equation}\label{halfM}
Y(q)= I\ \cdot {\rm T}\Big[\prod_{t_\mathrm{BPS}=0}^{\pi} O(\gamma,s,q)(t_\mathrm{BPS})\Big].
\end{equation}
where $Y$ is well defined up to conjugation.
From what was just said, it immediately follows that
$$Y^2(q)=M(q).$$

\medskip

\subsection{Fractional monodromy K(q)}\label{sec:frakMon}
In the above we have used the general CPT symmetries of $\cn=2$ to refine the monodromy
to a half-monodromy.  More generally, as already noted in \ref{YK}  this idea can be used to obtain fractional monodromies for special
theories with extra R-symmetries.  Suppose
we have an $\cn=2$ system which has an extra R-symmetry, say a $\mathbb{Z}_k$ symmetry generated
by an element $h$,
which acts on the $\cn=2$ central charge $Z$ by a $\mathbb{Z}_k$ action:
$$h Z h^{-1}={\rm exp} (2\pi i/k) \,Z$$
Such a symmetry also acts on the line operators $X_i$ by an order $k$ operation, 
$$h(X_i)=h X_i h^{-1}$$
which is not universal and depends on the symmetry in question.  
Let $S(t',t)$ be the time-ordered product of all $\Psi$'s for BPS states with
phase between $t$ and $t'$.  Then one has
$$S(t'+2\pi/m,t+2\pi/m)= h S(t',t) h^{-1}$$
Let
$$S=S(2\pi/m,0), \qquad K=h^{-1}S.$$
Then it immediately follows that the full monodromy $M$ is given by
$$M= S(2\pi,2\pi(1-1/m))S(2\pi(1-1/m),2\pi(1-2/m))\cdots S(4\pi/m,2\pi/m) S(\pi/m,0)$$
$$= [h^{k-1} S h^{1-k}][h^{k-2} S h^{2-k}]\cdots [h S h^{-1}] S= (h^{-1} S)^k,$$
that is,
$$M=K^k.$$
In other words we can take the $k$-th root of the monodromy operator
in such cases.  This operation will be useful in the context of some of the examples
that we will consider.

In the context of 4 dimensional geometry, just as in the case of half-monodromy, in order
to preserve supersymmetry, as we go around the R-twisted circle,  instead we now include the
action $h$ accompanied by rotation of the cigar around its tip by $-2\pi/k$.

\subsection{Action of BPS operators on the line operators}\label{action}

From the above expressions it is obvious that the action of the monodromy $M$ on any line operator, i.e. any operator $\co$ in the quantum torus algebra,
\begin{equation}\label{adj1}
\co\rightarrow M^{-1}\co M 
\end{equation}
can be obtained by a (time--ordered) sequence of `elementary' transformations of the form
\begin{equation}\label{adj2}
 \co\rightarrow \Psi(q^s\, \cu_\gamma; q)^{\mp 1}\, \co\, \Psi(q^s\, \cu_\gamma; q)^{\pm 1}.
\end{equation} 
In particular,
\begin{equation}\begin{split}
 \cu_{\gamma_1}&\: \rightarrow \Psi(q^s\, \cu_{\gamma_2}; q)^{\mp 1}\, \cu_{\gamma_1}\, \Psi(q^s\, \cu_{\gamma_2}; q)^{\pm 1}=\\
&=
\cu_{\gamma_1}\, \Psi(q^{\langle \gamma_2,\gamma_1\rangle+s}\, \cu_{\gamma_2};q)^{\mp1}\,\Psi(q^s\, \cu_{\gamma_2}; q)^{\pm 1}=\\
&=\begin{cases}
  \cu_{\gamma_1}\, (1-q^{s+1/2} \cu_{\gamma_2})^{\pm 1}(1-q^{s+3/2} \cu_{\gamma_2})^{\pm 1}\cdots (1-q^{s+\langle \gamma_2,\gamma_1\rangle- 1/2} \cu_{\gamma_2})^{\pm 1} & \langle\gamma_2,\gamma_1\rangle\geq 0\\
\cu_{\gamma_1}\, (1-q^{s-1/2} \cu_{\gamma_2})^{\mp 1}(1-q^{s-3/2} \cu_{\gamma_2})^{\mp 1}\cdots (1-q^{s-|\langle \gamma_2,\gamma_1\rangle|+ 1/2} \cu_{\gamma_2})^{\mp 1} & \langle\gamma_2,\gamma_1\rangle\leq 0.
 \end{cases}
\end{split}\label{eletransf}\end{equation}

In order to construct the action of the quantum monodromy $M$, and more generally the wall--crossing maps, one has to work out the combinatorics of many such non--commuting `elementary' operations. \textit{A priori,} for a generic $4d$ $\cn=2$ theory, this combinatorics may be quite intricate. However, there exists a large class of interesting models in which the combinatorics may be elegantly organized in terms of some recently developed mathematics known as cluster algebras.

In this special class of theories, the combinatorics of the wall--crossing jumps may be re--expressed in terms of the combinatorics of quivers, as discussed in section \ref{quivers}. It is possible that this elegant class actually \emph{exhausts} all 4d $\cn=2$ theories.

\section{Quivers, $ADE$, and TBA}\label{quivers}

Let us first recall the description of a quiver $Q$ attached to a 4d $\cn=2$ theory.  
The vertices of $Q$ are labelled by
basis elements $\{\gamma_i\}$ of the charge lattice $\Gamma$.  If $\langle \gamma_i,\gamma_j\rangle > 0$ we draw $\langle \gamma_i,\gamma_j\rangle$ arrows $i\rightarrow j$, whereas if $\langle \gamma_i,\gamma_j\rangle<0$ we draw
$|\langle \gamma_i,\gamma_j\rangle|$ arrows $i\leftarrow j$.
The quiver $Q$ is not uniquely defined; it depends on a choice of basis in the charge lattice $\Gamma$. 
Changing the basis $\{\gamma_i\}\rightarrow \{\gamma^\prime_i\}$ we get a different-looking diagram 
which encodes the same quantum torus algebra.

Quivers have occurred in the $\cn=2$ literature before:  indeed, for many theories one can write down a
supersymmetric quiver quantum mechanics which captures the spectrum of BPS states.
See \cite{Denef:2002ru} for a discussion close to our current perspective.
This quiver quantum mechanics has a gauge group for each node and 
a bifundamental matter field for each arrow, as well as (possibly, in the case
where the quiver has closed loops) a superpotential.
In other words, we can view the nodes as building blocks for BPS bound states,
and that every BPS states can be labeled by a {\it positive} linear combination of nodes
(related to the rank of the gauge group at each node). 
These quivers are generically more complicated than the $Q$ we defined above,
which by construction never 
contains 1-cycles (adjoint fields) or 2-cycles (pairs of arrows $i \to j$ and $j \to i$).
However, there is a sense in which we can reduce any quiver quantum mechanics to one
governed by a simple $Q$:  namely, we deform the superpotential of the model to give mass
to as many bifundamentals as possible.  This process eliminates all 1-cycles and 2-cycles,
by ``cancelling'' pairs of arrows running in opposite directions.
This process may change the BPS spectrum, but
since the monodromy $M$ is topological, we expect that it will not be changed.
So as long as we are only interested in $M$ we may as well reduce to the $Q$.
Note that
 we may thus associate to such a quiver an integer--valued skew--symmetric matrix $B_{ij}$, counting
the number of arrows between the nodes, taking into account orientation.  $B_{ij}$ is called the \textit{exchange matrix} of the quiver.
The quantum torus algebra \eqref{op-algebra-2} is conveniently encoded exactly by such a quiver $Q$. 

As already noted, $Q$ is not uniquely defined:  there is some freedom to choose the basis of charges.  
If we are interested
in using $Q$ to compute the BPS spectrum, though, we cannot make completely arbitrary changes of basis:  rather we
should restrict our attention to a special class of basis changes, the so-called
{\it quiver mutations}, which physically get interpreted as Seiberg duality (albeit reduced to 1 dimension),
as used e.g. in \cite{Chuang:2008aw}.
Concretely, one defines a basic mutation $\mu_k(Q)$ of the quiver $Q$ at the $k$--th vertex by performing the following two operations \cite{cluster-intro}:
\begin{enumerate}
\item reverse all arrows incident with the vertex $k$;
\item for all vertices $i\neq j$ distinct from $k$, modify the numbers of arrows between $i$ and $j$ as 
shown in the box
\begin{center}
 \begin{tabular}{|c|c|}\hline
$Q$ & $\mu_k(Q)$\\\hline
 $\begin{diagram}
  \node{i} \arrow[2]{e,t}{r}\arrow{se,b}{s} \node{} \node{j}\\
\node{}\node{k}\arrow{ne,b}{t}
 \end{diagram}$
 &$
   \begin{diagram}
  \node{i} \arrow[2]{e,t}{r+st} \node{} \node{j}\arrow{sw,b}{t}\\
\node{}\node{k}\arrow{nw,b}{s}
 \end{diagram}
$
\\\hline
$
 \begin{diagram}
  \node{i} \arrow[2]{e,t}{r} \node{} \node{j}\arrow{sw,b}{t}\\
\node{}\node{k}\arrow{nw,b}{s}
 \end{diagram}
$
& 
$
  \begin{diagram}
  \node{i} \arrow[2]{e,t}{r-st}\arrow{se,b}{s}\node{} \node{j}\\
\node{}\node{k}\arrow{ne,b}{t}
 \end{diagram}
$
\\\hline
\end{tabular}\label{boxquivemut}
\end{center}
where $r$, $s$, $t$ are non-negative integers, and an arrow $i\xrightarrow{l} j$ with $l\geq 0$ means that $l$ arrows go from $i$ to $j$ while an arrow $i\xrightarrow{l} j$ with $l\leq 0$ means $|l|$ arrows going in the opposite direction.
 \end{enumerate} 
Notice that the definition implies that $\mu_k$ is an involution:
\begin{equation}\label{muinv}
 (\mu_k)^2=\text{identity}.
\end{equation} 
Two quivers are said to be in the same mutation--class (or mutation--equivalent) if one can be transformed into the other by a finite sequence 
of such quiver mutations.

It is interesting that exactly the same type of quiver appears in the discussion of massive
$\cn=2$ theories in 2d \cite{Cecotti:1993rm}, and this is one of our motivations to conjecture a general 4d/2d correspondence in 
\S.\,\ref{4d2d}.

%



\subsection{Mutation operators and cluster algebras}

In the $4d$ case, to each vertex $i$ of the quiver $Q$ there is associated an element of the quantum torus algebra namely $\cu_{\gamma_i}$. Since $\{\gamma_i\}$ is a basis of $\Gamma$, the $\cu_i$ generate the quantum torus algebra.  
We focus on the special class of $\cn=2$ models discussed at the end of Section \ref{strucmon}, 
and make the change of notation from the complexified line operators $\cu_i\rightarrow -X_i$, which
is more customary in the context of cluster algebras.

The mutation of the quiver at the vertex $k$, $\mu_k$, corresponds to a change of basis in $\hat\Gamma$, hence a change of the associated generators of the torus algebra which explicitly reads\footnote{\ The overall power of $q^{1/2}$ in the \textsc{rhs} of eqn.\eqref{firstchbas} corresponds to the definition of the `normal ordered product', so that the \textsc{rhs}  is just $X_{\gamma_i+ [\langle \gamma_i,\gamma_k\rangle]_+ \gamma_k}$.}
\begin{align}
 &X_i\rightarrow X^\prime_i= q^{-\langle \gamma_i,\gamma_k\rangle\, [\langle \gamma_i,\gamma_k\rangle]_+/2}\, X_i X_k^{[\langle \gamma_i,\gamma_k\rangle]_+} & & i\neq k\label{firstchbas}\\
&X_k\rightarrow X^\prime_k=X_k^{-1},\label{firstchbas2}
\end{align} 
where $[a]_+=\max\{a,0\}$.

We will be considering `cluster mutations in the context of cluster algebra' \cite{MR1887642, MR2004457} (see refs.\,\cite{MR2383126,MR2132323,cluster-intro} for reviews), which we will momentarily define. For us it is especially important to consider 
the \emph{quantum} version of these algebras \cite{MR2146350,MR2567745,qd-pentagon,qd-cluster,clqd2}. 
Begin with some quiver $Q$ and the associated generators of the quantum torus algebra $\{X_k\}$.
A \emph{cluster algebra} is the algebra generated by the $X_k$ and all its mutations by the monodromy
operators, as discussed in \S.\,\ref{action}.  For various applications it is convenient
to define combinations of the above change of basis mutation of the quiver, and the action of the
BPS states on the line operators.  Namely,
the (quantum) cluster--mutation at the $k$--th vertex is defined by  \cite{MR2567745}
\begin{equation}
 \cq_k\colon X_j\mapsto \Psi(-X_k;q)^{-1}\, \mu_k(X_j)\, \Psi(-X_k;q)
\end{equation}
(and so it is --- up to the change of coordinates $\mu_k$ --- precisely our `elementary' quantum transformation \eqref{eletransf}.)
The combination of `elementary' quantum transformation and change of coordinates has the property that $(\cq_k)^2=\mathrm{identity}$, \cite{MR2567745}. In fact, using the $\cq_k$'s instead of our old `elementary' quantum transformations simplifies the
combinatorics and will make manifest their relation with the Thermodynamic Bethe Ansatz (TBA). For instance, consider a model with a quiver $Q$ whose vertices are either sinks (no outgoing arrows) or sources (no ingoing arrows). The monodromy has typically the form (up to conjugation) 
\begin{equation}\begin{split}\label{sinkssources}
 M &= 
\prod_\mathrm{sources} \Psi(-X_k^{-1};q)\,\prod_\mathrm{sinks} \Psi(-X_j^{-1};q)\,\prod_\mathrm{sources} \Psi(-X_k;q)\,\prod_\mathrm{sinks} \Psi(-X_j;q)=\\
&= \Big(\prod_\mathrm{sources} \hat\mu_k\,\Psi(-X_k;q)\, \prod_\mathrm{sinks}\hat\mu_j\, \Psi(-X_j;q)\Big)^2=\Big(\prod_\mathrm{sources}\cq_k\, \prod_\mathrm{sinks}\cq_j\Big)^2,
\end{split}\end{equation}  
where $\hat\mu_k$ is the operator in the quantum torus algebra such that
\begin{equation}
 \hat\mu_k X_j\hat \mu_k=\mu_k(X_j),
\end{equation} 
(which just inverts $X_k$ if the $k$--vertex is a source).

The `square--root' of the monodromy, $\prod_\mathrm{sources}\cq_k\, \prod_\mathrm{sinks}\cq_j$, may be identified with the half--monodromy $Y(q)$. Below we shall see how the combinatorics of the cluster--mutations will automatically organize the monodromy $M(q)$ in the form of the right power of the fractional monodromy $K(q)$ whenever the 4d SCFT has a discrete symmetry of the kind discussed in \S.\,\ref{sec:frakMon}.
Thus, the cluster--algebra formalism seems to capture some of the essential features of the physical system. 

\medskip

\subsection{Simply--laced quivers}\label{simplyla}

Assume the quiver $Q$ is simply--laced, that is $|\langle \gamma_i,\gamma_j\rangle|\leq 1$ for all $i$, $j$.
In this case, the quantum monodromy has a simpler structure.   The case of $A_{n-1}$-quiver, is already
known to arise in the context of Argyres-Douglas theories, of the form $y^2=x^n$.  We suggest that the
D and E quivers arise similarly.  For the $A_{n-1}$ case
there is one chamber of moduli space where the BPS spectrum can be described elegantly as follows \cite{Shapere:1999xr}:
the BPS states correspond simply to the nodes of the Dynkin diagram $A_{n-1}$, and the inner products between 
their charges are given by the (antisymmetric) reduction of the Cartan matrix. Thus the quiver $Q$ is the $A_{n-1}$ Dynkin graph with the arrows oriented in such a way that even (resp.\! odd) nodes are sources (resp.\! sinks). The monodromy then should have the `cluster' form \eqref{sinkssources};
the order of factors in eqn.\eqref{sinkssources} agrees with the BPS phase assignements found in ref.\,\cite{Shapere:1999xr} (for $y^2=W(x)$ with real roots for $W(x)=0$).
 
We conjecture that for the $D_n$, and similarly the three
$E$ cases,
there is a chamber where the BPS states are in 1-1 correspondence
with the quiver nodes, and the ordering of their central charge is such that the BPS
states associated to even nodes appear
together, and odd nodes together.  For the BPS spectrum in such cases,
one can say something using the approach of \cite{Kachru:1996fv,Shapere:1999xr}:
in the dual type IIB setup, 
the BPS states are special Lagrangian 3-cycles in the local CY 3-fold geometry given by
$f(x,y)+u^2+v^2=0$, and
viewing the local threefold as a $G$ singularity fibered over $\C$ (with the generic fiber resolved), 
one sees that the charges of such 3-cycles belong to the 
root lattice of $G$.  What is not established is whether there is some chamber of moduli space
in which the only charges supporting BPS states are the {\it simple} roots.  

The $A_{n-1}$ theories were also studied in \cite{Gaiotto:2009hg}.  In that
context the fact that the classical monodromy has the right order, namely $M^{n+2} = 1$ is a relatively easy
consequence of the description given there (it follows from the geometric realization of $M$
as a rotation in the plane.)  Below we will see that this extends to the quantum monodromy as well, confirming
our prediction.  We will also show that with the assumption of the degeneracy for the D and E series,
the quantum monodromy works as expected also in these cases.\footnote{We thank Bernhard Keller
for informing us that this is in fact an example of a more general statement, which holds for any bipartite
quiver whose underlying graph is a tree.}

In the quantum torus algebra we have introduced a `normal ordered product' $N[\cdots]$ 
\begin{equation}
 \cu_{\gamma+\gamma^\prime}\equiv N[\cu_\gamma \cu_{\gamma^\prime}]\equiv (q^{-1/2})^{\langle\gamma,\gamma^\prime\rangle}\cu_\gamma \cu_{\gamma^\prime},
\end{equation} 
which is associative and commutative
\begin{align}
	&N\big[\cu_\gamma\, N[\cu_{\gamma^\prime}\cu_{\gamma^{\prime\prime}}]\big]=N\big[N[\cu_\gamma \cu_{\gamma^\prime}]\,\cu_{\gamma^{\prime\prime}}\big],
	&&N[\cu_\gamma\cu_{\gamma^\prime}]=N[\cu_{\gamma^\prime}\cu_\gamma].
\end{align}

For $|\langle \gamma_i,\gamma_j\rangle|\leq 1$, eqn.\eqref{eletransf} reduces to
\begin{equation}\label{quantclas}
 \cu_{\gamma_1} \rightarrow N\big[(1-q^s \cu_{\gamma_2})^{\pm \langle \gamma_2,\gamma_1\rangle}\, \cu_{\gamma_1}\big],
\end{equation}  
where the rational map inside the bracket is the classical `elementary' symplectomorphism generated by the element $q^s\cu_{\gamma_2}$ in the classical torus algebra \cite{ks1}. Since the normal ordered product is associative and commutative, this relation between classical and quantum remains valid for any composition of such `elementary' transformation, and in particular for the monodromy $M$. Hence, for models associated to simply laced quivers, the quantum monodromy $M$ has the same action as the classical monodromy up to the replacement  $\cu_\gamma\rightarrow q^{s_\gamma}\cu_\gamma\equiv -X_\gamma$ and the normal ordered prescription on the quantum operators.

In particular, if the classical monodromy, seen as a rational map $X_\gamma\rightarrow X^\prime_\gamma$, has order $r$, the quantum monodromy must also have order $r$.   Thus the $A_{n-1}$ case works as predicted.
Below, we will provide an alternative derivation of it which applies to all the ADE cases using results known
for cluster algebras.
In order to do this we will need some machinary.

In the ADE case, let us consider the chamber where we propose that the BPS states
are in 1-1 correspondence with the nodes of the quiver.
Let $X_\ell\equiv X_{\gamma_\ell}$ ($\ell=1,2,\dots, m$) be the operators associated to a basis of $\Gamma$. Write
\begin{displaymath}
R_k\colon  X_\ell \rightarrow R_{k}(X_j)_\ell 
\end{displaymath}
for the map induced by the adjoint action of $\Psi(-X_k;q)$
\begin{equation}
 \Psi(-X_k;q)^{-1}\, X_\ell\, \Psi(X_k;q)= N\big[R_{k,\ell}(-X_j)\big],
\end{equation} 
and $I$ for the inversion rational map $I\colon X_\ell\rightarrow 1/X_\ell$. Then the quantum monodromy
of a simply laced quiver in this basis acts as
\begin{equation}\label{ratmap}
M^{-1} X_\ell M = N\Big[(I\circ R_{k_n}\circ R_{k_{n-1}}\circ\cdots \circ R_1)^2\Big]_\ell=N[Y^2],
\end{equation} 
where the $R_{k_i}$ are time--ordered according to the BPS phase 
and the square stands for the reiteration of the rational map in parentheses. 
This connection between half-monodromy $Y$ and full monodromy $M$ was 
already explained in \S.\ref{explicit}.\footnote{The quantum monodromy for the non-simply laced
case can also be done using a trick known as  \emph{diagram folding}, which will
be discussed in Appendix \ref{sec:diagramfol}.}


We will first study the simplest Argyres-Douglas theory given by the $A_2$ quiver, and then generalize it
to all the ADE cases. 

\subsubsection{Example: the $A_2$ model}  

As a first example, we consider the quiver $Q_{A_2}$ whose underlying graph is the $A_2$ Dynkin diagram
\begin{equation}\label{quiA2}
 Q\colon\qquad
\begin{diagram}
 \node{1}\node{}\node{2}\arrow[2]{w}
\end{diagram}.
\end{equation} 
Thus $X_1X_2= q^{-1}X_2X_1$, while the quantum monodromy, up to conjugacy, reads\footnote{\ We use the fact that
the spin is zero for all BPS multiplets, as follows from  \cite{Shapere:1999xr}.}
\begin{equation}\label{eq:A2M}
M= \Psi(-X_1;q)\,\Psi(-X_2;q)\,\Psi(-X_1^{-1};q)\, \Psi(-X_2^{-1};q). 
\end{equation} 
The corresponding actions of $\Psi$'s on $X_i$ are given by
\begin{align}
 R_1&\colon (X_1,X_2)\rightarrow \big(X_1,X_2/(1+X_1)\big)\\
R_2&\colon (X_1,X_2)\rightarrow \big(X_1(1+X_2),X_2\big)\\
I\circ R_2\circ R_1&\colon (X_1,X_2)\rightarrow \big((X_2+1)/(X_1X_2),X_1/(1+(X_1+1)X_2))\big)\\
M&\colon (X_1,X_2)\rightarrow \big((1+X_1)X_2,X_1^{-1}\big).\label{mondpen5}
\end{align}
The map \eqref{mondpen5} associated to $M$ is a celebrated rational map appearing in many contexts. To set it in a more canonical-looking form, let us define a sequence of rational functions $u_k(X_1,X_2)$ (where $k\in\mathbb{Z}$) by iterating the monodromy transformation:
\begin{equation}
M^{-(k-1)} X_2^{-1} M^{k-1}\equiv u_k(X_1,X_2) \quad \text{(normal ordered)}.
\end{equation}
One has $u_1=X_2^{-1}$, $u_2=X_1$, and the recursion relation
\begin{gather}\label{eq:recreA2}
 u_{k+2}u_k=(1+u_{k+1}).\\
\intertext{The sequence}
\cdots, u_1, u_2, \frac{1+u_2}{u_1}, \frac{1+u_1+u_2}{u_1u_2}, \frac{1+u_1}{u_2}, u_1, u_2,\cdots
\end{gather}
repeats after $5$ steps, $u_{k+5}=u_k$. Hence the quantum monodromy has order $5$, and $M^5$ is a central element in the $A_2$ quantum torus algebra.  This is as expected for the Argyres-Douglas theory given by the singularity
$y^2=x^3$.

Define ($a=1,2$)
\begin{equation}
 Y_a(k)=\begin{cases}
         u_k & \text{if } k=a \mod 2\\
u_{k+1} & \text{if } k\neq a \mod 2.
        \end{cases}
\end{equation} 
$Y_a(k)$ is a solution to the Zamolodchikov $Y$--\textit{system} associated to the thermodynamical Bethe ansatz (TBA) for the $A_2$ solvable $2d$ model \cite{Zamolodchikov:1991et}.

The identification between the solution to the Zamolodchikov $Y$--system for the solvable model associated to a Dynkin diagram of the $ADE$ type \cite{Zamolodchikov:1991et} and the rational map whose normal ordered version gives the quantum monodromy $M$ extends to all examples as we show in \S.\,\ref{ADEgen} and \S.\,\ref{sec:pairsofD}.

Write $X_k$ ($k\in\mathbb{Z}$) for the normal--ordered quantum operator corresponding to the rational function $u_k$.   (These five functions also play a privileged role in the physics of the corresponding $\cn=2$
theories, namely they correspond to five distinguished line operators \cite{Gaiotto:2010be}.)
One has the commutation relation $X_{k+1}X_k=qX_kX_{k+1}$ and, from eqn.\eqref{eq:A2M},
\begin{equation}
 M=\Psi(-X_k)\,\Psi(-X_{k-1})\, \Psi(-X_{k-2})\,\Psi(-X_{k-3})
\end{equation} 
where the \textsc{rhs} is independent of $k$ thanks to the (quantized version of the) recursion relations \eqref{eq:recreA2}.
Using the independence on $k$, it is elementary to check
\begin{equation}
 X_kM=MX_{k-4}\equiv M X_{k+1},\qquad \text{since } X_{k+5}\equiv X_k.
\end{equation} 
The two last equations give an alternative, and more symmetric, way of understanding the action of the monodromy $M$ on the quantum torus algebra.

\subsection{The $ADE$ models in the canonical BPS chamber}\label{ADEgen}

The above analysis for the $A_2$ model may be extended to a large class of theories whose quiver is based on an $ADE$ (simply--laced) Dynkin diagram, which arise from the M5 brane having the corresponding singularity.  In a given
 theory in general one gets a full class of mutation--equivalent quivers; they correspond to different BPS chambers separated by walls of marginal stability; one passes from one to the other with repeated application of elementary mutations. The mutation--class is finite precisely for the classes of the $ADE$ quivers. In this case there is a
 `canonical' chamber in which the quiver is the Dynkin graph with only sinks and sources\footnote{\ We say that a node $k$ is a \textit{source} (resp.\! a \textit{sink}) if there are no ingoing (resp.\! outgoing) arrows to $k$.} to which eqn.\eqref{sinkssources} applies. Formulae for $M$ valid in an alternative chamber are presented in Appendix \ref{app:linearchamber}.

An $ADE$ quiver $Q$ is, in particular, a tree and hence a bipartite graph\footnote{\ $Q_0$ stands for the set of vertices of the quiver $Q$. }.  In other words we can assign a parity to each node.
Most of the following considerations hold for any such bipartite quiver. We number the vertices of $Q$ in such a way that even (resp.\! odd) ones correspond to $V_{+1}$ (resp.\! $V_{-1}$).  We have $Q_0=V_{+1}\cup V_{-1}$. We orient the quiver in such a way that the even nodes are sources, while the odd ones are sinks. Hence the exchange matrix $b_{ij}$ has the form
\begin{equation}
	b_{ij}=\begin{cases}
	\geq 0 & i\ \text{even and }j\ \text{odd}\\
	\leq 0 & i\ \text{odd and }j\ \text{even}\\
	0 & \text{otherwise.}
\end{cases}
\end{equation}
The order of the elementary factors in the quantum monodromy is first the $\Psi(q^{s_k}\cu_k;q)$ with even $k$ and then those with odd $k$ (notice that even/odd $\cu_k$'s commute between themselves, so there is no need to further specify the order). To simplify the comparison with the TBA $Y$--systems, it is convenient to set
\begin{displaymath}
X_k\equiv -q^{s_k}\cu_k. 
\end{displaymath}
Then
\begin{equation}\label{monbiparite}
	M= \Big(\!\!\!\prod_{k=0\atop\mod 2}\!\!\!\Psi(-X_k;q)\!\!\!\prod_{k=1\atop\mod 2}\!\!\!\Psi(-X_k;q)\,I\Big)^2.
\end{equation}
To simplify this expression, we enlarge our system, making the central element $q$ of the quantum torus algebra a dynamical variable (which we shall fix to its numerical value at the end of the computation). 
Then the inversion automorphism of the enlarged algebra, $I$, may be written as
\begin{gather}
I=I_{+1}\cdot I_{-1}	\\
\intertext{where $I_\varepsilon$ ($\varepsilon=\pm 1$) is the enlarged algebra automorphism}
I_\varepsilon \colon (X_k,q)\mapsto \big(X_k^{-\varepsilon (-1)^k}, q^{-1}\big),
\end{gather}
namely, $I_{+1}$ (resp.\! $I_{-1}$) inverts just the even (resp.\! odd) variables, while making $q\rightarrow q^{-1}$ to preserve the quantum torus relations. Up to conjugacy, the quantum monodromy \eqref{monbiparite} may be rewritten as
\begin{equation}
M= \Big( L_{+1}\, L_{-1}\Big)^2,	
\end{equation}
with
\begin{equation}
	L_\varepsilon = I_{-\varepsilon}\,\prod_{(-1)^k=\varepsilon} \Psi(-X_k;q)^\varepsilon
\end{equation}
where we used the identity \eqref{genidentity}. We define the \textit{classical} rational maps $\tau_\varepsilon$  by
\begin{equation}\begin{split}
	\tau_{-\varepsilon}\colon X_k &\rightarrow L_\epsilon^{-1} X_k L_\varepsilon\Big|_\mathrm{classical\ limit}=\\
&=	\begin{cases} X_k^{-1}\prod_{(-1)^j=\varepsilon}(1+X_j)^{\varepsilon\, \langle \gamma_j,\gamma_k\rangle} & \text{if }(-1)^k=-\varepsilon\\
	X_k & \text{otherwise}\end{cases}\\
	&=	\begin{cases}X_k^{-1} \prod_{(-1)^j=\varepsilon}(1+X_j)^{-C_{kj}} & \text{if }(-1)^k=-\varepsilon\\
	X_k & \text{otherwise},
	\end{cases}\end{split}
\end{equation}
where we used eqn.\eqref{eletransf}, and  $C_{kj}$ is the (symmetric) Cartan matrix for the $ADE$ Dynkin diagram
\begin{equation}
 C_{kj}=2\,\delta_{kj} -(-1)^j \langle \gamma_j,\gamma_k\rangle. 
\end{equation}

Then the quantum monodromy $M$ for the canonical $ADE$ quiver is
\begin{equation}
	M^{-1} X_k M =N\big[(\tau_{-1}\tau_{+1})^2(X_k)\big]
\end{equation}
Setting, for all $s\in \mathbb{Z}_{\geq 0}$,
\begin{equation}\label{eq:Yk(s)}
	Y_k(s)= \underbrace{\tau_{-1}\tau_{+1}\tau_{-1}\tau_{+1}\cdots \tau_{\pm 1}}_{s\ \text{times}}(X_k)
\end{equation}
we get a solution to the Zamolodchikov $Y$--system associated to the given Dynkin diagram \cite{MR2031858,Zamolodchikov:1991et, MR2004457,Frenkel:1995vx}
\begin{equation}
	Y_k(s+1)\, Y_k(s-1)= \prod_{j\neq k}\big(1+Y_k(s)\big)^{-C_{kj}}.
\end{equation}
\smallskip

It is known \cite{MR2031858,MR2004457,Zamolodchikov:1991et} that the order of the rational map $\tau_-\tau_+$ is $(h+2)/2$ if $\omega_0=-1$ and $h+2$ otherwise, where $h$ is the Coxeter number of the given Dynkin diagram and $\omega_0$ is the element of the Weyl group of maximal length. Given that the physical monodromy $M$ is the \textit{square} of $\tau_{-1}\tau_{+1}$, and that a Lie algebra has $\omega_0=-1$ iff $h$ is even and all the exponents $m_j$ are odd \cite{MR1890629}, we have the periods in table \ref{Table}.  This agrees with the predictions made
in \S.\,\ref{rtwist}.  This not only supports our conjecture for the BPS structure of the D and E series, but
it is also a confirmation of our general picture for the order of the monodromy group and its relation to R-charges.

\begin{table}
\begin{center}
 \begin{tabular}{|c|c|c|c|}\hline
algebra & $h+2$ & exponents & order $M$\\\hline
$A_n$ $n\geq 2$ & $n+3$ & $1,2,3,\dots, n$ & 
$\begin{cases}
  n+3 & n\ \text{even}\\
(n+3)/2 & n\ \text{odd}
 \end{cases}$\\\hline 
$D_n$ $n\geq 3$ & $2n$ & $n-1, 1,3,7,\dots, 2n-3$ &
$\begin{cases}
  n/2 & n\ \text{even}\\
n & n\ \text{odd}
 \end{cases}$\\\hline
$E_6$ & $14$ & $1,4,5,7,8,11$ & $7$\\\hline
$E_7$ & $20$ & $1,5,7,9,11,13,17$ & $5$\\\hline
$E_8$ & $32$ & $1,7,11,13,17,19,23,29$ & $8$\\\hline 
\end{tabular} 
\caption{\label{Table} Order of the quantum monodromy $M$ for the $ADE$ theories.}
\end{center}
\end{table}

\section{Generalization to Hypersurface CY Singularities}\label{sec:HyCY}

In Section \ref{rtwist} we focused on $\cn=2$ theories which can be
viewed as the theories on M5-branes with worldvolume $\Sigma \times \R^4$.  
The conformal points in moduli space correspond to $\Sigma$
developing singularities.  

The same theories could also be obtained as in \cite{Klemm:1996bj,Ooguri:1996wj}
by compactifying type IIB on local Calabi-Yau 3-folds of the form
$$f(x,y)+u^2+v^2=0.$$
However, this is not the most general form that a local Calabi-Yau threefold
can have, even if we restrict to hypersurfaces in $\C^4$.
In particular, we are interested in conformal $\cn=2$ theories.  Such a theory would come from a
quasi-homogeneous hypersurface singularity, and more specifically, one which can appear
at finite distance in the moduli space of a compact Calabi-Yau.  
(As previously, the
need for quasi-homogeneity follows from the existence of the 
R-symmetry at the conformal fixed point.)

So what are the possible such singularities?
The answer to this question is known \cite{MR92j:32028}
and also derived in \cite{Gukov:1999ya} (see also \cite{Shapere:1999xr}).
Consider a threefold given locally by a hypersurface
$$W(x_i)=0$$
where $i = 1, \dots, 4$.
Suppose $W$ is quasi-homogeneous, so that for some $q_i$ we have
$$W(\lambda^{q_i}x_i)=\lambda W(x_i).$$
Then this singularity is at finite distance in Calabi-Yau moduli space if and only if
$${\hat c} := 4 - 2\sum_{i=1}^4 q_i < 2, \qquad \textrm{i.e.} \qquad \sum_{i=1}^{4} q_i >1.$$
Note that any quasi-homogeneous singularity of the type $f(x_1,x_2)+x_3^2+x_4^2 = 0$
satisfies this condition (since we will have $q_3 = q_4 = 1/2$ in this case).  
This recovers the cases we already discussed.  But there are more general possibilities.
For example, we can take any pair of A-D-E singularities:  letting $G$ stand for some
simply-laced Lie algebra, and $W_G$ the corresponding quasihomogeneous polynomial, we may choose
$$W=W_G(x_1, x_2)+ W_{G'}(x_3,x_4)$$
since $W_G$ and $W_{G'}$ each separately contribute $\hat c < 1$.  So
for each pair $(G, G')$ we expect an $\cn=2$ SCFT in 4d.
These are only a subset of the possibilities; for example,
$$W = x_1^3+x_2^3+x_3^3+x_4^N$$
is not of this type for generic $N$ but still satisfies $\hat c <2$.

Let us now discuss the R-charges in these examples.  Since BPS masses are periods
of the holomorphic 3-form $(\prod_i dx_i) / dW$, we see that this 3-form must have
dimension 1, so we can write
$$[\prod_i dx_i] - [W]=1, \qquad [x_i]=a\cdot q_i, \qquad [W]=a,$$
which implies $a((\sum_i q_i) -1)=1$, so that
$$a={1\over (\sum_iq_i)-1}.$$
Now suppose
$$q_i={r_i\over d}$$
(and choose the minimal possible $d$).  Then it immediately follows that
$$[x_i]={q_i\over  (\sum_iq_i)-1}={r_i\over \sum_i r_i - d}.$$
Thus all dimensions, hence all R-charges, have denominator
$$r=\sum_i r_i - d.$$
But from our previous discussion in Section \ref{deforming} this implies that the BPS monodromy
should obey
$$
M^r = 1.
$$

For example, in the $(G, G')$ theories just described, 
we find that the denominator of R-charges involves $h+h'$, where $h, h'$ denote the dual
Coxeter numbers of $G, G'$.   This implies that the order of $M$ is at most $h+h'$.
If the numerator of the R-charges involve factors which all divide $h+h'$ the order
will be smaller, as was discussed for example in the $(A_{n-1},A_{m-1})$ case
(where we found $r=(m+n)/\gcd(m,n)$).  
We leave it as an easy exercise to the reader to find the minimal $r$ for each pair.

Since $M$ is constructed from the BPS spectrum of the theory (after deforming away
from the conformal point), $M^r = 1$ gives
a strong constraint on what that BPS spectrum can look like.
It would be desirable to check this condition by explicitly 
determining the BPS spectrum of the deformation
of the above CFT's.

For the $(G, G')$ theories we have at least some partial information.
As already noted the case $(G, G') = (A_n, A_1)$ the full answer is known \cite{Shapere:1999xr},
and we have already conjectured the form of the answer for the $(G,A_1)$ case, where
we get in some chamber one soliton for each node of $G$, whose central charge is orderered according to the
parity of the Dynkin node.
 
The natural conjecture would be that for $(G, G')$ 
the representation theory of the corresponding tensor product of quivers,
will yield this information.
In later sections (leveraging recent work on cluster algebras \cite{keller-periodicity}) 
we will use this to study the BPS spectrum for these theories and verify $M^r = 1$.

 Note that the labelling of $(G,G')$ is not unique.  For example, we have 
the isomorphism $(G,G')\sim (G',G)$, which is manifest
from the viewpoint of the CY3 fold.  In fact we have many additional such equivalences
which follow from this picture.  For example we have, 
$$(D_4,A_3)\sim (E_6,A_2),\quad (E_8,A_3)\sim (E_6,A_4),\quad ...$$
These will turn out to be non-trivial facts in the context of the associated cluster algebras!

\subsection{The gauge theory perspective}

It is natural to ask how the SCFTs we are considering can be realized 
in terms of purely four-dimensional gauge theories.  Thanks to the construction of \cite{Witten:1997sc},
we know that M5-branes on Seiberg-Witten curves $x_1^{m+1} + x_2^{n+1} + \cdots = 0$ (where $\cdots$ denotes
deformations by relevant operators) yield
points on the Coulomb branch of quiver gauge theories, where the quiver is the $A_m$ Dynkin diagram
and each node is an $A_n$ gauge theory.
More generally, quivers of $A_n$ gauge theories on affine $D$ or $E$ Dynkin diagrams
were studied in \cite{Katz:1998eq}, where local Calabi-Yau 3-fold geometries were
identified which replace the Seiberg-Witten curve.  Many of the Calabi-Yau 3-fold singularities
we are considering can be identified with special points in the moduli spaces of these theories. 
These include examples which are not of the $(G, G')$ type.  For example, the singularity
\begin{equation}\label{extra}
x_1^3+x_2^3+x_3^3+x_4^N=0
\end{equation}
can be obtained as a special limit of a certain product of $U(a_iN)$ gauge theories on an affine $E_6$ quiver,
where $a_i$ are the Dynkin indices of the $E_6$ affine Dynkin diagram \cite{Katz:1998eq}.

\subsection{The 5-brane perspective}\label{5bp}

Finally it is interesting to ask if we can reformulate the above theories in terms
of the conformal $(2,0)$ 5-brane theories.   As already noted, we can obtain the
$(A_{n-1},A_{m-1})$ theory by considering a single M5 brane on a curve
given by $\{x^n=y^m\}\subset \C^2$.   More generally any singularity of the form 
$$f(x,y)+uv=0$$
can be viewed as a single M5 brane.  These include all the pairs of the form $(G,A_1)$.
 These same theories can also be obtained in a different way from multiple M5 branes.  For example, consider
the CY singularity of $(A_{n-1},A_{m-1})$ type:  
$$x^n+y^m+uv=0.$$
This can be viewed as an $A_{m-1}:y^m+uv=a$ singularity, which is dual to $m$ M5 branes fibered over the $x$-plane:
$$y^m+uv=x^n$$
where for each $x\not=0$ the $m$ M5 branes have been split, and they
all come together at $x=0$.
It is not possible to get all the rest in terms of only M5 branes, however.  
Moreover, M5 brane corresponds
only to the A-type $(2,0)$ CFT in 6 dimensions.  We also can consider D, E 5-brane theories.
Indeed the ADE (2,0) 5-brane theories are defined by type IIB theories compactified on ADE singularities down to 6 dimensions.  From the above construction of the CY 3-fold singularities we can take three coordinates
(say the $y,u,v$)
of the singularities as defining the 5-brane theory of $G$ type and the other one (say the $x$) as defining
how the $G$ type brane fibers over the $x$-plane as we go down from 6 to 4 dimensions.
These would lead to a description of all the $(G,A)$ theories.
The other three types, $(D,D),(D,E)$ and $(E,E)$, do not seem to admit such a description.

\section{Quivers associated to pairs of $ADE$ Dynkin diagrams}\label{sec:pairsofD}

We would like to predict the form of quivers for the pairs of ADE singularities discussed in the 
previous section.  Since we can view the nodes of each of the individual ADE nodes as
corresponding to a 3-cycle in the Calabi-Yau geometry, it is natural to expect
that the quiver associated to the pair of the ADE is made up of the tensor product of the two
quivers.  We would also need to know the number of arrows between the nodes.  It turns
out that these problems are isomorphic to the 2d problem of the tensor product of two associated
$\cn=2$  theories where we identify the LG superpotential
with the defining equation of the hypersurface.  We will explain in section \ref{4d2d} why this is to be expected.  Here we will
assume this is the case and use it to construct the corresponding quiver for the $\cn=2$ theory
in $d=4$.
 Consider first an $(A_n,A_m)$ singularity type of the CY 3-fold
\begin{equation}
 W(X,Y)= X^{m+1}+Y^{n+1}+\text{lower order monomials,}
\end{equation} 
which reduces to $A_n$ in the special case $m=1$. We use $W(X,Y)$ to define a $2d$ LG model and a $4d$ $\cn=2$ gauge theory sharing the same quiver $Q_{n,m}$.

\subsection{Square and triangle products of quivers} 

We construct the quiver $Q_{n,m}$ starting from the $tt^*$ geometry of the associated $2d$ model. Of course, the quiver is not unique, and we are interested in obtaining the simplest possible quiver in its mutation--class, namely the one which makes the physical properties most manifest.
To get this canonical quiver, we use a little trick. Assume we have two $2d$ LG models with superpotentials $W_1(X)$ and $W_2(Y)$. Consider now the model with superpotential
\begin{equation}
 W(X,Y,\lambda)=W_1(X)+\lambda\, W_2(Y)\qquad \lambda\in \mathbb{C}.
\end{equation}   
The $X$ and $Y$ sectors are decoupled, and the physical Hilbert space is just the tensor product of the original ones. However, for generic $\lambda$, not all the tensors products of the BPS states of the original theories are BPS states for the diagonal supersymmetry with supercharges $Q_1^A+Q_2^A$. Indeed, the central charge of the resulting theory is related to the central charges of the original ones as $Z=Z_1+\lambda\, Z_2$.
Hence the mass and the central charge of a state $|\mathrm{BPS}_1\rangle \otimes |\mathrm{BPS}_2\rangle$ are
\begin{multline}
 \hskip 0.6cm M=M_1+M_2,\qquad Z=Z_1+\lambda\, Z_2,\ \Rightarrow\\
\Rightarrow\ M\not=|Z|\ \ \text{unless } \lambda\, \frac{Z_2}{Z_1}\ \text{is real positive}.\hskip 3.5cm 
\end{multline} 
So, if $\lambda$ is generic enough, the only BPS states (with respect to the diagonal $\cn=2$ superalgebra) are of the form
\begin{equation}
 |\mathrm{BPS}_1\rangle \otimes | k_2\rangle\quad \text{or}\quad
 |k_1\rangle \otimes |\mathrm{BPS}_2\rangle,
\end{equation}  
where $|k_i\rangle$ stands for the $k$--th  \textsc{susy} vacuum of the original $i$--th theory.

The BPS quiver $Q$ corresponding to such a generic $\lambda$ is called the tensor product quiver. However we are still free to redefine the \textsc{susy} vacua $|k_i\rangle \rightarrow \pm |k_i\rangle$, getting a different exchange matrix $B_{(k,l)(k^\prime,l^\prime)}$. Starting with the canonically oriented $G$, $G^\prime$ quivers, it is convenient to redefine the signs of vacua
as 
\begin{equation}
	|k_i\rangle \rightarrow (-1)^{k_i+i}\,|k_i\rangle,
\end{equation}
with the effect
\begin{align}
	&B_{(j,k)(j\pm 1,k)}\rightarrow (-1)^{k-1}\,B_{(j,k)(j\pm 1,k)}\label{arrred1}\\
	&B_{(j,k)(j,k\pm 1)}\rightarrow (-1)^{j}\,B_{(j,k)(j,k\pm 1)}.\label{arrred2}
\end{align}
The resulting quiver is called the \emph{square tensor product} 
of the original quivers $G$ and $G^\prime$  \cite{keller-periodicity}, written 
\begin{equation*}
G\,\square\, G^\prime.
\end{equation*} 

From eqns.\eqref{arrred1}\eqref{arrred2} we see that $G^\prime\,\square\, G$ is the \emph{dual} quiver to $G\,\square\, G^\prime$ (namely the quiver obtained by reversing all the arrows). The two quivers are in the same mutation--class; indeed one passes from one to the other by applying $\prod_{(k+l= \mathrm{odd})}\mu_{k,l}$. Physically, reversing all the arrows is equivalent to taking $q\rightarrow q^{-1}$. 
\smallskip

\begin{figure}
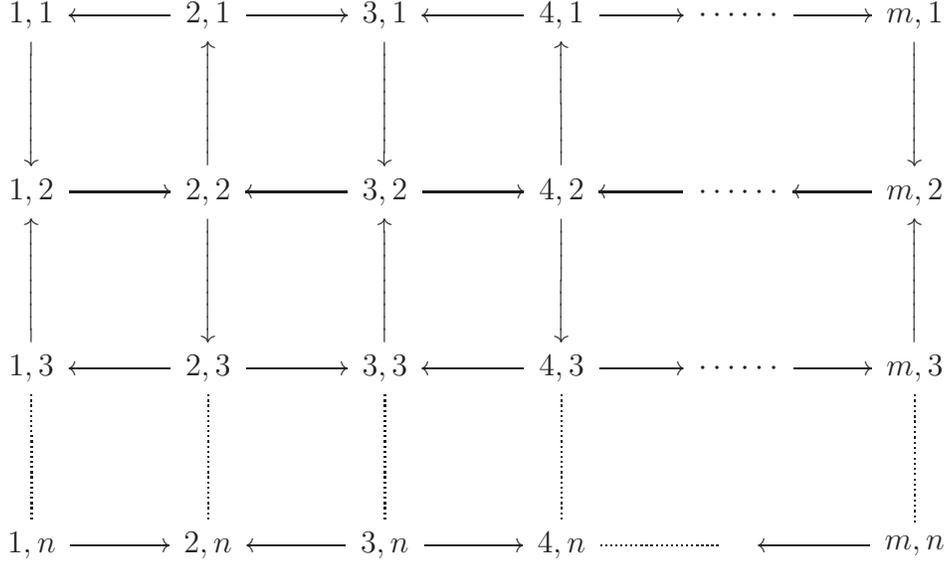

 \begin{equation*}
 \begin{diagram}
  \node{1,1}\arrow{s}\node{2,1}\arrow{w}\arrow{e}\node{3,1}\arrow{s} \node{4,1}\arrow{w} \arrow{e} \node{\cdots\cdots}\arrow{e}\node{m,1}\arrow{s}\\
 \node{1,2} \arrow{e} \node{2,2}\arrow{n}\arrow{s} \node{3,2}\arrow{w}\arrow{e} \node{4,2}\arrow{n}\arrow{s}
  \node{\cdots\cdots }\arrow{w}\node{m,2}\arrow{w}\\ 
 \node{1,3}\arrow{n} \node{2,3}\arrow{w}\arrow{e}\node{3,3}\arrow{n} \node{4,3}\arrow{w}\arrow{e} \node{\cdots\cdots}\arrow{e}\node{m,3}\arrow{n}\\
 \node{1,n}\arrow{e}\arrow{n,..,-} \node{2,n}\arrow{n,..,-} \node{3,n}\arrow{n,..,-}\arrow{w}\arrow{e} \node{4,n}\arrow{n,..,-} \arrow{e,..,-} \node{\ }\node{m,n}\arrow{w}\arrow{n,..,-}
 \end{diagram} 
 \end{equation*} 
\caption{\label{squaremnquiver} The $A_m\,\square\,A_n$ quiver.}
\end{figure}

Explicitly, for the $(A_m,A_n)$ case we may consider the superpotential
\begin{equation}\label{supprod}
 W(X,Y)= \frac{1}{2^m}\,T_{m+1}(X)+\frac{\mu^{n+1}}{2^n}\, T_{n+1}(\mu^{-1} Y),
\end{equation} 
whose $tt^*$ equations can be explicitly solved in terms of PIII transcendents \cite{Cecotti:1991me}.
For generic $\lambda\equiv \mu^{n+1}$ we get the quiver $A_m\,\square\,A_n$ represented in figure \ref{squaremnquiver}.
From the figure it is manifest that $A_n\,\square\,A_m$ differs from $A_m\,\square\,A_n$ only in the overall orientation.
\smallskip

\begin{figure}
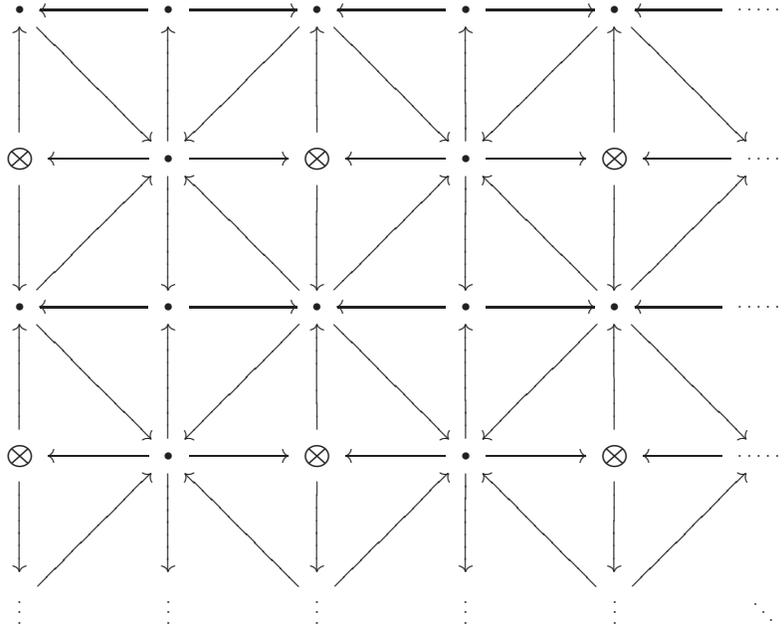

  \begin{equation*}\begin{tiny}                   
   \begin{diagram}
   \node{\bullet} \arrow{se} \node{\bullet} \arrow{w}\arrow{e} \node{\bullet} \arrow{sw}\arrow{se} \node{\bullet}\arrow{w}\arrow{e} \node{\bullet} \arrow{sw}\arrow{se} \node{\cdots\cdots}\arrow{w}\\
\node{\bigotimes}\arrow{n}\arrow{s} \node{\bullet}\arrow{n}\arrow{s}\arrow{w}\arrow{e} \node{\bigotimes} \arrow{n}\arrow{s}
\node{\bullet} \arrow{n}\arrow{e}\arrow{s}\arrow{w} \node{\bigotimes}\arrow{n}\arrow{s}\node{\cdot\cdots}\arrow{w}\\
\node{\bullet} \arrow{ne}\arrow{se} \node{\bullet}\arrow{w}\arrow{e} \node{\bullet}\arrow{nw}\arrow{ne}\arrow{sw}\arrow{se}
\node{\bullet}\arrow{w}
\arrow{e} \node{\bullet} \arrow{nw}\arrow{ne}\arrow{sw}\arrow{se}\node{\cdots\cdots}\arrow{w}\\
\node{\bigotimes}\arrow{n}\arrow{s} \node{\bullet}\arrow{w}\arrow{e} \arrow{n}\arrow{s}\node{\bigotimes}\arrow{n}\arrow{s}\node{\bullet} \arrow{n}\arrow{e}\arrow{s}\arrow{w} \node{\bigotimes}\arrow{n}\arrow{s} \node{\cdots\cdots}\arrow{w}\\
\node{\vdots}\arrow{ne}\node{\vdots}\node{\vdots}\arrow{nw}\arrow{ne}\node{\vdots}\node{\vdots}\arrow{ne}\arrow{nw}\node{\ddots}
  \end{diagram}
 \end{tiny} \end{equation*} 
\caption{\label{fig:triangle} The $A_m\,\boxtimes\, A_n$ quiver. Mutation at the nodes $\otimes$ gives back $A_m\,\square\,A_n$.}
\end{figure}

Special values of $\lambda$ will give different quivers. In particular, the quiver obtained by taking
$\lambda$ real (and orienting the resulting $3$--loops) is called the \emph{triangle tensor product} of the $A_m$, $A_n$ quivers \cite{keller-periodicity} 
\begin{equation}
 A_m\boxtimes A_n.
\end{equation} 
(see figure \ref{fig:triangle}).

A result in quiver algebras \cite{keller-periodicity} states that square and triangle tensor products of two quivers with only sources and sinks are mutation equivalent, that  is, related by a chain of $2d$ wall--crossings. From our point of view, this result is obvious, since the two quivers $G\boxtimes G^\prime$ and $G\,\square\, G^\prime$ are related by a continuous deformation of the parameter $\lambda$ in the superpotential.
\medskip

Given a pair of $ADE$ singularities $(G,G^\prime)$
we define the \emph{canonical chamber} for either the corresponding $4d$ and $2d$ theories to be the chamber in which the quiver is given by the square tensor product of the  canonical quivers for the two minimal singularities, $G\,\square\, G^\prime$.

In particular, our \textit{single} $ADE$ models of section \ref{quivers} are to be identified with the pair of Dynkin diagram
model $(G,A_1)$ which correspond to the same singularity.

\subsection{Grassmannian coordinate rings}

A remark that will be relevant below is the following: the quiver $Q_{m,n}\equiv A_m\square A_n$ is the quiver which defines a very specific cluster algebra of the geometric type, isomorphic to the \textit{homogeneous coordinate ring of the Grassmannian}
$\mathbb{G}(m,n+m)$ \cite{MR2205721}. From the geometric duality
\begin{equation}
 \mathbb{G}(m,n+m)\sim \mathbb{G}(n,n+m)
\end{equation} 
we infer a duality $(A_m,A_n)\sim (A_n,A_m)$. Moreover, comparing the corresponding singularities, we have the identifications in table
\ref{table2} for the Grassmannian cluster--algebras of \emph{finite}--type \cite{MR2132323,cluster-intro,MR2205721}
which naturally follows from the CY 3-fold description.

\begin{table}
 \begin{center}
  \begin{tabular}{|c|c|c|c|}\hline
singularity & Grassmannian & single $ADE$ & pair of $ADE$'s\\\hline
$Y^{n+1}+X^2$ & $\mathbb{G}(2,n+3)$ & $A_n$ & $(A_n,A_1)$\\\hline
$Y^{3}+X^3$ & $\mathbb{G}(3,6)$ & $D_4$ & $(A_2,A_2)$\\\hline
$Y^4+X^3$ & $\mathbb{G}(3,7)$ & $E_6$ & $(A_3,A_2)$\\\hline 
$Y^5+X^3$ & $\mathbb{G}(3,8)$ & $E_8$ & $(A_4,A_2)$\\\hline
  \end{tabular} 
\caption{\label{table2} Isomorphisms between the cluster algebras of finite--type.  }
 \end{center}
\end{table} 

\subsection{Quantum monodromy and the associated $Y$--systems}\label{sec:YY}

The quantum monodromy $M(q)$ for a more general $\hat c<2$ singularity is given, in principle, by the phase--ordered product of the elementary transformations $\Psi(\cu_\gamma;q)$ associated with each BPS particle (in some reference chamber). Unfortunately, contrary to the case of a $\hat c<1$ singularity, we have no \emph{a priori} knowledge of the BPS data needed to construct $M(q)$ directly. 

However, in the particular case of a $(G,G^\prime)$ singularity, we may still use an indirect strategy to guess $M(q)$. Indeed, TBA and cluster--algebra theories present operators which are natural monodromy candidates: They are the generalization of the ones we found above in the single diagram case, and reduce to them for $(G,A_1)$. These operators are canonically defined by the quiver $Q$ of the theory, and their conjugacy class is an invariant of the mutation--class.  

The obvious guess is that (a power of) the canonical operator for a given $Q$ is the quantum monodromy for a $4d$ $\cn=2$ model associated with the same quiver. Therefore we shall proceed in two steps: First we present the evidence for the identification of a power of the canonical operator with $M(q)$, and then expand the resulting expression for $M(q)$ in products of elementary factors $\Psi(\cu_\gamma;q)$ to \emph{extract} the putative BPS spectrum (in the canonical chamber) which was not known in advance.  

By the previous discussion,  
the $(G,G^\prime)$ theory in the canonical chamber should correspond to the canonical square quiver $G\,\square\,G^\prime$. This quiver is simply--laced, and so the arguments of section \ref{quivers} apply. In particular, under the normal ordered symbol, we may identify the quantum and the classical monodromies.

\subsubsection{The $(G,G^\prime)$ $Y$--system and the operator $\widehat{\boldsymbol{m}}_\square$}

We start by reviewing the $Y$--system associated to a pair of Dynkin quivers  \cite{Zamolodchikov:1991et,Ravanini:1992fi,Kuniba:1993cn,MR2031858,Frenkel:1995vx,Volkov:2006zu,keller-periodicity}. We write $(k,l)$ for the vertex of the quiver $G\,\square\,G^\prime$ corresponding to the vertices $k\in G$ and $l\in G^\prime$.\smallskip

Following ref.\,\cite{keller-periodicity}, for $\epsilon$, $\epsilon^\prime=\pm 1$, we set
\begin{equation}
 \boldsymbol{m}_{\epsilon,\epsilon^\prime}= \prod_{(-1)^k=\epsilon\atop (-1)^l=\epsilon^\prime}\cq_{(k,l)},
\end{equation} 
that is, $\boldsymbol{m}_{\epsilon,\epsilon^\prime}$ is the product of the elementary mutations $\cq_{(k,l)}$ having indices of fixed parity. Notice that the $\boldsymbol{m}_{\epsilon,\epsilon^\prime}$ are well--defined since the factors mutually commute. One defines the cluster--mutation
\begin{equation}\label{monsquare}
 \boldsymbol{m}_{G\,\square\,G^\prime}\equiv \boldsymbol{m}_{-1,-1}\,\boldsymbol{m}_{+1,+1}\,\boldsymbol{m}_{-1,+1}\,\boldsymbol{m}_{+1,-1}.
\end{equation} 
When there is no danger of confusion, we simplify our notation writing $\boldsymbol{m}_\square$ for $\boldsymbol{m}_{G\,\square\,G^\prime}$.
We write $\widehat{\boldsymbol{m}}_\square$ for the quantum operator whose adjoint action induces the (normal--ordered) mutation $\boldsymbol{m}_\square$
\begin{equation}
	\widehat{\boldsymbol{m}}_\square^{\ -1}\,X_\ell\, \widehat{\boldsymbol{m}}_\square= N\big[\boldsymbol{m}_\square(X_\ell)\big].
\end{equation}
Notice that this condition fixes $\widehat{\boldsymbol{m}}_\square$ only up to an overall factor which may be a non--trivial function of $q$.
\smallskip
 
In the single Dynkin graph case, $(G,A_1)$, we have the equality
\begin{equation}\label{mvsMa1}
	M(q)=\widehat{\boldsymbol{m}}_\square^{\ 2}\equiv \widehat{\boldsymbol{m}}_{G\,\square\, A_1}^{\ \ \ h(A_1)},
\end{equation}
between the \emph{square} of the canonical operator $\boldsymbol{m}_\square$ and the physical monodromy $M(q)$. In \S.\,\ref{msquareMq} below we argue that a similar relation is true for arbitrary $(G,G^\prime)$.

The operator $\boldsymbol{m}_\square$ acts on the quantum algebra by the normal--ordered version of the classical rational map
defined by the $(G,G^\prime)$ $Y$\textit{--system} \cite{Zamolodchikov:1991et,Ravanini:1992fi,Kuniba:1993cn,MR2031858,Frenkel:1995vx,Volkov:2006zu,keller-periodicity}
\begin{equation}\label{pairTBA}
 Y_{k,a}(s+1)\, Y_{k,a}(s-1)= \frac{\prod_{j\neq k} \big(1+Y_{j,a}(s)\big)^{-C_{kj}}}{\prod_{b\neq a} \big(1+Y_{j,a}(s)^{-1}\big)^{-C^\prime_{ab}}},
\end{equation} 
where $C_{jk}$ and $C^\prime_{ab}$ are, respectively, the Cartan matrices of the Dynkin diagrams $G$ and $G^\prime$.
Notice that the interchange $G\leftrightarrow G^\prime$ is equivalent to the rational map
\begin{equation}
 Y_{ka}(s)\longleftrightarrow \frac{1}{Y_{ka}(s)}.
\end{equation} 
The relation between the operator $\boldsymbol{m}_\square$ and the solution to the $Y$--system is \cite{keller-periodicity}
\begin{equation}\label{Ymcha}
	Y_{k,a}(s)=\begin{cases}
	\boldsymbol{m}_{+1,-1}\,\boldsymbol{m}_{-1,+1}\cdot Y_{k,a}(s-1) & s\ \text{odd}\\
	\boldsymbol{m}_{+1,+1}\,\boldsymbol{m}_{-1,-1}\cdot Y_{k,a}(s-1) & s\ \text{even}.
\end{cases}
\end{equation}

It has been conjectured by Zamolodchikov \cite{Zamolodchikov:1991et} in the $(G, A_1)$ cases,
by Kuniba-Nakanishi \cite{Kuniba:1992ev} for $(G, A_n)$, and finally by Ravanini-Tateo-Valleriani
\cite{Ravanini:1992fi} for $(G, G')$, that the $Y$--system \eqref{pairTBA} is periodic of period
$2(h+h^\prime)$ where $h$, $h^\prime$ are the Coxeter numbers of $G$ and $G^\prime$, respectively. From eqn.\eqref{Ymcha} we see that this is equivalent to
$\boldsymbol{m}_\square$ having order dividing $h+h^\prime$.
This conjecture has now been proven in refs.\!\cite{Ravanini:1992fi,Kuniba:1993cn,MR2031858,Frenkel:1995vx,Volkov:2006zu,keller-periodicity}, 
In fact, one has a more precise result \cite{keller-periodicity,kellerprivate}:
the order of $\boldsymbol{m}_\square$ is exactly $(h+h^\prime)/2$ if both Lie algebras $G$ and $G^\prime$ have 
$\omega_0=-1$, and $h+h^\prime$ otherwise.

\subsubsection{$\widehat{\boldsymbol{m}}_\square$ and $M(q)$}\label{msquareMq}

On physical grounds, we know from the analysis in section \ref{sec:HyCY} that the order $r$ of the monodromy $M(q)$ of the $(G,G^\prime)$ theory should divide $h+h^\prime$. Thanks to the above periodicity theorem for $(G,G^\prime)$ $Y$--systems, we know that any power of the canonically defined operator $\widehat{\boldsymbol{m}}_\square$ has a finite order which divides $h+h^\prime$. Then the natural guess is that $M(q)$ is some power of $\widehat{\boldsymbol{m}}_\square$. 

We already know that this is the case if $G$ or $G^\prime$ is $A_1$, see eqn.\eqref{mvsMa1}.

More precisely, it was shown in section \ref{sec:HyCY} that for $(A_{n-1},A_{m-1})$ the order of the monodromy must be precisely
$(m+n)/\gcd(m,n)$. Thus we expect the right relation to be
\begin{equation}\label{mMAm}
 M(q)=(\widehat{\boldsymbol{m}}_\square)^m,
\end{equation} 
or, more generally,
\begin{equation}\label{genMGG}
 M(q)=(\widehat{\boldsymbol{m}}_{G\,\square\,G^\prime})^{h(G^\prime)},
\end{equation} 
which has the right order.
Moreover, equation \eqref{genMGG} has the correct symmetry under $G\leftrightarrow G^\prime$, if we recall that the replacement $G\,\square\,G^\prime \leftrightarrow G^\prime\,\square\,G$ inverts the orientation of the quiver, with the effect
$$\boldsymbol{m}_{G\,\square\,G^\prime}\ \longleftrightarrow\, \boldsymbol{m}_{G^\prime\,\square\, G} \ \equiv\ (\boldsymbol{m}_{G\,\square\, G^\prime})^{-1}.$$
 Indeed, writing explicitly the quiver dependence,
\begin{equation}\label{GG1duality}
(\boldsymbol{m}_{G\,\square\,G^\prime})^{h(G^\prime)}\equiv \boldsymbol{m}_\square^{\ h(G^\prime)-h(G)-h(G^\prime)}=(\boldsymbol{m}_\square^{\ -1})^{h(G)} \equiv (\boldsymbol{m}_{G^\prime\,\square\, G})^{h(G)}, 
\end{equation}
which is manifestly symmetric under $G\leftrightarrow G^\prime$.

Equation \eqref{GG1duality} is true with $\boldsymbol{m}_\square$ replaced by the quantum operator $\widehat{\boldsymbol{m}}_\square$ with the modification that, in this case, the equality means that the two sides have the same adjoint action, but they may differ by an overall `trivial' $q$--dependent factor.

We shall return to the 
$(G,G^\prime)\leftrightarrow (G^\prime,G)$ duality in \S.\,\ref{subsub:leftright}, after developping the necessary tools.
 \medskip

We may also understand equation \eqref{mMAm} from the point of view of the fractional monodromy introduced in \S.\,\ref{sec:frakMon}. Consider the singular (\textit{i.e.}\! conformal) $(A_{n-1},A_{m-1})$ SW curve $X^n+Y^m=0$. Switching on suitable Coulomb branch parameters we may deform it to the non--singular curve
\begin{equation}
 X^n+Y^m=\mu,
\end{equation} 
which has a $\mathbb{Z}_n\times \mathbb{Z}_m$ symmetry.  Focusing on the second factor,
and using the results of \S.\,\ref{sec:frakMon},
we deduce that the $M(q)$ can be written as an $m$--th power of a natural operator, in agreement with eqn.\eqref{mMAm}.  Furthermore, \eqref{GG1duality} shows that $M(q)$ can also be written as an $n$-th power of another natural operator, as is expected. 
\medskip

There is a more conclusive argument showing that the power in the \textsc{rhs} of eqn.\eqref{genMGG} must be $h(G^\prime)$. 
The point is that not all (ordered) products of $\cq_k$'s may be consistently identified with the monodromy of some (unknown) $4d$ $\cn=2$ theory. To be a candidate monodromy, an operator needs to have the particular structure described in section \ref{strucmon}: a phase--ordered product of $\Psi(\cu_\gamma;q)$ satisfying the constraints following from PCT as well as the other discrete symmetries the model may have.

We claim, in particular, that $\widehat{\boldsymbol{m}}_\square$ is not, in general, a product of only $\Psi(\cu_k;q)$ operators, while 
$(\widehat{\boldsymbol{m}}_\square)^{\pm h(G^\prime)}$ can always be written in this way. For many pairs $(G,G^\prime)$ also the PCT structure of $M$ is obvious for the operator $(\widehat{\boldsymbol{m}}_\square)^{\pm h(G^\prime)}$. For the other cases the PCT property is also expected to hold, but not manifest from the explicit expression of $\widehat{\boldsymbol{m}}_\square$; to fully establish PCT requires repeated use of the higher identities for products of quantum dilogarithms.

In fact, we are informed by Bernhard Keller\footnote{\ We thank Bernhard Keller for discussions on cluster--algebras and especially for having repeatedly assured us that many of our physically--motivated conjectures were indeed true mathematical facts.} \cite{kellerprivate} that at the level of adjoint actions, 
eqn.\eqref{genMGG} is known to experts in cluster--algebras and representation theory as a true mathematical fact.

In the next subsection we explain in down--to--earth terms our claim about the properties of $M(q)= (\widehat{\boldsymbol{m}}_\square)^{h(G^\prime)}$.
 Then in \S.\,\ref{subsub:leftright} we show how the $G\leftrightarrow G^\prime$ duality works concretely. 

 \subsubsection{Cluster--mutations \textit{vs.} the quantum KS group}\label{CMvsQKSG}
 
 A necessary condition for the $k$--th power of $\widehat{\boldsymbol{m}}_\square$ to be an $\cn=2$ monodromy is that there exists an ordered sequence of operators $\cu_\gamma$ such that for all $\gamma^\prime\in \Gamma$,
\begin{equation}\label{clustvsKS}
N\big[{\boldsymbol{m}}_\square^{\ k}(\cu_{\gamma^\prime})\big]= \Big(T\prod \Psi(\cu_\gamma)\Big)^{-1} \cu_{\gamma^\prime}\,\Big(T\prod \Psi(\cu_\gamma)\Big)
\end{equation}
 that is, the cluster--mutation ${\boldsymbol{m}_\square}^{k}$ must be an \textit{integral} element of the quantum KS group. 
 
 $\boldsymbol{m}_\square$ is a product of the elementary cluster--mutations $\cq_k$ at each node of the quiver $G\,\square\,G^\prime$ (in a specific order, see eqn.\eqref{monsquare}). The $\cq_k$ are the product of a quiver--mutation $\mu_k$ and the adjoint action of $\Psi(-X_k;q)$. Were it not for the insertions of the $\mu_k$'s, $\boldsymbol{m}_\square$ would be trivially the adjoint action of an operator of the form $\prod\Psi(-X_k)$. Thus eqn.\eqref{clustvsKS} requires the various $\mu_k$'s to combine into the identity map; then their net effect is to change some of $X_k$ into more general monomials in the quantum algebra generators, $\cu_\gamma$, corresponding to BPS particles with composite charges $\gamma\in\Gamma$.\smallskip
 
 To illustrate the idea, 
let us review the $(A_m,A_1)$ case in this language. We have $\boldsymbol{m}_\square=\boldsymbol{m}_{-1}\,\boldsymbol{m}_{+1}$ with
 $\boldsymbol{m}_\varepsilon=\prod_{(-1)^k=\varepsilon}\cq_k$. We start from the initial quiver $Q_m$
\begin{equation}\label{iniquiv}	
\begin{diagram}
	\node{X_1}\node{X_2}\arrow{w}\arrow{e}\node{X_3}\node{X_4}\arrow{w}\arrow{e}\node{\cdots\cdots}
\end{diagram}
\end{equation}
The action of $\boldsymbol{m}_{+1}$ produces two effects: it enforces the rational map
\begin{align}
	&X_\ell \rightarrow \Big(\prod_\mathrm{even}\Psi(-X_k)\Big)^{-1}\,\mu_\mathrm{even}(X_\ell)\,\Big(\prod_\mathrm{even}\Psi(-X_k)\Big),\\
	&\text{where}\quad \mu_\mathrm{even}(X_\ell)=X_\ell^{(-1)^{\ell-1}},
\end{align}
 and changes the quiver (and basis in $\Gamma$) as
 \begin{equation}	
\begin{diagram}
	\node{X_1}\arrow{e}\node{X_2^{-1}}\node{X_3}\arrow{w}\arrow{e}\node{X_4^{-1}}\node{\cdots\cdots}\arrow{w}
\end{diagram}
\end{equation}
Let us apply $\boldsymbol{m}_{-1}$ to the result. The composite operator $\boldsymbol{m}_{-1}\,\boldsymbol{m}_{+1}$ then gives the rational map
\begin{align}
	&\begin{aligned}\label{exmu}X_\ell \rightarrow \Big(\prod_\mathrm{odd}\Psi(-X_j)\Big)^{-1}&\Big(\prod_\mathrm{even}\Psi(-\mu_\mathrm{odd}(X_k))\Big)^{-1}\,\mu_\mathrm{odd}\,\mu_\mathrm{even}(X_\ell)\,\times \\
	&\times\Big(\prod_\mathrm{even}\Psi(-\mu_\mathrm{odd}(X_k))\Big)\Big(\prod_\mathrm{odd}\Psi(-X_j)\Big),\end{aligned}\\
	&\text{where}\quad \mu_\mathrm{odd}(X_\ell)=X_\ell^{(-1)^{\ell}},
\end{align}
while the quiver and basis become
\begin{equation}\label{quifinA1}	
\begin{diagram}
	\node{X_1^{-1}}\node{X_2^{-1}}\arrow{w}\arrow{e}\node{X_3^{-1}}\node{X_4^{-1}}\arrow{w}\arrow{e}\node{\cdots\cdots}
\end{diagram}
\end{equation}
that is, we return to the initial quiver $A_m$, but with an inverted system of generators $X_k^{-1}$. If we apply $\boldsymbol{m}_\square$ twice, we return to the original quiver $A_m$ and lattice basis, and the net effect of the $\mu_k$'s is to change the arguments in the $\Psi$'s, so that the action of $\boldsymbol{m}_\square^{\ 2}$ equals the adjoint action of the operator
\begin{equation}
	\prod_\mathrm{even}\Psi(-X_k^{-1})\,\prod_\mathrm{odd}\Psi(-X_j^{-1})\,
	\prod_\mathrm{even}\Psi(-X_k)\,\prod_\mathrm{odd}\Psi(-X_j),
\end{equation}
which we recognize for the correct physical monodromy (computed in the canonical chamber) with its correct PCT structure.
\smallskip

The general case is similar. The crucial point is that the $\mu_k$'s will act also on the arguments $X_k$ of the previous $\Psi$'s, as in eqn.\eqref{exmu}. Concretely, one writes a cluster--mutation as an ordered product $\vec\prod\cq_k$. Assuming the corresponding ordered product $\vec\prod\mu_k$ is the identity map, the given cluster--mutation  will be equal to the (normal--ordered) adjoint action of the operator (schematically)
\begin{equation}
	\vec\prod_k\Psi\Big(-\prod_{j>k}\mu_j(X_k)\Big),
\end{equation}
which is an integral element of the KS group.

For a $(G,G^\prime)$ quiver, the cluster--mutation $\boldsymbol{m}_\square$ does not have the property that the corresponding quiver--mutation $\mu_\square\equiv\vec\prod\mu_k$ is the identity. As in the $(A_m,A_1)$ example above, it is true that the mutation $\mu_\square$ reproduces the original quiver $G\,\square\,G^\prime$, but with a \textit{different} choice of basis in the lattice $\Gamma$ (that is, a different set of generators of the quantum algebra). However, we claim that  $(\boldsymbol{m}_\square)^{h(G^\prime)}$ has the desidered property, namely
\begin{equation}\label{musquareh}
 \mu_\square^{\ h(G^\prime)}=\text{identity}.
\end{equation} 
\medskip

To prove eqn.\eqref{musquareh}, we start with a general $(G,A_m)$ theory ($G=ADE$) whose quiver is $G\,\square\, A_m$.
The corresponding quantum algebra is generated by the invertible operators $X_{k,l}$, where $k=1,\dots, \mathrm{rank}\,G$, and $l=1,2,\dots, m$.

As shown in appendix \ref{ape:agymnastics}, the quiver--mutation $\mu_\square$ maps the quiver $G\,\square\, A_m$ to itself, and hence should correspond to an inner automorphism of the quantum algebra given by the adjoint action of some quantum operator $\hat\mu_\square$. More precisely, in the appendix it is shown that the adjoint action of $\hat\mu_\square$ generates the (normal ordered version of) the following classical rational map:
\begin{align}
& X_{2l,2k+1} \mapsto X_{2l,2k-1}X_{2l,2k}X_{2l,2k+1}X_{2l,2k+2}X_{2l,2k+3}\label{ksquare1}\\
&X_{2l,2k+2}\mapsto X^{-1}_{2l,2k+1}X^{-1}_{2l,2k+2}X^{-1}_{2l,2k+3}\\
&X_{2l+1,2k+1}\mapsto X^{-1}_{2l+1,2k}X^{-1}_{2l+1,2k+1}X^{-1}_{2l+1,2k+2}\\
&X_{2l+1,2k+2}\mapsto X_{2l+1,2k}X_{2l+1,2k+1}X_{2l+1,2k+2}X_{2l+1,2k+3}X_{2l+1,2k+4}\\
\intertext{with the \textit{convention}}
&X_{l,k}=1\ \text{for } k=0\ \text{or } k>m,\\
\intertext{and the \textit{exceptions}}
&\bullet \quad X_{2l,1}\mapsto X_{2l,2}X_{2l,3}\\
&\bullet\quad \begin{cases}
               X_{2l,m}\mapsto X_{2l,m-2}X_{2l,m-1} & m\ \text{odd}\\
X_{2l+1,m}\mapsto X_{2l+1,m-2}X_{2l+1,m-1} & m\ \text{even.}\label{ksquaren}
              \end{cases}
\end{align}
 
A remarkable property of $\mu_\square$ is that it maps $X_{l,k}$ into a rational function of the $X_{l,k^\prime}$'s with a \emph{fixed} $l$. Thus, to show that 
$$(\mu_\square)^{m+1}\equiv\mu_\square^{\ h(A_m)}=\mathbf{1},$$
 we may work at fixed $l$. In other words, we may effectively replace $G$ by the trivial quiver $A_1$. Then, 
 writing 
  `$\mapsto$' for the action of $\mu_\square$, we see that repeated applications of $\mu_\square$ produce the following chains of transformations
\begin{align*}
 &\underline{m\ \text{even},\ l\ \text{even}}\\
&\begin{aligned}
  X_{l,1}&\mapsto X_{l,2}X_{l,3}\mapsto X_{l,4}X_{l,5}\mapsto \cdots \mapsto X_{l,m-2}X_{l,m-1}\mapsto X_{l,m} \mapsto\\
&\mapsto X_{l,m-1}^{-1}X_{l,m}^{-1}\mapsto
X^{-1}_{l,m-3}X^{-1}_{l,m-2}\mapsto \cdots \mapsto X^{-1}_{l,1}X_{l,2}^{-1}\mapsto X_{l,1}
 \end{aligned}\\
&\underline{m\ \text{odd},\ l\ \text{even}}\\
&\begin{aligned}
  X_{l,1}&\mapsto X_{l,2}X_{l,3}\mapsto X_{l,4}X_{l,5}\mapsto \cdots \mapsto X_{l,m-1}X_{l,m}\mapsto X_{l,m}^{-1} \mapsto\\
&\mapsto X_{l,m-2}^{-1}X_{l,m-1}^{-1}\mapsto
X^{-1}_{l,m-4}X^{-1}_{l,m-3}\mapsto \cdots \mapsto X^{-1}_{l,1}X_{l,2}^{-1}\mapsto X_{l,1},
 \end{aligned}
\end{align*}
\begin{align*}
 &\underline{l\ \text{odd}}\\
&\ \ \text{same chain of rational maps as for }l \ \text{even and }m\ \text{of the same}\\
&\ \ \text{parity, but in the \textit{inverse} order:}\\
&\hskip 1.75cm X_{l,1}\mapsto X^{-1}_{l,1}X^{-1}_{l,2}\mapsto \cdots \mapsto X_{l,2}X_{l,3}\mapsto X_{l,1}.
\end{align*}
so, in all cases, $(\mu_\square)^{m+1}$ is the identity map\footnote{\ Notice that the above chain of transformations shows that $\mu_\square^{\ m+1}$ is the identity acting on $X_{l,1}$ and $X_{l,j}X_{l,j+1}$; this is enough to conclude that it acts as the identity on all variables $X_{l,k}$.}, and no smaller power of $\mu_\square$ has this property. This establishes eqn.\eqref{musquareh} for $(G,A_m)$.
\smallskip 

Finally, we extend our result to arbitrary Dynkin pairs $(G,G^\prime)$.
The above analysis shows that $\mu_\square$ is a rational map of the form
\begin{equation}
	\begin{aligned}
	 &\mu_\square(X_{2\ell,k})= R_k\big(X_{2\ell,1}, X_{2\ell,2},\cdots, X_{2\ell,\,\mathrm{rank}\,G^\prime}\big)\\
&\mu_\square(X_{2\ell+1,k})= \tilde R_k\big(X_{2\ell+1,1}, X_{2\ell+1,2},\cdots, X_{2\ell+1,\,\mathrm{rank}\,G^\prime}\big)
	\end{aligned}
\qquad \text{for a \underline{fixed} }\ell,
\end{equation}
where the maps $R_k$, $\tilde R_k$ do not depend on $\ell$ and are each other inverses. 
Then  the property $\mu_\square^{h(G^\prime)}=\boldsymbol{1}$ is true for $G\,\square\, G^\prime$ if and only if it holds for $A_1\,\square\, G^\prime$, that is, if is true for the $ADE$ Dynkin quivers with the \textit{inverted} arrows.

Let $\Lambda_\mathrm{root}$ be the weight lattice of $G^\prime$. We identify the monomials in the quantum algebra of $A_1\,\square\,G^\prime$ with elements of $\Lambda_\mathrm{root}$ according to
\begin{equation}
 \sum_i n_i\,\alpha_i \mapsto N\Big[\prod_i X_{\alpha_i}^{n_i}\Big],
\end{equation}  
where $\alpha_i\in\Lambda_\mathrm{root}$ are the simple roots. As always, we identify quantum transformation and classical rational maps \emph{via} the normal ordered product $N[\cdots]$. Then the elementary mutations $\mu_k$ of $A_1\,\square\,G^\prime$ act on the quantum algebra as the elementary reflections $s_k$ generating the Weyl group $W$ of $G^\prime$. The product $\mu_\square$ is then identified with the Coxeter element of $W$ whose order, by definition, is $h(G^\prime)$. This completes the proof of eqn.\eqref{musquareh}.

Thus
\begin{equation*}
	\widehat{\boldsymbol{m}}_\square^{\,\ h(G^\prime)}\equiv (\widehat{\boldsymbol{m}}_{G\,\square\,G^\prime})^{h(G^\prime)}
\end{equation*}
 is an integral element of the KS group, and no smaller power has this property.

\subsubsection{The $G\leftrightarrow G^\prime$ duality again}\label{subsub:leftright}

The duality $(G,G^\prime)\leftrightarrow (G^\prime,G)$ is easily understood in the language of the previous section. First of all, we observe that from $\cq_k^2=\mathbf{1}$, it follows that $\boldsymbol{m}_\square^{\ -1}$ is given by the \textsc{rhs} of eqn.\eqref{monsquare} with the factors in the inverse order
\begin{equation}\label{msquareminus}
\boldsymbol{m}_{G^\prime\,\square\,G}\equiv\boldsymbol{m}_\square^{\ -1}= \boldsymbol{m}_{+1,-1}\,\boldsymbol{m}_{-1,+1}\,\boldsymbol{m}_{+1,+1}\,\boldsymbol{m}_{-1,-1}. 
\end{equation}  
The powers of the \textsc{rhs} of eqn.\eqref{msquareminus} are again products of $\cq_k$'s; they may be written as products of quantum dilogarithms (not \emph{inverse} quantum dilogarithms\,!) provided the corresponding products of $\mu_k$'s are the identity rational map. We claim that this property holds for the power
\begin{equation*}
 (\boldsymbol{m}_\square^{\ -1})^{h(G)},
\end{equation*} 
therefore establishing equation \eqref{GG1duality}.

Indeed, the two transformations $\boldsymbol{m}_\square$ and $\boldsymbol{m}_\square^{\ -1}$ are \textit{equivalent up to conjugacy}
\begin{equation}
	(\boldsymbol{m}_\square)^{-1}=(\boldsymbol{m}_{+1,-1}\,\boldsymbol{m}_{-1,+1})\,
	\boldsymbol{m}_\square\,(\boldsymbol{m}_{+1,-1}\,\boldsymbol{m}_{-1,+1})^{-1},
\end{equation}
while the quiver--mutation $\boldsymbol{m}_{+1,-1}\,\boldsymbol{m}_{-1,+1}$ has precisely the effect
\begin{equation}\label{dualGGGG}
	G\,\square\,G^\prime \leftrightarrow G^\prime\,\square\,G
\end{equation}
(up to some change of basis). Thus, under $G\leftrightarrow G^\prime$, $\boldsymbol{m}_\square^{\ h(G^\prime)}\leftrightarrow (\boldsymbol{m}_\square^{\,\ -1})^{h(G)}$, and
\begin{equation}\label{moneqmon}
	M(q) = \widehat{\boldsymbol{m}}_\square^{\,\ h(G^\prime)}\ \longleftrightarrow\ 
	(\widehat{\boldsymbol{m}}_\square^{\,\ -1})^{h(G)}= M(q) 
\end{equation}
so the physical monodromy $M(q)$ is invariant under the $G\leftrightarrow G^\prime$ duality, as it is obvious from the geometric description of, say,
the SW curve.

Again we stress that \eqref{moneqmon} holds as equality of adjoint actions; the operators has to be normalized in some canonical way in order to have strict equality.\smallskip

The two quantum operators $(\widehat{\boldsymbol{m}}_{G\,\square\,G^\prime})^{h(G^\prime)}$
and $(\widehat{\boldsymbol{m}}_{G\,\square\,G^\prime})^{h(G^\prime)}$, while having the same adjoint action on the quantum torus algebra, differ in a crucial way: When written as ordered products of elementary wall--crossing operators $\Psi(\cu_\gamma;q)$, they have, respectively, $r(G)\, r(G^\prime)\, h(G^\prime)$ and $r(G)\, r(G^\prime)\, h(G)$ factors (where $r(G)=\mathrm{rank}\, G$). Equating the two expressions give non--trivial quantum dilogarithm identities which, under the appropriate circumstances, may be interpreted as wall--crossing formulae between two dual canonical chambers.  
\smallskip

The duality \eqref{dualGGGG} amounts to inverting all the arrows of the quiver $G\,\square\,G^\prime$. This is the same as
\begin{equation}
 q\leftrightarrow q^{-1},
\end{equation}  
so one expects to relate the two dual pictures as analytic continuations in opposite half--planes.
This will turn out to be essentially correct.
\smallskip

\noindent\textbf{Remark.} In the language of algebra representation theory \cite{MR2197389,auslander,gabriel}, the arguments developed in the present section correspond to the relations between the suspension functor, the Serre functor, and the Auslander--Reiten translation
\cite{keller-periodicity,kellerprivate}.

  \subsection{Example: the putative BPS spectra of $(A_m,A_3)$ theories}

To compile exhaustive lists of (putative) BPS spectra of the $4d$ $\cn=2$ theories is outside the purposes of the present paper. 
However, to illustrate the idea, we briefly discuss the $(A_m,A_3)$ example.

Using the formulae from the preceding section, we have
\begin{equation*}\begin{split}
 \widehat{\boldsymbol{m}}_\square =&\hat\mu_\square\, \prod_k \Psi\big(-X_{2k+1,1}^{-1}X_{2k+1,2}^{-1}X_{2k+1,3}^{-1}\big)\:
\prod_k \Psi\big(-X_{2k,1}^{-1}X_{2k,2}^{-1}\big)\,\Psi\big(-X_{2k,2}^{-1}X_{2k,3}^{-1}\big)\times\\
&\times\prod_k \Psi\big(-X_{2k,1}X_{2k,2}X_{2k,3}\big)\: \prod_k \Psi\big(-X_{2k+1,2}X_{2k+1,3}\big)\,
\Psi\big(-X_{2k+1,1}X_{2k+1,2}\big),
\end{split}\end{equation*} 
where {all monomials in the arguments of the $\Psi$'s are meant to be \emph{normal--ordered}.}
Then
\begin{equation*}
 \begin{split}
 Y(q)&=\widehat{\mathbf{m}}_\square^{\ 2}=\\
 =&I'\cdot \prod_{k} \Psi(X_{2k+1,2})\, \prod_k \Psi(X_{2k,1})\,\Psi(X_{2k,3})\,
\prod_k\Psi(X_{2k,2})\,\prod_k\Psi(X_{2k+1,1})\,\Psi(X_{2k+1,3})\times\\
 &\times \prod_k \Psi\big(X_{2k+1,1}^{-1}X_{2k+1,2}^{-1}X_{2k+1,3}^{-1}\big)\:
\prod_k \Psi\big(X_{2k,1}^{-1}X_{2k,2}^{-1}\big)\,\Psi\big(X_{2k,2}^{-1}X_{2k,3}^{-1}\big)\times\\
&\times\prod_k \Psi\big(X_{2k,1}X_{2k,2}X_{2k,3}\big)\: \prod_k \Psi\big(X_{2k+1,2}X_{2k+1,3}\big)\,
\Psi\big(X_{2k+1,1}X_{2k+1,2}\big),
 \end{split}
\end{equation*} 
so the (putative) BPS spectrum in the reference chamber corresponds to one state for each of the following charge vectors in
$\Gamma$ (plus their PCT conjugates)
\begin{align}
 &\gamma_{k,a} && a=1,2,3,\ \ k=1,2,\dots, m\\
 & \gamma_{k,1}+\gamma_{k,2} \\
 &\gamma_{k,2}+\gamma_{k,3}\\
& \gamma_{k,1}+\gamma_{k,2}+\gamma_{k,3},
\end{align}
where $\{\gamma_{k,a}\}$ is the standard basis in $\Gamma$.  

Note that the above expression for $Y(q)$ also gives information about the BPS phases of these states.\footnote{It is
tempting to extrapolate from the above example that the BPS degeneracy for the $(A_m,G)$ case is given,
in some chamber, by charge vectors of the form  $\gamma_{k,b}$ where $b$ denotes a positive root
of $G$ and $k=1,2,\dots,m$ corresponds to a root of $A_m$, which suggests the following derivation:
Compactifying IIB to 6  dimensions with a $G$ singularity leads to tensionless strings
in 1-1 correspondence with the positive roots of $G$.  Moreover, the tension of the strings
vary over the extra complex dimension where $A_m$ singularity lives;  the strings can end
on any of the adjacent zeroes of the corresponding (Chebyshev type) polynomial representing
$A_m$, leading to this BPS degeneracy.

This idea is also consistent with the BPS state counting in the canonical $A_m\,\square\, G$ chamber. Indeed, the number of $\Psi$ factors in the operator $(\widehat{\boldsymbol{m}}_{A_m\,\square\,G})^{h(G)}$ is 
\begin{equation*}
	m\cdot r(G)\cdot h(G)= 2\, m\cdot \#\{\text{positive roots of }G\},
\end{equation*}
which is precisely the number of states predicted by the above scenario (counting also the PCT conjugate states).}

 \section{Trace of the monodromy for $(G,G^\prime)$ theories I: The irrational case}
 
 In this section we compute $\mathrm{Tr}\,M(q)$ (and $\mathrm{Tr}\,Y(q)$, $\mathrm{Tr}\,K(q)$) for some of the theories of interest as an illustration of the general principles and a check of our conjectures. The explicit form of $\mathrm{Tr}\,M(q)$ will depend, of course, on the particular representation of the associated quantum torus algebra we consider. When $q$ is an $N$--th root of unity (a case to be discussed in the next section) we have finite dimensional representations of the quantum algebra and the definition of the trace is obvious; in the irrational case the representations are necessarily infinite dimensional, and we must specify both the representation and the precise definition of the trace. In section \ref{subsub:generalq} we saw that $\mathrm{Tr}\,M(q)$ should be a holomorphic quasi-modular functions of $q$; correspondingly, we shall find that the various possible choices of representations and/or traces will lead to different linear combinations of the same modular blocks. We look at these blocks as being the fundamental traces, associated to the irreducible realizations of the system, while the particular linear combination one finds computing the trace in a specific representation should correspond to a direct sum of the fundamental ones.   We shall find that the $\mathrm{Tr}\,M(q)$ blocks correspond to characters of some $2d$ RCFT. 
On the other hand we already saw that the monodromy opeartion (in the classical limit) gets related to the TBA
systems arising from relevant deformations of 2d CFT's.  It will turn out that the characters we obtain from the
traces of the 4d monodromy operators are characters of the RCFT's associated to the  UV limit of the {\it same} 
2d theory!  In other words,  the trace of $q$--deformed version of the TBA monodromy  leads to characters
of the UV theory they arise from.  To the best of our knowledge this appears not to have
been known.  We use the path-integral formulation of the traces we have developed to offer an explanation of
this surprising 4d/2d connection.  As we discuss in section \ref{speculate}, it suggests
that a particular reduction of the 4d theory leads to the corresponding CFT.

\subsection{TBA's for ADET pairs}\label{TBAADET}

In preparation for computation of monodromy traces to be done later in this section, here we review some known facts about the TBA systems associated to pairs of Dynkin diagrams
and the predicted
conformal theories they arise from.
\smallskip

The quantum $\frac{1}{h(G^\prime)}$--th fractional monodromy $K(q)\equiv \widehat{\boldsymbol{m}}_\square$ has the general structure
$$K(q)=\hat\mu\,\prod \Psi(\cu_\gamma;q),$$ 
with $\hat\mu$ an operator with only a trivial kinematic $q$--dependence. From the Euler identity
\begin{equation}
 \Psi(X;q)=\sum_{k\geq 0}\frac{(-1)^k q^{k^2/2}}{(q)_k}X^k,\qquad \text{where } (q)_k=\prod_{m=1}^k(1-q^m),
\end{equation} 
it is obvious that a general trace
$$\mathrm{Tr}\big[\,K(q)\:\co\,\big],$$
with $\co$ an element of the quantum algebra, should be a linear combination of functions having the general form
\begin{equation}\label{genexpr}
 \chi_A(q;B,C)= \sum_{\mathbf{m}\in\mathbb{N}^s} \frac{q^{\tfrac{1}{2}\mathbf{m}\cdot A\cdot \mathbf{m}+B\cdot\mathbf{m}+ C}}{(q)_{m_1}\, (q)_{m_2}\cdots (q)_{m_s}},
\end{equation} 
where $s=\mathrm{rank}\,G\cdot \mathrm{rank}\,G^\prime$, $\mathbf{m}=(m_1,m_2,\dots, m_s)$ is a vector of non--negative integers, $A$ is a positive--definite symmetric $s\times s$ matrix, $B$ a vector and $C$ a scalar. In particular, $B$ and $C$ have to be seen as `sources' for the operator insertions. 

It turns out that, in many models, $\mathrm{Tr}\, M(q)$ can be also written in terms of $\chi_A(q;B,C)$ and $\eta(q)$ functions.

Remarkably, there is a large class of $2d$ RCFT which have characters of precisely this form (with $A$, $B$, $C$ rational in general) \cite{nahm}. 
For a given RCFT, the matrix $A$ is unique, while $B$ and $C$ depend on the particular character we consider. Only a subset of the functions $\chi_A(q;B,C)$ are actually independent, since 
\begin{equation}\label{trirec}
\chi_A(q;B,C)=q^C\, \chi_A(q;B,0),	
\end{equation}
 and we have the recursion relations (for $j=1,2,\cdots, s$)
\begin{equation}\label{recursion}
 \chi\big(q; B+A\cdot \mathbf{e}_j, C+B\cdot \mathbf{e}_j+\tfrac{1}{2}\mathbf{e}_j\cdot A\cdot\mathbf{e}_j\big) =
\chi(q; B,C)-\chi(q; B+\mathbf{e}_j,C),
\end{equation} 
where $\mathbf{e}_j=(0,\dots, 0,1,0,\dots ,0)$ with $1$ in the $j$--th position. Eqn.\eqref{recursion} just says that the functions
$\chi_A(q;B,C)$ may be expressed as continuous fractions \textit{\' a la} Ramanujan.
\smallskip

The expression \eqref{genexpr} for the CFT characters naturally relate to the TBA systems. One considers an integrable massive deformation of the given CFT, and computes the partition function using the known elastic $S$ matrix
$S_{ij}(\theta)=\exp(i\,\delta_{ij}(\theta))$; in the UV limit one gets \eqref{genexpr} with\footnote{\ If the $2d$ theory has ordinary statistics, the entries of $A_{ij}$ are integral. To get more general rational values one has to incorporate exotic statistics \cite{nahm}.} \cite{nahm}
$$A_{ij}=-\delta_{ij}(-\infty)/(2\pi).$$
Moreover the expression (\ref{genexpr}) defines the characters of the corresponding CFT that the TBA system arises from.   Indeed it is generally believed \cite{nahm} that the $\chi_A(q; B,C)$'s may have good modular properties only under some special circumstances which are essentially equivalent to the requirement that the $\chi_A(q;B,C)$'s arise as traces of quantum operators having the general structure of a $4d$ fractional monodromy $K(q)$ having finite order.

Many examples of $2d$ physical systems having these properties are known \cite{Zamolodchikov:1991et,Dasmahapatra:1993pw,Kedem:1993ze,Terhoeven:1992zf, nahm}. The known models are labelled by a pair $(G,G^\prime)$ of $ADET$ Dynkin diagrams. Here $ADE$ stands for the usual simply--laced Lie algebra graphs, while the tadpole diagrams, $T_r$ ($r\in \mathbb{N}$), are obtained from the Dynkin diagrams of $A_{2r}$ by folding in the middle and identifying the vertices pairwise. The Cartan matrix of $T_r$ is the same one for $A_r$ except that $(C_{T_r})_{rr}=1$ instead of $2$, and has determinant $1$. The Coxeter number of $T_r$ is the same as that of $A_{2r}$, \textit{i.e.}\! $2r+1$.

The $Y$--system solving the TBA for the $(G,G^\prime)$ theory is given by eqn.\eqref{pairTBA}.
The matrix $A$ giving the characters of the UV fixed point of the integrable model described by the pair $ADET$ of Dynkin diagrams $(G,G^\prime)$ is \cite{Zamolodchikov:1991et,Dasmahapatra:1993pw,Kedem:1993ze,Terhoeven:1992zf, nahm}
\begin{equation}
 A=C_{G}\otimes C_{G^\prime}^{-1}
\end{equation} 
where $C_G$ stands for the Cartan matrix of $G$. Notice that $A\leftrightarrow A^{-1}$ under $G\leftrightarrow G^\prime$;
thus the interchange of the two diagrams establishes a kind of duality between the corresponding pair of CFT's.

Examples of known RCFT with characters of this form are:
\vglue 9pt

\begin{center}
\begin{tabular}{|c|c|}\hline
RCFT & $(G,G^\prime)$\\\hline\hline
coset $G_{n+1}/U(1)^{\mathrm{rank}\, G}$ & $(G,A_n)$\\\hline\hline
minimal models $W^G(2,2n+3)$&$(G,T_n)$\\\hline
minimal models $W^G(2n+1,2n+3)$&$(T_n,G)$\\\hline
\end{tabular} 
\end{center}\vglue 12pt

\noindent notice that the pairs of RCFT in each block are dual in the present sense.   In the
above table $W^G(p,q)$ refers to the minimal model for $W^G$ algebra of type $(p,q)$ \cite{Caldeira:2004jy} (the case
$G=sl(2)$ is the standard Virasoro minimal models).  Note also that the coset
$G_{n+1}/U(1)^{\mathrm{rank}\, G}$, is also known as the CFT of parafermions of level
$(n+1)$ for the group $G$.

It is a longstanding conjecture, recently proven in \cite{Nakanishi:2009ye,tsyst4,tsyst5},
that the corresponding RCFT for $(G,G')$ type has central charge (see also \cite{nahm} and
references therein)
\begin{equation}\label{effcencharge}
 c^{eff}={r_Gr_{G'}h_G\over h_G+h_{G'}}
\end{equation} 
where $r$ and $h$ refer to rank and dual Coxeter numbers respectively (with
the definition of $r(T_n)=n,h(T_n)=2n+1$  for the $T$ series).  Moreover
the notion of the `effective' central charge is the same as the central charge $c$ in the unitary
case, and $c-24d$ where $d$ is the lowest dimension operator, in the non-unitary case.
Notice the following relation:
\begin{equation}c^{eff}(G,G')+c^{eff}(G',G)=r_Gr_{G'}\label{c+cformula}\end{equation}
suggesting that a suitable tensor product of the two theories gives a free theory involving
$r_Gr_{G'}$ bosons.  This suggests that the two theories $(G,G')$ and $(G',G)$  are `dual' 
CFT's in the sense of level-rank duality and in the known cases they are.

\smallskip

Given the relationship of the $4d$ monodromy with the TBA's $Y$--systems, eqn.\eqref{pairTBA}, and the general structure \eqref{genexpr},
it is natural to expect that
$\mathrm{Tr}\, M(q)$ and $\mathrm{Tr}\, K(q)$ will be related to RCFT character of this class. The actual computations confirm the expectation in many of the cases above, as we will see in the next section.
\smallskip

\subsection{Evidence for $\mathrm{Tr}[K(q)\,\co]$ $\equiv$ Characters of the $(G,G^\prime)$ RCFT}\label{evidence}

From the discussion in section \ref{sec:pairsofD}, we know that the monodromy $M(q)$ for the $(G,G^\prime)$ theory can be written either as
$(\widehat{\boldsymbol{m}}_\square)^{h(G)}$ or $(\widehat{\boldsymbol{m}}_\square^{\ -1})^{h(G^\prime)}$, the two expressions being possibly different by overall trivial factors. Comparing with the structure of eqn.\eqref{effcencharge}, it looks natural to identify 
\begin{gather*}
 \mathrm{Tr}_{(G,G')}\, M(q)^{1/h(G^\prime)} \sim \text{characters of }(G,G^\prime)\ {\rm TBA}\\
\mathrm{Tr}_{(G,G')}\, M(q^{-1})^{1/h(G^\prime)} \sim \text{characters of }(G^\prime,G)\ {\rm TBA},
\end{gather*}  
modulo some overall trivial factor corresponding, plausibly, to free theories (as the formula \eqref{c+cformula} for the central charges suggests).  
Note that on the left side of the above identifications the $\cn=2$ theories obviously
satisfy $(G,G')=(G',G)$ and lead to the same $M$;  the differences on the RHS arise by which fractional powers of $M$ we take and whether we consider
$q$ or $q^{-1}$ in the argument for the traces.
Furthermore we find evidence for
\begin{gather*}
 \mathrm{Tr}_{(A_n,A_1)}\, M(q)\sim \text{characters of }(A_{n-1},T_1)\ {\rm TBA}\\
\mathrm{Tr}_{(A_n,A_1)}\, M(q^{-1})\sim \text{characters of }(T_1,A_{n-1})\ {\rm TBA}.
\end{gather*}  
In particular the trace of the full monodromy $M(q)$ for the 4d Argyres-Douglas CFT given by $y^2=x^3$,  leads to characters of 2d Lee-Yang
edge singularity given by the $(2,5)$ minimal model.
The goal of the present subsection is to present some general evidence for this identification. In the rest of the section we shall check the proposal in some concrete examples.\medskip

 Consider a quantum trace of the form
\begin{equation}\label{gentraceGG}
	\mathrm{Tr}\Big[K(q)\, \prod X_i^{b_i}\Big].
\end{equation}
It may be written as a periodic path integral. The details of the path integral will depend on the particular realization of the quantum torus algebra one considers (see discussion below), but certain aspects are `universal', i.e., representation independent.
 
From the schematic structure of $K(q)$ we know that eqn.\eqref{gentraceGG} must be of the general form $\chi_A(q;B,C)$ with $B$ and $C$ some functions of the $b_i$'s. In terms of the periodic path integral, the recursion relations \eqref{recursion}\eqref{trirec} are interpreted as analog of `Schwinger--Dyson equations'.  We will provide evidence that these equations fix the quantum amplitudes \eqref{gentraceGG} up to an overall $b_i$ independent normalization.
Hence, the recursion relations determine the traces up to an overall function of $q$ which is, typically, a product of free partition functions.

Therefore, up to elementary factors, what really characterizes the quantum traces is the matrix $A$ which fixes the form of the recursion relations $\equiv$ Schwinger--Dynson equations \eqref{recursion}\eqref{trirec}. Thus, to compute the traces is to determine $A$. 

The identification we are suggesting sets the $s\times s$ matrix $A$ appearing in the quantum traces of $K(q)$ equal to the $s\times s$ matrix appearing in the TBA representation for the CFT characters of the UV limit of the corresponding $(G,G^\prime)$ $2d$ integrable model, namely
\begin{equation}\label{guessGG}
A =C_G\otimes C_{G^\prime}^{-1}.	
\end{equation}
The purpose of this subsection is to present convincing evidence for eqn.\eqref{guessGG}.

\subsubsection{Quiver duality and $\chi_A$ characters}

To be viable, the identification \eqref{guessGG}  has, in particular, to be consistent with the $G\leftrightarrow G^\prime$  duality.  
This boils down to explaining why, under $G\,\square\,G^\prime\leftrightarrow G^\prime\,\square\,G$ (which is equivalent to $q\leftrightarrow q^{-1}$), one has
\begin{equation*}
	A\longleftrightarrow A^{-1}.
\end{equation*}

 Define the functions
\begin{equation}\label{newchis}
	\widetilde{\chi}_A(q;B,C) \equiv \chi_A\big(q^{-1};BA, -C+\tfrac{1}{2}B\cdot A^{-1}\cdot B^t\big)
\end{equation}
 It is easy to check that
 \begin{gather}\begin{aligned}\label{dualrel1}
		\widetilde{\chi}_A\big(q; B+A^{-1}\mathbf{e}_j&,C+B\cdot\mathbf{e}_j+\tfrac{1}{2}\mathbf{e}_j\cdot A^{-1}\cdot \mathbf{e}_j\big)=\\
		&=\widetilde{\chi}_A(q;B,C)-\widetilde{\chi}_A(q;B+\mathbf{e}_j,C)\end{aligned}\\
		\intertext{and}
		\widetilde{\chi}_A(q;B,C)= q^C\: \widetilde{\chi}_A(q;B,0).\label{dualrel2}
\end{gather}
Eqns.\eqref{dualrel1}\eqref{dualrel2} are precisely the recursion relations \eqref{recursion}\eqref{trirec} satisfied by the character 
$$\chi_{A^{-1}}(q; B, C).$$

 By the `Schwinger--Dyson argument', i.e. that the traces
are characterized effectively by the above recursion relations we expect  $\widetilde{\chi}_A(q;B,C)$ to be equal to
$\chi_{A^{-1}}(q; B, C)$ up to some `elementary', $B$ and $C$ independent, overall normalization factor. 

On the other hand, from eqn.\eqref{newchis}, it is obvious that the $\widetilde{\chi}_A$'s are related to the $\chi_A$'s, apart for a relabelling of the sources $B$ and $C$, by the operation $q\leftrightarrow q^{-1}$ which invertes all the arrows of the quiver mapping
$G\,\square\, G^\prime\leftrightarrow G^\prime\,\square\,G$.

Thus we learn that the `level--rank' duality $G\leftrightarrow G^\prime$ must have the following effects on the traces of $K(q)$:
\begin{align*}&\bullet\qquad A\leftrightarrow A^{-1}\\
 &\bullet\qquad \text{a redefinition of the fields/sources }B, C
\end{align*}
up to  (possibly) `elementary' normalization factors,
in agreement with the proposed identification. This is (in general) the only way to preserve the basic recursion relations under `level-rank' duality.  In the next section we give further evidence of this picture.  
\smallskip

\subsubsection{Semi--classical limit}\label{semi}
  
In this section we provide further evidence of the above picture, but considering the semiclassical
limit.  In doing so we discover important facts about where the partition function localizes.
We have already predicted that this should correspond to expectation values
of the line operators corresponding to fixed points of the R-transformations;  furthermore
we shall see later that these classical solutions correspond to diagonalization of the Verlinde algebra
associated to the corresponding RCFT.

  Recall that our Chern-Simons theory is characterized by the quantum parameter $q$
and that the semi-classical limit corresponds to the limit $q\rightarrow 1$.
The recursion relations (our `Schwinger--Dyson equations') are the quantum analog of the classical equations of motion. Thus, in the limit $q\rightarrow 1$ they should reduce to the equations of some classical system. Which system will we get for our case?

Since $K(q)=\widehat{\boldsymbol{m}}_\square\equiv e^{-iH_\square}$ is the quantum operator generating a (discretized) time evolution whose classical limit is precisely the trajectory of the $Y$--system, the $q\rightarrow 1$ limit of the traces 
$$\mathrm{Tr}\Big[\,e^{-iH_\square}\, \co\,\Big]\equiv \mathrm{Tr}\Big[\,K(q)\, \co\,\Big]$$
should satisfy the $Y$--system evolution equations. 

The quantum traces of $K(q)$ are expected to be exactly given by the semi--classical approximation (parallel to the fact that its quantum action is the normal--ordered version of the classical one).  Indeed, from the viewpoint of the original $4d$ $\cn=2$ theory, $\mathrm{Tr}\, K(q)$ is a supersymmetric index. Thus, the (semi)classical limit should be strong enough to uniquely characterize the quantum traces, and hence to establish our identification \eqref{guessGG}.   This follows
because $q$--dependent corrections can be viewed as `gravitational corrections' and should be equivalent
if the underlying gauge systems are; the latter can be checked by studying the $q\rightarrow 1$ limit.

\smallskip

$e^{-iH_\square}$ is the quantum version of the symplectomorphism associated to the $Y$--system. Using the Duistermaat--Heckman formula \cite{Duis1,MR85e:58041,Duis3}, we get (schematically)
\begin{equation}
 \mathrm{Tr}\Big[\, e^{-iH_\square}\: \co\,\Big]\Big|_{q\sim 1} \sim\ \text{(functional determinant)} \sum_{\text{fixed points of}\atop Y\text{--evolution}}\, \co,
\end{equation} 
where we argue that the functional determinant is independent of the fixed point due to the underlying supersymmetry of the computation.
\smallskip  

On the other hand, $\mathrm{Tr}\big[\,e^{-iH_\square}\, \co\,\big]$ is of the form
\eqref{genexpr}.
Writing $Z_j=\exp(z_j)$, where $z_j$ is the field coupled to the source $B=\mathbf{e}_j$, and taking the $q\rightarrow 1$ limit of the recursion relations \eqref{recursion},
we get the classical equations
\begin{equation}\label{classlimmm}
	(1-Z_j)=\prod_k Z_k^{A_{jk}}.
\end{equation}
It is well known that these classical equations are equivalent to the fixed point equations for the $Y$--system if $A$ is equal to $C_G\otimes C_{G^\prime}^{-1}$ (or, dually, to $C_{G^\prime}\otimes C_G^{-1}$). To show this, we introduce the following notation: $(f(\mathbf{Y}))$ stands for the $\mathrm{rank}\,G\, \times\, \mathrm{rank}\,G^\prime$ matrix whose $(k,l)$ component is
$f(Y_{k,l})$. Then the fixed point equations for the $Y$--system read
\begin{equation}\label{Yfix1}
 \left(1+\frac{1}{\mathbf{Y}}\right)^{-\mathbf{1}\,\otimes\, C_{G^\prime}}=\Big(1+\mathbf{Y}\Big)^{-C_G\,\otimes\,\mathbf{1}}.
\end{equation} 
Write\footnote{\ This redefinition may seem unnatural at first sight. However, as better explained in section \ref{sec:quantFrob}, one has a full modular trajectory of classical limits corresponding to $q\rightarrow e^{2\pi i\tau}$ with $\tau\in\mathbb{Q}$. If $q$ is a non--trivial root of unity, we get a discrete dilogarithm correction which makes the change of variables $\mathbf{Y}\rightarrow\mathbf{Z}$ look natural, in view of the properties of the discrete dilogarithm \cite{baxter}. 
}
\begin{equation}
 \mathbf{Y}=\left(\frac{1-\mathbf{Z}}{\mathbf{Z}}\right),
\end{equation} 
then eqn.\eqref{Yfix1} becomes
\begin{equation}
(1- \mathbf{Z})^{\mathbf{1}\,\otimes\,C_{G^\prime}}= \mathbf{Z}^{C_G\,\otimes\,\mathbf{1}},
\end{equation} 
or, equivalently,
\begin{equation}
(1- \mathbf{Z})= \mathbf{Z}^{\,C_G\,\otimes\,C^{-1}_{G^\prime}},
\end{equation} 
which is of the form \eqref{classlimmm} with
\begin{equation}
 A=C_{G}\otimes C_{G^\prime}^{-1}, 
\end{equation} 
in agreement with the identification \eqref{guessGG}. Making $\mathbf{Z}\leftrightarrow 1-\mathbf{Z}$, corresponding to
$\mathbf{Y}\leftrightarrow \mathbf{Y}^{-1}$, implements the `level--rank' duality $A\leftrightarrow A^{-1}$.

It is also clear that the `level--rank' duality must be supplemented by a non--trivial field/source redefinition in order to keep the traces in the canonical $\chi_A$ form. 
\smallskip

We consider this argument as further evidence for the correctness of the identification \eqref{guessGG}.
\medskip

Later in this section, we shall check some of these expectactions in simple models. 
In particular, we shall see elementary examples of the $A\leftrightarrow A^{-1}$ duality.
However, to check the level--rank duality in full generality is probably quite hard. 
Indeed,
the relation between the corresponding characters, $\chi_A$ and $\chi_{A^{-1}}$, seen as an identity between $q$--series, is in itself a remarkable theorem in combinatorics and Number Theory which may or may be not known to the math literature. Below we shall establish some \textit{new} deep identity of this kind just by considering the very simplest $\cn=2$ theories. Even to state the precise mathematical equality one has to check is a rather complicated issue, due to the fact that the `trivial' overall (relative) normalizations appearing in the identifications, while \emph{dynamically} trivial, as functions of $q$ are higher transcendental objects.

Luckily, however, there are some examples in which the $q$--series equalities have already been established. For instance, for the dual pair
of the $(2,m)$/$(m-2,m)$ minimal models it is known that the respective characters are related by a $q\leftrightarrow q^{-1}$ duality
\cite{Berkovich:1994es}. In these cases one can be very explicit.

In the next subsection we argue that the Verlinde algebra of the corresponding RCFT is 
generated by the $X_i$ in the classical limit, in a canonical way.  Moreover the above fixed semi-classical
loci diagonalize the Verlinde ring.

\subsection{Line operators and Verlinde ring for $(G,G')$ models}\label{verl}

In the previous sections we have seen that the path integral
that computes the trace of the monodromy operators will give rise
at least in the $(G,G')$ theories to characters of RCFT's. 
 Moreover the insertion of the line operators
$X_i$
in the path integral corresponds to changing the characters of the RCFT (i.e. shifting the lattice sum by
linear terms in the lattice momentum). 
$${\rm Tr}\, K(q)\big(\prod X_r\big)\rightsquigarrow \sum_\alpha c_\alpha \chi_\alpha$$
and insertion of an extra $X_i$ changes the characters, and so we have roughly the structure
$$X_i \cdot \chi_j =C_{ij}^k\,\chi_k$$
The question is whether this can be made precise, and in particular if the $C_{ij}^k$ can be
made to be suitable positive integers as is required for the Verlinde algebra \cite{Verlinde:1988sn}.
Though this fact is not known in the full generality, in many cases (e.g. $(A,A')$ cases)  this turns out to
be true, as we will show using the work of Nahm and collaborators \cite{nahm,Nahm:1992sx}, in relation to TBA systems.
To make this more clear we consider the limit $q\rightarrow 1$ discussed in the previous section.
In this limit we expect $X_i$ to be represented by c-numbers, consistent with the recursion
relations.  In particular a suitable basis for $X_i$'s were discussed and denoted by $\mathbf{Z}$ and were
shown to satisfy
\begin{equation}
(1- \mathbf{Z})= \mathbf{Z}^{\,C_G\,\otimes\,C^{-1}_{G^\prime}},
\end{equation} 
which can be viewed as the equations defining the values of $\mathbf{Z}$ fixed by the monodromy.
That they should be fixed by the monodromy to contribute to the trace is clear, as the pieces
not fixed by it will give zero contribution, by the trace property.
The solution to these equation, a question motivated  by TBA,  has been studied in the literature
and in particular in \cite{nahm,Nahm:1992sx}.   Following the approach in \cite{nahm} we define a new set of variables
$$\Phi^{1\,\otimes\,C_{G^\prime}}=\mathbf{Z}^{-1}$$
which leads to
$$(1-\Phi^{-\, 1\,\otimes\,C_{G^\prime}})=\Phi^{-\,C_G\,\otimes\,1}.$$
Multiplying both sides by $\Phi^2$ and rearranging terms yields
\begin{equation}\label{verver}
\Phi^2= \Phi^{2 (1\,\otimes\,1)-\, 1\,\otimes\,C_{G^\prime}}+\Phi^{2 (1\,\otimes\,1)-\,C_G\,\otimes\,1}.
\end{equation} 
Note that since $2-C$ is a positive integral matrix, this equation is a relation with
integral powers and positive terms, which could in principle be a consequence of the Verlinde algebra,
Indeed this relation between the fixed points of the $Y$--system and the corresponding Verlinde algebra 
works as pointed out by Nahm and collaborators, at least for the $(A_{m-1},A_{n-1})$ theories.  In that context
the fields $\Phi_{a,i}$, with $1\leq a \leq (m-1)$ and $ 1\leq i \leq (n-1)$, get related to the $SU(m)_n$ characters 
(whcih are the main building blocks of the corresponding coset $SU(m)_n/U(1)^{m-1}$ fields) and are labeled by the $A_{m-1}$ and $A_{n-1}$ nodes respectively and
correspond to representations $ \rho_{ai}$ of $SU(m)$ in the $i$-th power of the $a$-th anti-symmetric fundamental representation \cite{nahm}.  In particular the solutions of the above system are given by 
$$\Phi_{a,i}={\rm Tr}_{\rho_{ai}}(g)$$
where
$g$ are particular elements of order $(m+n)$ (or $2(n+m)$ if $m$ is odd) in $SU(m)$, as is familiar in the context of diagonalization
of the Verlinde algebra.  We will see in the context of some examples, how similar Verlinde algebras
arise.  For example, we will show how the Verlinde algebra for the $(2,5)$ model arises for the full mondromy
in the $(A_2,A_1)$ models corresponding to the $(A_1,T_1)$ TBA systems.

\subsubsection{The cluster--algebra interpretation}

In view of future extensions to theories more general than the $(G,G^\prime)$ ones, we briefly comment on the interpretation of eqn.\eqref{verver} in the context of cluster--algebras. Let $Q\: (\equiv G\,\square\,G^\prime)$ be the quiver associated to the given model. We attach a variable $\Phi_k$ to each vertex $k\in Q$. Then eqn.\eqref{verver}  may be written in the form
\begin{equation}\label{exhangerelation}
 \Phi_k\cdot \Phi_k =\prod_{\text{arrows}\atop i\rightarrow k}\Phi_i + \prod_{\text{arrows}\atop k\rightarrow j}\Phi_j,
\end{equation}  
which extends to any quiver $Q$ (without $1$-- or $2$--cycles).

In fact, eqn.\eqref{exhangerelation} is the fixed--point version of the basic defining relation of a cluster--algebra, namely the \emph{exchange relation}
\cite{MR2383126,MR2132323,MR2295199,cluster-intro}, which generalizes the Ptolemy relation of Euclidean geometry. From the algebraic viewpoint, the exchange relation is more fundamental than the one associated to the $Y$--system \cite{MR2295199,cluster-intro}.

Indeed, in the math literature \cite{MR2295199,cluster-intro} the definition of the quiver mutation at the vertex $k$, $\mu_k$, is supplemented by the instruction of changing the associated variable $\Phi_k$ into $\Phi_k^\prime$ defined by
\begin{equation}
 \Phi_k^\prime\cdot \Phi_k =\prod_{\text{arrows}\atop i\rightarrow k}\Phi_i + \prod_{\text{arrows}\atop k\rightarrow j}\Phi_j,
\end{equation} 
while $\Phi_j$ remains invariant for $j\neq k$. These relations can be also be interpreted as $T$--system 
whose relations with TBA and RCFT are known (see ref.\,\cite{tsyst1,tsyst2,tsyst3,tsyst4,tsyst5} and references therein).

In the special case of the $\mathbb{G}(k,n)$ cluster--algebras the exchange relations reduce to the Pl\"ucker ones, as we shall see in the explicit examples below.

\subsection{Representations and traces for quantum torus algebras}
 
In preparation for the explicit computations, we briefly discuss the meaning of the trace in our quantum algebras.

 There is a canonical definition of a trace over the quantum torus algebra, namely the unique normalized trace--state in the sense of $\mathbf{C}^*$--algebras and Non--Commutative geometry \cite{MR1303779}. For the quantum algebra with (invertible) generators $X_i$ and relations $X_iX_j=q^{\epsilon_{ij}}X_jX_i$, and any Laurent series
\begin{equation*}
 F(X_i)=\sum_{\ell_i\in\mathbb{Z}} a_{\ell_1,\ell_2,\cdots,\ell_s}\, X_1^{\ell_1}\,X^{\ell_2}_2\cdots X^{\ell_s}_s
\end{equation*}
 the canonical trace is
\begin{equation}
	\mathrm{Tr}_\mathrm{can}\, F(X_i)\equiv a_{0,0,\cdots,0}.
\end{equation}
\smallskip

When the torus algebra is associated to a bi--partite simply--laced quiver $Q$ with $V$ vertices (as it happens for the $(G,G^\prime)$ theories), a convenient Hilbert space realization of  the algebra can be given\footnote{\ Actually, only for $|q|=1$ the $X_i$'s are bounded operators defined on the whole $L^2(\mathbb{R})$. Otherwise they are only densely--defined.} in\footnote{\ We write $V_e$ (resp.\! $V_o$) for the number of \emph{even} (resp.\! \emph{odd}) vertices in the bi--partite quiver $Q$.} $L^2(\mathbb{R}^{V_e})$. One writes
\begin{equation}\label{hilbertrep}
	X_i = \begin{cases}
	\exp(i x_i) & i\ \text{even}\\
	\lambda_i\,\exp\big(2\pi\,\tau\, \epsilon_{ij}\,\partial/\partial x_j\big) & i\ \text{odd}. 
\end{cases}
\end{equation}
 where $q=\exp(2\pi i\,\tau)$ and $\lambda_i \in (\mathbb{C}^*)^{V_o}$. The conjugation
\begin{equation}
 X_{2k+1}\rightarrow \Big(\prod_j X_{2j}^{n_{2j}}\Big)^{-1}\, X_{2k+1}\,\Big(\prod_j X_{2j}^{n_{2j}}\Big),
\end{equation} 
sends $\lambda_{2k+1}$ into $q^{\epsilon_{2k+1,2j}\,n_{2j}}\, \lambda_{2k+1}$, so (up to conjugacy\footnote{\ If $|q|=1$, as in the physical theory, this means up to unitary equivalence.}) the $\lambda_i$'s take value in the $V_o$--dimensional complex torus\footnote{\ If the skew--symmetric pairing $\langle\gamma_i |\gamma_j\rangle$ is degenerate, $J$ has the form $T\times \mathbb{C}^m$, with $T$ a compact complex torus. } $J$
\begin{equation}\label{JacTorus}
J\colon\quad \lambda_{2k+1}\sim q^{\epsilon_{2k+1,2j}\,n_{2j}}\, \lambda_{2k+1}\qquad \forall\, n_{2j}\in \mathbb{Z}.
\end{equation}

In the Schroedinger picture \eqref{hilbertrep}, an operator $\co$ is represented by the integral kernel
 $\langle x_2,x_4,\cdots\,|\mathcal{O}|x_2^\prime,x_4^\prime, \cdots \rangle$, and the trace is simply the integral of the 
 kernel along the diagonal subspace, $x_{2\ell}=x_{2\ell}^\prime$.
 
 However, $L^2(\mathbb{R}^{[V/2]})$ is far from being an irreducible representation of the quantum algebra. In fact the dual torus algebra generated by the operators \cite{qd-cluster,qd-pentagon}
\begin{equation}
	X_i^\vee = \begin{cases}
	\exp\big(2\pi\,\partial/\partial x_i\big) & i\ \text{even}\\
	\exp\big(i\,\tau^{-1}\, \epsilon_{ji}^{-1}\,x_j\big) & i\ \text{odd}. 
\end{cases}	
\end{equation}
 commutes with the original torus algebra\footnote{\ This doubled quantum torus algebra
would naturally arise if we complete the cigar in our construction to an $S^2$.  It is the algebra
associated with the `anti-topological sector' given by the other half of $S^2$.}. Then, in order to extract the irreducible trace--blocks, it is convenient to consider more general expressions of the form
\begin{multline}\label{deftraL2}
	\hskip 0.9cm\mathrm{Tr}_{L^2}\Big( M(q)\, F(X_i^\vee)\Big)=\\
	=\int dx_{2i}\,\langle x_2,x_4,\cdots |\,M(q)\, F(X_i^\vee)\,|x_{2},x_4,\cdots\rangle,\hskip 1.7cm
\end{multline}
which we see as `generating functions' for the trace--blocks of $M(q)$.

Finally, one has a path integral representation of the form
\begin{equation}
 \mathrm{Tr}\big[\cdots\big]= \int\limits_\mathrm{periodic} \left[\prod d\phi^i\right] \Big(\cdots\Big)\:\exp\left[\frac{i}{4\pi \tau}\int dt\:\epsilon_{ij}\, \phi^i\, \frac{d\phi^j}{dt}\right],
\end{equation} 
which is convenient for general arguments as those in the previous subsection.

\subsection{The trivial quiver}

The simplest possible example is the trivial quiver, $A_1\,\square\,A_1$, consisting of just one node and no arrows.
From the point of view of TBA, $(A_1,A_1)$ corresponds to a free fermion (trivial $S$--matrix). In this case, the quantum algebra is generated by a single invertible operator $Y$, and hence it is commutative.
The monodromy acts trivially since all adjoint actions are trivial in a commutative algebra. This makes the $(A_1,A_1)$ example somewhat \textit{degenerate.} In particular, in an irreducible representation $Y$ acts as a $c$--number. 

However, $M(q)$, while commuting, is not a trivial function of $q$. Indeed\footnote{\ For convenience, we redefines $M(q)$ by multiplying it by $q^{-1/24}$, which is a pure phase in the physical situation $|q|=1$.},
\begin{equation}
 M(q)=q^{-1/24}\,\Psi(Y;q)\, \Psi(Y^{-1};q)= \frac{1}{\eta(q)} \sum_{k\in\mathbb{Z}} q^{k^2/2}\,(-Y)^k\equiv \frac{\Theta(-Y;q)}{\eta(q)},
\end{equation}
which is the partition function of a complex massless free fermion. Since $M(q)=Y(q)^2$, the half $R$--twist $Y(q)$ should correspond to a single (real) massless free fermion which is the CFT limit of the $(A_1,A_1)$ integrable theory. 

The canonical trace,
\begin{equation}
 \mathrm{Tr}_\mathrm{can}\, M(q)= \frac{1}{\eta(q)}.
\end{equation} 
is just a particular sum over the different irreducible representations. The same is true for the trace of $Y(q)$
\begin{equation}
\mathrm{Tr}\,Y(q;\lambda)= \Psi(\lambda;q)\equiv \prod_{k=0}^\infty (1-q^{n+1/2}\lambda).
\end{equation}
This example clarifies which kind of `trivial' factors we shall expect to appear in the monodromy traces.

\subsection{The $(A_2,A_1)$ theory}

The first non--trivial example is $A_2\,\square\, A_1$. We shall first check the main themes of this section on this model, where detailed computations are doable.  In this case we shall find that ${\rm Tr}\,M(q) $ lead to characters of the
$(A_1,T_1)$ TBA, \textit{i.e.}\! $(2,5)$ minimal model corresponding to Lee-Yang edge singularity.  Furthermore
we show $q\rightarrow q^{-1}$ leads to characters of $(3,5)$ model.  We also consider the trace ${\rm Tr}\, Y(q)$ of the
half-monodromy for the $(A_2,A_1)$ theory and find that it leads to characters of the $A_2$ level 2 parafermionic system 
corresponding to the $SU(3)_2/U(1)^2$ coset.
  
The quantum algebra in this case has just two generators, $X$, $Y$, satisfying the relation $XY=q\,YX$. The monodromy is
\begin{equation}\label{mon:2,5}
	M(q)=\Psi(Y;q)\,\Psi(X;q)\, \Psi(Y^{-1};q)\,\Psi(X^{-1};q),
\end{equation}
with adjoint action
\begin{align}\label{adact2,5}
&M(q)^{-1}\,X\,M(q)=Y^{-1}\\	
&M(q)^{-1}\,Y\,M(q)=(1-q^{1/2}Y)X,\label{adact2,5,2}
\end{align}
of order $r=5$.

\subsubsection{Canonical trace $\rightarrow$ characters of the $(2,5)$ minimal model}\label{sec:cantrace}
The simplest way to compute the canonical trace is to exploit the following basic identities for the quantum dilogarithm\footnote{\ The first identity is the usual Euler product \cite{andrews}, the second one the Jacobi triple product identity, while the last one is proven in ref.\cite{Faddeev:1993rs}.} (valid for $|q|<1$),
\begin{align}
	\label{Eulerid}&\Psi(X;q)\equiv\sum_{k\geq 0}\frac{(-1)^k q^{k^2/2}}{(q;q)_k} X^k \qquad \text{where } (a;q)_m=\prod_{k=0}^{m-1}(1-aq^{k})\\
	&\Psi(X;q)\,\Psi(X^{-1};q)\equiv q^{1/24}\,\eta(q)^{-1}\,\sum_{k\in \mathbb{Z}}q^{k^2/2}(-X)^k\label{Eulerid2}\\
	&\Psi(X;q)\,\Psi(Y^{-1};q)=\Psi(Y^{-1},q)\, \Psi(-q^{-1/2}Y^{-1}X;q)\,\Psi(X;q)\label{Eulerid3}
\end{align}
 Using the third identity to commute the two central factors in eqn.\eqref{mon:2,5}, we get
\begin{multline}\label{monA2id}
	q^{-1/12}\,\eta(q)^2\, M(q)=\\
	= \sum_{m_1,m_2\in\mathbb{Z}}\sum_{\ell\geq 0} q^{(m_1^2+m_2^2)/2}\frac{q^{\ell^2/2}}{(q;q)_\ell} (-Y)^{m_1}(XY^{-1})^{\ell}
	(-Y)^{m_2}=\\
	=\sum_{a,b\in\mathbb{Z}} (-1)^{a-b}\,G_{a-b}(q)\: q^{(a-b)^2/2}\, X^b\,Y^a,\hskip 2.5cm
\end{multline}
where the functions $G_\ell(q)$ are defined by
\begin{equation}
 G_\ell(q)=\sum_{k=0}^\infty \frac{q^{k^2+k\ell}}{(q;q)_k}\qquad k\in\mathbb{Z}.
\end{equation} 
satisfy the three--terms recursion rule, as an example of \ref{recursion},
\begin{equation}\label{3termrec}
 G_{\ell+2}(q)= q^{-(\ell+1)}\Big( G_\ell(q)-G_{\ell+1}(q)\Big),
\end{equation} 
and give the sum of one of the celebrated Ramanujan continuous fractions \cite{MR1117903}
\begin{equation}\label{Ramcontfract}
 \frac{G_{k+1}(q)}{G_k(q)}= \frac{1}{1+\frac{q^{k+1}}{1+\frac{q^{k+2}}{1+\frac{q^{k+3}}{1+\cdots}}}}
\end{equation}
One has
\begin{equation}
	G_0(q)= G(q),\qquad G_1(q)=H(q),
\end{equation}
where $G(q)$ and $H(q)$ are the Rogers--Ramanujan functions \cite{MR1117903}, which express the sum of the basic Ramanujan continuous fraction (namely eqn.\eqref{Ramcontfract} with $k=0$). The celebrated Rogers--Ramanujan identities \cite{rogers,rama3,slater2}
allow to rewrite them as infinite products (for $\ell=0,1$)\footnote{We are informed by Ole Warnaar that
this may in fact be done for arbitrary $\ell$ using results of \cite{MR1722235}.}
\begin{equation}
 G_\ell(q)=\prod_{j=1}^\infty\big(1-q^{5j-1-\ell}\big)^{-1}\,\big(1-q^{5j-4-\ell}\big)^{-1}.
\end{equation} 

From eqn.\eqref{monA2id} we have
\begin{equation}
 \mathrm{Tr}_\mathrm{can}\big[ M(q)\, X^m\big]= (-1)^m\,\frac{q^{m^2/2+1/12}}{\eta(q)^2}\, G_m(q).
\end{equation} 
In particular, for $m=0$
\begin{equation}\label{trcan25}
	\mathrm{Tr}_\mathrm{can}\, M(q)= q^{1/12}\,\frac{G(q)}{\eta(q)^2},
\end{equation}
which, again, may be written as an infinite product.

To correctly interpret the above equation, we have to recall that the adjoint action \eqref{adact2,5}\eqref{adact2,5,2} defines $M(q)$ only up to an overall normalization which may be a non--trivial function of $q$. We wish $M(q)$ to be unitary for $q$ a phase (real couplings of the CS theory), so the overall function should be a power of $q$. If we redefine

\begin{equation}
	M(q)\rightarrow q^{-1/10}\, M(q)
\end{equation}
 we get
\begin{equation}\label{firtchara2,5}
	\mathrm{Tr}_\mathrm{can}\, M(q)= \frac{q^{-60}\,G(q)}{\eta(q)^2}.
\end{equation}
which is in fact a $\Gamma_1(5)$--modular function \cite{MR1021804}.

More physically, apart from the `trivial' factor $\eta(q)^{-1}$, the function is the modular character $\chi_{1,3}(q)$ of the $(2,5)$ minimal model. This model has two independent characters, namely
\begin{equation}
	\chi_{1,3}^{(2,5)}(q)=q^{-1/60}\, G(q)\hskip 1.5cm \chi_{1,1}^{(2,5)}(q)=q^{11/60}\, H(q),
\end{equation}
which are transformed one into the the other under modular transformations. Apart from trivial factors, the two characters are precisely given by
\begin{equation}
 \chi_{1,3}^{(2,5)}\propto \mathrm{Tr}_\mathrm{can}\big[ M(q)\big],\qquad  \chi_{1,1}^{(2,5)}\propto \mathrm{Tr}_\mathrm{can}\big[ M(q)\, X\big],
\end{equation} 
while the Ramanujan three--terms recursion relation \eqref{3termrec} just means that all the traces
\begin{displaymath}
 \mathrm{Tr}_\mathrm{can}\big[ M(q)\, X^m Y^n\big]\qquad m,n\in\mathbb{Z},
\end{displaymath}
are expressed as linear combinations of the two basic characters $\chi_{1,3}^{(2,5)}$ and $\chi^{(2,5)}_{1,1}$.
Both characters enter in the expression of $\mathrm{Tr}\,M(q)$ on a general representation of the quantum torus algebra. (Notice that the second $(2,5)$ character is, essentially, the coefficient of $X$ in the expansion in the \textsc{rhs} of eqn.\eqref{monA2id}).

The general recursion relations \eqref{recursion} which, as discussed in \S.\,\ref{evidence} define the traces,
 in the $(A_2,A_1)$ case reduces to the Ramanujan one \eqref{3termrec}.  In fact the arguments of \S.\,\ref{evidence} give us a lot more:  Let 
$$\Phi =-X$$
and define
$$\langle\cdots\rangle ={\Tr}\big[\cdots M(q)\big]$$
The three term recursion relation then implies that as $q\rightarrow 1$:
$$\langle \Phi^2 \cdots\rangle =\langle (\Phi +1)\cdots\rangle$$
where $\cdots$ stands for any line operators. Mathematically, this is, of course, just the fixed point equation for $(A_2,A_1)$ classical monodromy. Physically, this equation says that the line operator $\Phi$
is localized on a subspace which realizes the Verlinde algebra of the $(2,5)$ model:
$$\Phi \times \Phi=1+\Phi$$
This is indeed remarkable\,!  Not only we are getting the characters of the $(2,5)$ model,
but also we are finding a natural realization of the generators of the Verlinde algebra
in terms of the line operators!

\subsubsection{$\mathrm{Tr}_\mathrm{can}\,M(q)$ and the quiver of $(A_2,A_1)$}

There is another (equivalent) way of writing $\mathrm{Tr}_\mathrm{can}\,M(q)$. The equality of the two expressions may be regarded as a new identity of the Rogers--Ramanujan type. This second formulation has the advantage of shedding light on the relation between the characters of the $(2,5)$ minimal model and the Dynkin diagram of the $A_2$ Lie algebra.

To get the alternative expression, one starts from the definition \eqref{mon:2,5} of $M$, uses the Euler identity
\eqref{Eulerid} to write each elementary factor as a Laurent series, and takes the canonical trace. One finds
\begin{equation}\label{secondform}
 \mathrm{Tr}_\mathrm{can}\,M(q)=\sum_{m_1,m_2\geq 0} \frac{q^{m_1^2+m_2^2-m_1m_2}}{(q;q)^2_{m_1}\,(q;q)^2_{m_2} }\equiv \sum_{\textbf{m}\in \mathbb{N}^2} \frac{q^{\mathbf{m}^t C_2\otimes C_1^{-1} \mathbf{m}}}{(q;q)^2_{m_1}\:(q;q)^2_{m_2}},
\end{equation} 
where $C_2$ (resp.\! $C_1$) is the Cartan matrix of $A_2$ (resp.\! $A_1$). 

One checks that \eqref{secondform} coincides with \eqref{trcan25}.

\subsubsection{Duality between the $(2,5)$ and $(3,5)$ minimal models} 
The `level--rank' duality $A_2\,\square\, A_1\,\leftrightarrow\, A_1\,\square\,A_2$ maps the $(2,5)$ minimal model into the $(3,5)$ one.  
The duality of the characters between the minimal models    
$(p,p^\prime)$ and $(p^\prime-p,p^\prime)$ has beeen established, from a purely CFT viewpoint, in ref.\cite{Berkovich:1994es} (see also \cite{MR1669957} Section 5.6): In a sense, the two set of characters are interchanged by the formal operation $q\rightarrow q^{-1}$.
Of course, we have to expect a relation of the characters only up to overall `trivial' factors.

The operation $q\rightarrow q^{-1}$ maps the elementary factors $\Psi(X^{\pm 1};q)$, $\Psi(Y^{\pm 1};q)$ into their
inverses $\Psi(X^{\pm 1};q)$, $\Psi(Y^{\pm 1};q)$. To be a symmetry of the torus algebra, the inversion of $q$ should be supplemented by $X\leftrightarrow Y$. Then the operation has the effect\footnote{\ As always, $\sim$ means equality up to conjugacy.}
\begin{multline}
 M(q)=\Psi(Y;q)\, \Psi(X;q)\, \Psi(Y^{-1};q)\, \Psi(X^{-1};q) \rightarrow\\
\rightarrow\Psi(X;q)^{-1}\, \Psi(Y;q)^{-1}\, \Psi(X^{-1};q)^{-1}\, \Psi(Y^{-1};q)^{-1}\sim\\
\sim \Psi(X^{-1};q)^{-1}\,\Psi(Y^{-1};q)^{-1}\, \Psi(X;q)^{-1}\, \Psi(Y;q)^{-1}\equiv\\
\equiv M(q)^{-1}.\hskip 7cm 
\end{multline}
Therefore, we expect to find the characters of the $(3,5)$ minimal model in the trace $\mathrm{Tr}\,M(q)^{-1}$.
This would exactly correspond to the analytic relation between the characters of the $(2,5)$ and $(3,5)$ minimal models discussed in ref.\cite{Berkovich:1994es}.

We compute the trace in the $L^2(\mathbb{R})$ (reducible) representation with $\lambda_1=1$
\begin{equation}
 X=\exp(ix),\qquad Y=\exp(-2\pi i\tau d/dx)
\end{equation}  We first note that the operator
\begin{equation}
	\mathcal{M}=\mathfrak{F}\, \Psi(Y;q)^{-1}=\Psi(X^{-1};q)^{-1}\,\mathfrak{F},
\end{equation}
 where $\mathfrak{F}$ is the Fourier transform normalized as\footnote{\ Here and elsewhere $q=\exp(2\pi i\tau)$ with $\mathrm{Im}\,\tau>0$.}
\begin{equation}
	\mathfrak{F}\psi(y)= \frac{1}{2\pi \sqrt{\tau}}\int\limits_{-\infty}^{+\infty} dx\, e^{-i xy/2\pi \tau}\,\psi(x),
\end{equation}
has the properties
\begin{align}
 \cm^{-1} X\cm&= Y^{-1}\\
\cm^{-1} Y\cm&= (1-q^{1/2}Y)X,
\end{align}
and so it may be regarded as a representation of the quantum monodromy in the $L^2(\mathbb{R})$ Hilbert space. 

We compute the trace of $\cm^{-1}$ following the strategy outlined around eqn.\eqref{deftraL2}.
In order to get more symmetric--looking formulae, we make the special choice for the function $F(X^\vee,Y^\vee)=\Psi(Y^\vee; \tilde q)$, where $\tilde q=\exp(-2\pi i/\tau)$ is the $S$--modular transform of $q$. Then
\begin{equation}
 \mathrm{Tr}_{L^2}\Big(\Psi(Y^\vee; \tilde q))\,\cm^{-1}\Big)=\frac{1}{2\pi\, \sqrt{\tau}}
\int\limits_{-\infty}^{+\infty} dx\, \Psi(e^{i x/\tau};\tilde q)\, \Psi(e^{ix};q)\, \exp(i x^2/2\pi \tau).
\end{equation}
The Gaussian factor in the integrand has norm $|\exp(i x^2/2\pi\tau)|=\exp\big(\frac{\mathrm{Im}\,\tau}{2\pi |\tau|^2}x^2\big)$, so it is absolutely convergent for $\mathrm{Im}\,\tau <0$ (as is natural, since we obtained this expression by making, formally, $q\rightarrow q^{-1}$). Expanding the $\Psi$ functions in powers using the Euler identity \eqref{Eulerid} and performing the resulting Gaussian integrals, we get
\begin{equation}
 \mathrm{Tr}_{L^2}\Big(\Psi(Y^\vee; \tilde q))\,\cm^{-1}\Big)=
\frac{e^{i\pi/4}}{\sqrt{2}}\sum_{k,l\geq 0}(-1)^{kl} \frac{(-1)^k q^{l^2/4}}{(q;q)_l}\;\frac{(-1)^l \tilde q^{(k^2+2k)/4}}{(\tilde q;\tilde q)_k},
\end{equation}
where the sums now are convergent in the usual \emph{upper} half--plane. The \textsc{rhs} can be written as a bilinear in the four blocks (here $(q)_m\equiv (q:q)_m$)
\begin{align}
 &\chi^{(3,5)}_{1,2}(q)=\sum_{m\geq 0\atop \text{even}}\frac{q^{m^2/4}}{(q)_m}
& &q^{1/4}\,\chi^{(3,5)}_{1,3}(q)=\sum_{m\geq 0\atop \text{odd}}\frac{q^{m^2/4}}{(q)_m}\\
&\chi^{(3,5)}_{1,1}(\tilde q)=\sum_{m\geq 0\atop \text{even}}\frac{\tilde q^{(m^2+2m)/4}}{(\tilde q)_m}
&& \tilde q^{3/4}\,\chi^{(3,5)}_{1,4}(\tilde q)=\sum_{m\geq 0\atop \text{odd}}\frac{\tilde q^{(m^2+2m)/4}}{(\tilde q)_m},
\end{align}
which are precisely the four conformal blocks of the $(3,5)$ minimal model \cite{Kedem:1993ze}.
Each of these characters has an expression as an infinite product, thanks to the generalized Rogers--Ramanujan identities of Slater \cite{slater2}.

On the other hand, using the Rogers identities \cite{slater2,alladi} we may rewrite the above functions in terms of the Rogers--Ramanujan functions $G(\cdot)$, $H(\cdot)$ of argument $q^{1/4}$
\begin{equation*}
 \sum_{n\geq 0} \frac{(\pm 1)^m q^{n^2/4}}{(q)_n}=\frac{G(\pm q^{1/4})}{(-q^{1/2};q^{1/2})_\infty},\qquad
\sum_{n\geq 0} \frac{(\pm 1)^m q^{(n^2+2n)/4}}{(q)_n}=\frac{H(\pm q^{1/4})}{(-q^{1/2};q^{1/2})_\infty}.
\end{equation*}

\subsubsection{Verlinde algebra for $W^{sl(n)}(2,5)$ minimal models, $(A_n,A_1)$ theory and Hyperkahler space}\label{sec:Verlinde}

In section \ref{sec:cantrace} we have shown that the fixed--point equations for the $(A_2,A_1)$ model reproduces the Verlinde algebra of the corresponding $(2,5)$ minimal model. This is just an example of a general pattern 
as discussed in \S.\ref{verl}. 
In this section we will focus more generally on the $( A_n,A_1)$ theories, in the context of the full monodromy.  We will give evidence later in this paper that this corresponds to the $(A_{n-1},T_1)$ TBA system which in turn are believed
to correspond to deformations of the $(2,5)$ minimal models of $W^{sl(n)}$. 
As we have already noted the expectation values of the line operators
contributing to the trace correspond to R-symmetry invariant configurations, and on this
subspace they should realize the Verlinde algebra.
On the other hand the line operators can be viewed as coordinates of Hyperkahler manifold
corresponding to compactifications from 4 to 3 dimensions, and R-symmetry should
act on this space.  Thus these distinguished coordinates evaluated at the R-symmetric points should
realize the Verlinde algebra.  Moreover these points form a basis
where the Verlinde algebra is diagonalized.
The Verlinde algebra for the $(2,5)$ minimal models of Virasoro is well known
and we will recover the algebra for it.  We are not aware of the Verlinde algebra for the
$(2,5)$ minimal models of the $W^{sl(n)}$ model for the higher $n$'s, but the
result we find suggests that they should have at least a Verlinde sub-algebra realizing the algebra of the
ordinary
$(2,n+3)$ minimal model for the Virasoro algebra.

 We begin by recalling a few basic facts from \cite{Gaiotto:2009hg}
about the hyperkahler moduli space $\cm$ which arises upon compactifying $(A_1,A_n)$ theory,
from four to three dimensions on the circle (with no twist).
In this case, in its generic complex structure, $\cm$ can be described as a moduli space of $SL(2,\C)$ connections on 
$C = \C\PP^1$, regular everywhere except for a certain irregular singularity at $\infty$.  Temporarily fix a point on $\cm$, i.e. 
a particular flat connection.  There is then a two-dimensional space of flat sections, which is acted on by the $SL(2,\C)$ of
constant gauge transformations.  By studying the asymptotics 
of the flat sections near the singularity, one obtains $n+3$ distinguished lines in this space --- or said otherwise,
$n+3$ distinguished points $z_i$ on the $\C\PP^1$ of projectively flat sections.  As we vary the flat connection these 
$n+3$ points vary arbitrarily, except that consecutive 
points in the cyclic ordering never coincide.  Conversely such a configuration of $n+3$ points 
actually determines a flat connection up to gauge.  So altogether we find that $\cm$ is a dense open subset of 
$(\C\PP^1)^{n+3} / SL(2,\C)$.  The classical monodromy operator $M$ acts on $\cm$ simply, by the cyclic shift of
the $n+3$ points by $2$ units.

For simplicity we restrict to the 
case of $n$ even.  Write $z_{i,j} = z_i - z_j$, and then for $0 \le k \le n+1$, define the $SL(2,\C)$-invariant combinations
\begin{equation}
 L_k = \begin{cases} z_{1,k+2} \frac{z_{2,3} z_{4,5} \cdots z_{k,k+1}}{z_{1,2} z_{3,4} \cdots z_{k+1,k+2}} & k \textrm{ even,} \\ 
                        z_{1,k+2} \frac{z_{k+3,k+4} z_{k+5,k+6} \cdots z_{n-1,n}}{z_{k+2,k+3} z_{k+4,k+5} \cdots z_{n,1}} & k \textrm{ odd.}
          \end{cases}
\end{equation}

Here we consider what happens when we restrict these functions to the fixed locus of $M$.  
On this locus $z_{i,j}$ depends only on $i-j$ mod $n+3$.
Call this quantity $p_{i-j}$, and denote the restriction of $L_k$ to the fixed locus as $\Phi_k$; then we have simply
\begin{equation}
\Phi_k = \frac{p_{k+1}}{p_1}.
\end{equation}
Note in particular the relations
\begin{equation} \label{verrel}
\Phi_0 = 1, \qquad \Phi_{k} = - \Phi_{n+1-k}.
\end{equation}
So we have nontrivial independent functions $\Phi_0, \dots, \Phi_{\frac{n}{2}}$ on the $M$-fixed locus.
Moreover, applying the ``Pl\"ucker'' relations
\begin{equation}
z_{i,j} z_{k,l} + z_{k,i} z_{j,l} + z_{j,k} z_{i,l} = 0
\end{equation}
with $i = 1$, $j = r + 2$, $k = 2$, $l = \ell + 3$, gives
\begin{equation} \label{verlinde}
\Phi_r \Phi_\ell = \Phi_{\ell - r} + \Phi_{r - 1} \Phi_{\ell + 1}. 
\end{equation}
Note that for $r=l$ this is exactly the kind of relation we have already anticipated  (\ref{exhangerelation}).
For $n=2$ this gives the Verlinde ring of the $(2,5)$ model (modulo $\Phi_1\rightarrow -\Phi_1$).
For the general case, by induction this can also be written
\begin{equation} \label{verlinde2}
\Phi_r \Phi_\ell = \Phi_{\ell - r} + \Phi_{\ell - r + 2} + \cdots + \Phi_{\ell + r - 2} + \Phi_{\ell + r}.
\end{equation}

Remarkably, \eqref{verlinde2} is  the Verlinde algebra \cite{Verlinde:1988sn} of the $(2,n+3)$
minimal model.  It would be interesting to see if the algebra of $W^{sl(n)}(2,5)$,
which is the model we would expect, is ismorphic to this.

\subsubsection{The trace of the half monodromy}

In all the $(G,A_1)$ models, the quantum monodromy $M(q)$ is the square of the operator $Y(q)$ which generates the solution to the associated $Y$--system, $M(q)=Y(q)^2$. It is natural to consider $\mathrm{Tr}\, Y(q)$. 

In the particular case of the $(A_2,A_1)$ model, one has\footnote{\ As always in the present paper, the real meaning of this statement is that $Y(q)^5$ is a central element commuting with all generators of the quantum torus algebra, namely it is the adjoint action of $Y(q)^5$ which is the identity.} $Y(q)^5=\mathbf{1}$. Thus $Y(q)=M(q)^{-2}$. Going to the $L^2(\mathbb{R})$ representation, we consider
\begin{equation}
	\mathcal{M}^{-2}= \Big( \Psi(X^{-1};q)^{-1}\, \mathfrak{F}^2\,\Psi(Y;q)^{-1}\Big)^{-1}=\Psi(Y;q)\, P\, \Psi(X^{-1};q),
\end{equation}
where $P=\mathfrak{F}^2$ is the parity operation $P\colon x\rightarrow -x$. Then
\begin{multline}\label{tracehalf}
\mathrm{Tr}_{L^{2}}\Big[F_1(X^\vee)\,\mathcal{M}^{-2}\,F_2(Y^{\vee})\Big]=\\
=\int dx\  \langle x | \Psi(e^{-2\pi \tau d/dx};q)\, F_1(e^{-2\pi d/dx})|-x\rangle\: \Psi(e^{-ix};q)\, F_2(e^{i x/\tau})\,\hskip 0.6cm
\end{multline}
From the general identity,
\begin{multline*}
	\int dx\, \langle -x| \sum_{n,m}a_n b_m\, e^{-2\pi (n\tau+m) d/dx}|x\rangle\: H(e^{-ix})\, F(e^{ix/\tau})\,=\\
	=\frac{1}{2}\sum_{n,m} a_n b_m\, H((-1)^m\,q^{-n/2})\,F_2((-1)^n\,q^{\vee\ m/2})=\\
	= \text{linear combination of } \sum_{n\ \mathrm{even}} a_n\, H(\pm q^{-n/2}),\ 
\sum_{n\ \mathrm{odd}} a_n\, H(\pm q^{-n/2}),\hskip 1.5cm 
\end{multline*}
we see that the expression \eqref{tracehalf} may be written in terms of the four trace--blocks
\begin{equation}
 \chi(\vec Q; q)=\sum_{(m_1,m_2)\in \mathbb{N}^2\atop (m_1,m_2)=\vec Q\mod 2}\frac{q^{(m_1^2+m_2^2-m_1m_2)/2}}{(q;q)_{m_1}\: (q;q)_{m_2}}.
\end{equation} 
Writing $\mathbf{m}=(m_1,m_2)$ and introducing the matrix
\begin{equation}\label{charSU(3)}
 A= C_{A_2}\otimes C_{A_1}^{-1}= \begin{pmatrix} 1 & -\frac{1}{2}\\ -\frac{1}{2} & 1\end{pmatrix}
\end{equation} 
we write the trace--blocks of $\mathrm{Tr}_{L^2}\, Y(q)$ in the general form of \S\S.\ref{TBAADET}\ref{evidence}
(see refs.\cite{Dasmahapatra:1993pw,Kedem:1993ze}) namely
\begin{equation}\label{TRYA2}
 \chi(\vec Q;q)= \sum_{\mathbf{m}\in \mathbb{N}^2\atop \mathbf{m}= \vec Q\mod 2} \frac{q^{\tfrac{1}{2}\textbf{m}^tA\textbf{m}}}{(q;q)_{m_1}\, (q;q)_{m_2}}
\end{equation}  
which correspond to conformal characters of the coset model \cite{Dasmahapatra:1993pw,Kedem:1993ze} (see also \cite{Terhoeven:1992zf}) 
\begin{equation}
 SU(3)_2/U(1)^2,
\end{equation} 
which is precisely the UV fixed point of the $A_2$ reflectionless scattering theory whose TBA is given by the $A_2$ $Y$--system. 

Thus, in the $(A_2,A_1)$ theory all the expectations of \S.\ref{evidence} are verified. The traces of the $K(q)$ give the characters of the unitary CFT theory whose massive integrable deformation leads to the $A_2$ elastic $S$--matrix\,!
\smallskip

\textbf{Remark}. Comparing eqns.\eqref{charSU(3)} and \eqref{secondform} we see that $\mathrm{Tr}\,M(q)=\mathrm{Tr}\, K(q)^2$ is given by double series whose $(m_1,m_2)$--th term is the square of the corresponding term in the double series giving the trace of $K(q)$.

\subsection{$(G,A_1)$ models}

The quiver of the $(G,A_1)$ theories (where $G=ADE$) has the $G$ Dynkin diagram as underlying graph.
We shall write $Y_j$, $X_k$ for the generators of the quantum torus algebra associated, respectively, to
odd and even vertices of $G$.

\subsubsection{$\mathrm{Tr}\, Y(q)$ for the $(G,A_1)$ models}

One has
\begin{equation}
 Y(q)= I\,\Big(\prod \Psi(Y_j;q)\Big)\Big(\prod \Psi(X_k;q)\Big),
\end{equation} 
where $I$ is the inversion (or parity in the Schroedinger picture) automorphism of the quantum torus algebra.

In the Schroedinger representation of eqn.\eqref{hilbertrep} with the $\lambda_i$'s set to $1$, for any \emph{normal--ordered} Laurent monomial $N\big[\prod X_j^{m_j}\prod Y_k^{n_k}\big]$ one has the identity:
\begin{equation}
 \mathrm{Tr}_{L^2}\Big( I\cdot N\Big[\prod X_j^{m_j}\prod Y_k^{n_k}\Big]\Big)=C,
\end{equation} 
for a constant $C$ that we set to $1$ by rescaling the trace.
Then
\begin{equation}\label{TRYG}\begin{split}
 \mathrm{Tr}_{L^2}\big[Y(q)\big]\Big|_{\lambda_i=1}& = \sum_{\mathbf{m}\in \mathbb{Z}^r_+}
(-1)^{|\mathbf{m}|}\,
\frac{q^{\mathbf{m}^t\cdot C_{G}\cdot \mathbf{m}/4}}{(q)_{m_1}\, (q)_{m_2}\cdots (q)_{m_r}}\\
&=\sum_{\mathbf{m}\in \mathbb{Z}^r_+}
\frac{q^{\tfrac{1}{2}\mathbf{m}^t \cdot C_{G}\otimes C_{A_1}^{-1}\cdot \mathbf{m}+B^{(0)}\cdot\mathbf{m}}}{(q)_{m_1}\, (q)_{m_2}\cdots (q)_{m_r}}
\end{split}
\end{equation} 
where\footnote{\ The apparent discrepance in sign with respect to eqn.\eqref{TRYA2} for $G=A_2$, is due to a different convention for the sign of the square root $q^{1/2}$; the `natural' convention from the $4d$ viewpoint is the opposite one with respect to the usual one for $2d$ solvable systems; cfr.\! the sign redefinitions in the discussion of section \ref{quivers}).} $|\mathbf{m}|=\sum_i m_i$ and $2\tau\,B^{(0)}=(1,1,\dots, 1)$. 

Eqn.\eqref{TRYG} corresponds precisely to our identification \eqref{guessGG}. \smallskip

To get the full
 set of $\chi_A(q;B,C)$ characters with $A=C_G\otimes C^{-1}_{A_1}$, one must consider more general traces 
$$\mathrm{Tr}\Big[Y(q)\, \prod X_j^{n_j}\prod Y_k^{m_k}\Big].$$ 
The recursion relations \eqref{recursion} then give all such correlations in terms of a \textit{finite} number of linearly independent (over the field $\mathbb{C}(q)$) characters. 

TBA identifies the above $\chi_A$ characters with those of the coset model
\cite{Dasmahapatra:1993pw,Kedem:1993ze}
\begin{equation}
 (G^{(1)})_{2}/U(1)^r\qquad \text{where } r=\mathrm{rank}\, G,
\end{equation} 
namely the so--called \emph{generalized} parafermionic theory ($G=A_n$ gives the standard $\mathbb{Z}_{n+1}$ parafermions).

\subsubsection{$\mathrm{Tr}\,Y(q,\boldsymbol{z})$ as a function on the torus $J$}

We reintroduce the coordinates $\lambda_i$ of the complex torus $J$, eqn.\eqref{JacTorus}. We define the vector
$\boldsymbol{z}=\{z_i\}_{i=1,\dots, r}$ with components
\begin{equation}
 z_i=\begin{cases}
      \log\lambda_i/(2\pi i\tau) & i\ \text{odd}\\
0 & i\ \text{even},
     \end{cases}
\end{equation}
the $z_i$'s being well--defined up to the identifications
\begin{equation}
  z_i\sim z_i+m_i/\tau+ \Omega_{i\alpha}\,n_\alpha,\qquad m_i,\ n_\alpha\in \mathbb{Z},
\end{equation} 
with $\Omega_{i\alpha}$ the rectangular $V_o\times V_e$ matrix $\langle \gamma_i,\gamma_\alpha\rangle$.

Then eqn.\eqref{TRYG} generalizes to
\begin{equation}\label{TRYGz}
 \mathrm{Tr}_{L^2}\big[Y(q,\boldsymbol{z})\big] = \sum_{\mathbf{m}\in \mathbb{Z}^r_+}
\,
\frac{q^{\tfrac{1}{2}\mathbf{m}^t\cdot A\cdot \mathbf{m}/4+(B^{(0)}+\boldsymbol{z})\cdot\mathbf{m}}}{(q)_{m_1}\, (q)_{m_2}\cdots (q)_{m_r}}.
\end{equation} 

We see eqn.\eqref{TRYGz} as a nice confirmation of our general picture.

\subsubsection{The canonical trace of $M(q)$}\label{cantraceAn}

The monodromy operator for the $(G,A_1)$ theory reads
\begin{equation}\label{MGA1}
 M(q)= \Big(\prod \Psi(Y_j;q)\Big)\Big(\prod \Psi(X_k;q)\Big)\Big(\prod \Psi(Y_l^{-1};q)\Big)\Big(\prod \Psi(X^{-1}_i;q)\Big).
\end{equation} 

We replace each $\Psi$ in eqn.\eqref{MGA1} with its Euler expansion \eqref{Eulerid}, and take the term of order zero in $X_j$, $Y_k$. One gets\footnote{\ $\mathbf{Z}^r_+$ stands for the $r$--tupple of non--negative integers.}
\begin{equation}
 \mathrm{Tr}_\mathrm{can}\, M(q)=\sum_{\mathbf{m}\in \mathbb{Z}^r_+}\frac{q^{\mathbf{m}^t\cdot C_{G}\cdot \mathbf{m}/2}}{(q)^2_{m_1}\, (q)^2_{m_2}\cdots (q)_{m_r}^2},
\end{equation}  
where $r$ is the rank of $G$ and $C_{G}$ its Cartan matrix.

More generally, one has the formula
\begin{equation}\label{tracesGA1}
 \mathrm{Tr}_\mathrm{can}\big[M(q)\,\prod_i X_i^{a_i}\big] =(-1)^{\sum_i a_i}\, q^{\mathbf{a}\cdot\mathbf{a}/2}\,
\sum_{\mathbf{m}\in \mathbb{Z}^r_+}\frac{q^{\mathbf{m}^t\cdot C_{G}\cdot \mathbf{m}/2- \mathbf{a}\cdot \mathbf{m}}}{(q)^2_{m_1}\, (q)^2_{m_2}\cdots (q)_{m_r}^2}
\end{equation} 
where $\mathbf{a}$ is an integral vector with vanishing odd entries. Again, we may write a finite--number--of--terms recursion relation for the traces \eqref{tracesGA1}, meaning that they may all be written in terms of a finiter set of trace--blocks (namely characters).   For $G=A_n$, these correspond to characters of the $W^{sl(n)}(2,5)$ model, i.e. the $(2,5)$
minimal model for the $sl(n)$ W-algebra \cite{Caldeira:2004jy}.  We show this by relating it to the characters
of the $(A_{n-1},T_1)$ TBA in the next section.

As in the $(A_2,A_1)$ example, the terms in the sum \eqref{tracesGA1} are, term by term, the \emph{square} of the corresponding ones in the sum \eqref{TRYG}, for $(G,A_1)$ half-monodromy.

\subsection{Wall--crossing and new $q$--series identities}

Comparing the two expressions for the canonical trace of $M(q)$ for the $(A_2,A_1)$ theory, eqns.\eqref{trcan25} and \eqref{secondform}, we get the identity
\begin{equation}\label{identitiyn=1}
	\sum_{m_1,m_2\geq 0}\frac{q^{\mathbf{m}\cdot C_{A_2}\cdot \mathbf{m}/2}}{(q)^2_{m_1}\,(q)^2_{m_2}}= \frac{q^{1/12}}{\eta(q)^2}\, G(q).
\end{equation}
 This equation is already a remarkable identity of the Rogers--Ramanujan type which, to the best of our knowledge, was not known before. In fact, we can use the physical ideas of the present paper to generate many infinite families of such identities. Given the interest of $q$--series identities in Combinatorics and Number Theory, in this section we outline a strategy to generate many such identities and give some relevant example.\smallskip

We know from physics that the conjugacy class of the monodromy $M(q)$ is a wall--crossing invariant. Thus, the function $\mathrm{Tr}\, M(q)^k$ is independent of the BPS chamber in which we compute it. On the other hand, the explicit expression of $\mathrm{Tr}\, M(q)^k$ as a $q$--series varies enormously from one BPS chamber to the other. Equating the $q$--series obtained by computing the trace in different chambers, we get identities between $q$--series, as well as identities between $q$--series and infinite products, which generalize those of Rogers, Ramanujan, and many other authors. It is conceivable that all the known such identities are just special instance of $4d$ wall--crossing; certainly, using wall--crossing we generate many new identities.

As an example, we present an infinite family of such identities, generalizing eqn.\eqref{identitiyn=1}.

\subsubsection{The $(A_n,A_1)$ theory in the linear BPS chamber and $(A_{n-1},T_1)$ TBA}\label{sec:a2nlincham}

In \S.\,\ref{cantraceAn} we computed $\mathrm{Tr}\, M(q)$ for (in particular) the $(A_{n},A_1)$ theory using the BPS spectrum of the canonical chamber. In appendix \ref{app:linearchamber} we discuss an alternative BPS chamber for these models, namely the \textit{linear} one with quiver
\begin{equation}
 \begin{diagram}
 \node{X_1}\node{X_2}\arrow{w}\node{X_3}\arrow{w}\node{X_4}\arrow{w}\arrow{e,..,-}\node{}\node{X_n}\arrow{w} 
 \end{diagram}, 
\end{equation} 
corresponding to the $A_n$ quantum torus algebra in the form
\begin{gather}
	X_{k+1}X_{k}=q\, X_{k}X_{k+1}\\
	X_kX_j=X_jX_k\qquad \text{for }|k-j|>1.
\end{gather}
As a by-product, this will lead to the identification of the trace of the full monodromy for the $(A_n,A_1)$ theory with the UV characters
of the $(A_{n-1},T_1)$ TBA system.

The quantum monodromy reads
\begin{multline}
	M(q)= \Psi(X_1;q)\, \Psi(X_2;q)\, \Psi(X_3;q)\cdots \Psi(X_{n};q)\times\\
	\times\Psi(X_1^{-1};q)\, \Psi(X_2^{-1};q)\, \Psi(X_3^{-1};q)\cdots \Psi(X_{n}^{-1};q).\hskip 1.6cm
\end{multline}
Applying recursively the identities \eqref{Eulerid}\eqref{Eulerid2}\eqref{Eulerid3}, we get
\begin{multline*}
q^{-n/24}\, M(q)=\\
 =\frac{\Theta(-X_1,q)}{\eta(q)}\, \Psi(-q^{-1/2}X_1^{-1}X_2;q)\,	
\frac{\Theta(-X_2,q)}{\eta(q)}\, \Psi(-q^{-1/2}X_2^{-1}X_3;q)\,\frac{\Theta(-X_3,q)}{\eta(q)}\cdots\\
\hskip 1cm\cdots \Psi(-q^{-1/2}X_{n-2}^{-1}X_{n-2};q)\,
\frac{\Theta(-X_{n-1},q)}{\eta(q)}\, \Psi(-q^{-1/2}X_{n-1}^{-1}X_{n};q)\,\frac{\Theta(-X_{n},q)}{\eta(q)}
\end{multline*}
where $\Theta(x;q)=\sum_{k\in\mathbb{Z}}q^{k^2/2}x^k$. Expanding the functions in the \textsc{rhs},
\begin{equation}\begin{split}
	\big(q^{-1/24}\eta(q)\big)^{n}&\,M(q)
	=\sum_{k_1,\cdots, k_{n}\in \mathbb{Z}}\  \sum_{\ell_1,\cdots, \ell_{n-1}\geq 0}
	\frac{q^{\sum_i k^2_i/2}}{(q)_{\ell_1}(q)_{\ell_2}\cdots (q)_{\ell_{n-1}}}\times\\
	&\times X_1^{k_1-l_1}X_2^{l_1+k_2-l_2}
	X_3^{l_2+k_3-l_3}\cdots X_{n-1}^{l_{n-2}+k_{n-1}-l_{n-1}}X_{n}^{l_{n-1}+k_{n}}.
\end{split}\end{equation}
Taking the canonical trace of both sides we get
\begin{equation}
	\begin{split}
	q^{-n/24}\:\mathrm{Tr}_\mathrm{can}\,M(q)=\frac{1}{\eta(q)^{n}}\,
\sum_{\boldsymbol{\ell}\in \mathbb{Z}^{n-1}_+}
\frac{q^{\,\boldsymbol{\ell}\cdot C_{n-1}\cdot\boldsymbol{\ell}/2}}{(q)_{\ell_1}(q)_{\ell_2}\cdots (q)_{\ell_{n-1}}}	
\end{split}
\end{equation}
where $C_{n-1}$ stands for the Cartan matrix of $A_{n-1}$.
Then, for all $A_n$, we have the identity
\begin{equation}\label{newqidentity}
 \sum_{\mathbf{m}\in \mathbb{Z}^n_+}\frac{q^{\mathbf{m}^t\cdot C_n\cdot \mathbf{m}/2}}{(q)^2_{m_1}\, (q)^2_{m_2}\cdots (q)_{m_n}^2}\equiv \frac{q^{n/24}}{\eta(q)^{n}}\,
\sum_{\boldsymbol{\ell}\in \mathbb{Z}^{n-1}_+}
\frac{q^{\,\boldsymbol{\ell}^t\cdot C_{n-1}\cdot\boldsymbol{\ell}/2}}{(q)_{\ell_1}(q)_{\ell_2}\cdots (q)_{\ell_{n-1}}}.	
\end{equation} 
The $n=1$ case of this equality (corresponding to the trivial quiver) is an identity due to Euler \cite{andrews}. The identities for $n\geq 2$ seem not to have been known previously.\footnote{After posting the first
version of this preprint we were advised by Ole Warnaar that these identities can be proven using
standard hypergeometric series techniques.}
We have checked this equality using Mathematica for $n\leq 8$ (up to order $q^{101}$ for $n=2$) finding perfect agreement.
For instance, for $n=5$ both sides of eqn.\eqref{newqidentity} have the $q$--expansion
\begin{multline}
 1+15\, q+100\, q^2+500\, q^3+2070\, q^4+7546\, q^5+24935\, q^6+76320\,
   q^7+\\ +219285\, q^8+597655\, q^9+1556718\, q^{10}+3898485\,
   q^{11}+O\left(q^{12}\right)\hskip 1cm
\end{multline} 

It should be stressed that the series in the \textsc{rhs} of eqn.\eqref{newqidentity} correspond to the TBA characters associated to the pair of Dynkin diagram $(A_{n-1}, T_1)$ where $T_n$ are the tadpole graphs.

The identity \eqref{newqidentity} can be written in a more suggestive way.
Introduce a Hilbert space with basis $|0\rangle, |1\rangle,\cdots, |k\rangle,\cdots$,
and two operators $R(q)$, $L(q)$ acting on this space with matrix elements
\begin{align}
 & \langle m_1| R(q) |m_2\rangle = \frac{q^{(m_1-m_2)^2/2}}{q^{-1/24}\eta(q)\, (q)_{m_1}}\\
& \langle m_1| L(q) |m_2\rangle = \frac{q^{(m_1-m_2)^2/2}}{(q)_{m_1}\, (q)_{m_2}}.
\end{align}
Then, the  identity \eqref{newqidentity} can be expressed as
\begin{equation}
 \langle 0|L(q)^{n+1}|0\rangle= \langle 0| R(q)^n|0\rangle.
\end{equation}

\section{Trace of the monodromy for $(G, G')$ theories II: The rational case}\label{ratcase}

If $q$ is an $N$-th root of unity, the irreducible representations of the quantum torus algebra have (finite) dimension
$N^\varrho$, where $\varrho$ is half the rank of the skew--symmetric form $\langle\gamma_i,\gamma_j\rangle$. One may consider the quantum monodromy in this finite--dimensional setting. There is a subtlety in doing this, related to the`quantum Frobenius phenomenon' (\S.\,\ref{sec:quantFrob} below) which eventually will lead to the Verlinde algebras of the relevant $2d$ CTF theories (\S.\,\ref{sec:Verlinde} below).  In other words we have already noted that the
line operators $X_\gamma=\cu_\gamma^N$ are central elements of the quantum torus algebra
and not suprisingly they transform under $M$ according to the {\it classical} monodromy operation.  For
the supersymmetric amplitudes not to vanish they should localize on fixed points of the
R-symmetry action.  Each choice of a fixed point, corresponds to fixing the boundary
conditions on the Melvin cigar.
 Once localized, we find, in examples that $X_\gamma$ realize the Verlinde algebra,
and the representations of the cluster algebra are labeled by diagonlizations of the fusion ring.
We then go on to illustrate how the rational case works in various examples.
In some ways, the monodromy operator in the rational case is mathematically more 
well-defined as the relevant space is finite dimensional, due to the uniquness
for the choice of the  trace operation. \smallskip

\subsection{$q$ a root of unity: the quantum Frobenius property}\label{sec:quantFrob} 

We specialize the quantum dilogarithm operators we have been discussing in the irrational
case, to the case in which $q$ is an $N$--th root of unity, $q^N=1$. 

Consider first the particular case in which $q=1$. The quantum torus algebra reduces to the classical \textit{commuting} one, and naively all adjoint actions of the monodromy and its `elementary' factors, eqns.\eqref{adj1}\eqref{adj2}, become trivial. However, physically, we know that this is not correct: in the classical limit $q\rightarrow 1$ the quantum monodromy is replaced by the classical monodromy which is a non trivial symplectomorphism of the classical torus. In particular, the `elementary' transformation \eqref{adj2} is generated by an Hamiltonian which is given by classical dilogarithms $\mathrm{Li}_2(\cu_\gamma)$. Indeed, for $|q|<1$
\begin{equation}
 -\log\Psi(q^sx;q)=\sum_{k=1}^\infty \frac{1}{k}\, \frac{(q^{s+1/2}x)^k}{1-q^k}.
\end{equation} 
Writing $q=e^{-\epsilon}$ and taking $\epsilon\rightarrow 0$,
\begin{equation}\label{hamclas}
- \log\Psi(q^sx;q)\Big|_{\epsilon\rightarrow 0} = \frac{1}{\epsilon}\sum_{k=1}^\infty \frac{x^k}{k^2}+O(1)\equiv \frac{1}{\epsilon}\, \mathrm{Li}_2(x)+O(1), 
\end{equation} 
while the quantum commutators go to a classical Poisson structure on the commuting torus,
\begin{align}
 & [\cu_\gamma,\cu_{\gamma^\prime}]=\epsilon\, \{\cu_\gamma,\cu_{\gamma^\prime}\}+O(\epsilon^2)\label{eqcomm}\\
& \{\cu_\gamma,\cu_{\gamma^\prime}\}\equiv (\pm 1)^{\langle \gamma,\gamma^\prime\rangle}\, \langle\gamma,\gamma^\prime\rangle\, \cu_{\gamma+\gamma^\prime},
\end{align}
where the sign factor $(\pm 1)^{\langle \gamma,\gamma^\prime\rangle}$ depends on which of the two roots $q^{1/2}\rightarrow \pm 1$ one takes in the definition of the `normal ordered product'
\begin{equation}
 \cu_{\gamma+\gamma^\prime} \equiv (q^{-1/2})^{\langle \gamma,\gamma^\prime\rangle}\, \cu_\gamma \cu_{\gamma^\prime}.
\end{equation} 
Dirac's quantization together with $4d$ $\cn=2$ index theory imply that the correct classical limit, for a $4d$ gauge theory, corresponds to the non-trivial sign $q^{1/2}=-1$.

The $\epsilon$ in the commutator \eqref{eqcomm} will cancel against the $1/\epsilon$ in front the Hamiltonian \eqref{hamclas}, and in the limit $\epsilon\rightarrow 0$ we get a non--trivial classical symplectomorphism, and hence a non--trivial classical monodromy.
\medskip

The above discussion may be generalized to $q$ an $N$--th root of unity. From the Non--Commutative geometry of the quantum torus \cite{MR1303779} we know that the two quantum algebras $\cu_\gamma \cu_{\gamma^\prime}= q^{\langle\gamma,\gamma^\prime\rangle}\, \cu_{\gamma^\prime} \cu_\gamma$ and $\tilde \cu_\gamma \tilde \cu_{\gamma^\prime}= \tilde q^{\langle\gamma,\gamma^\prime\rangle}\, \tilde \cu_{\gamma^\prime} \tilde \cu_\gamma$, where
$q=\exp(2\pi i\tau)$ (resp.\! $\tilde q=\exp(2\pi i \tilde\tau$)) and
\begin{equation}
 \tilde\tau=\frac{a\tau+b}{c\tau+d},\quad\text{with}\ \  
\begin{pmatrix}
 a & b\\ c & d
\end{pmatrix}\in SL(2,\mathbb{Z}),
\end{equation}  
are \emph{Morita equivqalent}. In particular, a quantum torus algebra with $\tau\in\mathbb{Q}$ is Morita equivalent to the classical (commutative) torus algebra. Here we shall not pursue this line of thought, but rather use the formulae from the topological string theory, since they lead to stronger results than plain Morita equivalence.
\medskip

In order to do this, we write $q=\exp(2\pi i\tau)$ with
\begin{equation}\label{eq:tau}
 \tau= \frac{k}{N}+i\frac{\epsilon}{N^2},\qquad (k,N)=1,\ \text{and } \epsilon>0.
\end{equation} 
We define the \textit{reduced} $q$ as $q_r\equiv \exp(-2\pi\epsilon)$. We have
\begin{equation}\label{redalgebra}
 \cu_\gamma^N\, \cu_{\gamma^\prime}^N = q_r^{\langle \gamma,\gamma^\prime\rangle}\, \cu_{\gamma^\prime}^N\, \cu_{\gamma}^N,
\end{equation}  
so the variables $\cu_\gamma^N$ ($\gamma\in \Gamma$) generate their own \textit{reduced} quantum torus algebra which, as $\epsilon\rightarrow 0$, becomes the \textit{classical} torus algebra. In fact,  at $q^N=1$, the variables $\cu_\gamma^N$ belong to the center of the quantum torus algebra. The map from the quantum torus algebra to its center given by
\begin{equation}
 \cu_\gamma\mapsto \cu_\gamma^N \equiv \cu_{N\gamma}
\end{equation} 
will be called the \emph{quantum Frobenius map.}
\smallskip

In order to get the formulae for $q$ a root of unity, we have to consider $\Psi(q^s \cu_\gamma;q)$ with $q$ as in eqn.\eqref{eq:tau} and take the limit $\epsilon\rightarrow 0$.
Write
\begin{align}\label{psiacc}
 \Psi(x;q)&=\prod_{n=0}^\infty (1-x q^{n+1/2})=
\prod_{j\geq 0} \prod_{h=0}^{N-1} \Big(1-x\, e^{2\pi i k (h+1/2)/N}\, e^{-2\pi \epsilon (j N+h+(1/2))/N^2}\Big)  
\end{align}
If $\epsilon\ll N^2$, the \textsc{rhs} can be approximated to leading order as 
\begin{equation}\label{prodclasqdi}
\prod_{j\geq 0} (1- e^{\pi i k}x^N\, q_r^{j+1/2})\equiv \Psi(e^{\pi i k} x^N; q_r).
\end{equation} 
This result has a simple meaning: it is the quantum dilogarithm on the `reduced' quantum torus algebra \eqref{prodclasqdi}. Its effect, at the operator level, is to implement the adjoint action
\eqref{adj2} on the quantum Frobenius subalgebra generated by the operators $\cu_\gamma^N\equiv U_{N\gamma}$. As $\epsilon \rightarrow 0$, $q_r\rightarrow 1$, the Frobenius subalgebra becomes classical (commuting), and the action of the monodromy reduces to the classical one, as before. Moreover, the operators $\cu_{\gamma}^N$ become central elements of the algebra and hence, by Schur's lemma,
act as $c$--numbers in any irreducible representation of the torus algebra. The adjoint action \eqref{adj2}
 maps (generically) the irreducible representation corresponding to given numerical values of the central elements $\cu_\gamma^N$ to a different irreducible representation where the central elements $\cu_\gamma^N$ take different values.

More precisely, comparing the actions of the monodromy on the original quantum torus algebra and on its reduced subalgebra, we get the following

\textbf{The quantum Frobenius theorem} \textit{Assume $q=\exp(2\pi i k/N)$ with $k$, $N$ coprime integers. Then the quantum monodromy $M$ acts on the central elements of the quantum torus algebra $X_\gamma= e^{2\pi ik (s_\gamma+1/2)}\cu_\gamma^N$ as the \emph{classical} monodromy $M_\mathrm{clas}$ acts on the $\cu_\gamma$'s.}

Applying this result to simple $4d$ $\cn=2$ theories we reproduce the mathematical results which go under the name of quantum Frobenius identities \cite{kassel}\cite{clqd2}. The above result, which
we derived in \S.\ref{pure} using path-integral arguments, is a far--reaching generalization of these results, generating more such identities.


The Frobenius property is not the end of the story. At $q$ an $N$--th root of unity, the quantum monodromy has two effects:
it changes (classically) the values of the central elements $\cu_\gamma^N$, mapping one irreducible representation into (generically) a different one, and it acts by an ordinary adjoint action on the finite matrices representing the operators $\cu_\gamma$ in the given irreducible representation. This is the \emph{discrete} part of the quantum monodromy at a root of unity.

To get the \textit{discrete} part, we have just to compute the subleading terms in eqn.\eqref{psiacc} as $\epsilon\rightarrow 0$.
We start from the identity
\begin{equation}
 -\log\Psi(x;q)= \sum_{l=0}^{N-1} \sum_{r\geq 1}\frac{1}{r}\,\frac{e^{2\pi k r l/N} \tilde q^{lr} \big(e^{\pi i k/N} \tilde q^{1/2} x\big)^r}{1-\tilde q^{Nr}}
\end{equation} 
where $\tilde q= e^{-2\pi i k/N} q$, which is true for all $N$'s, $k$'s and $|q|<1$. Setting $\tilde q= \exp(-2\pi\epsilon/N^2)$,
we have
\begin{equation}\begin{split}\label{1overeps}
 -&\log\Psi(x;q)=\\ &=\frac{1}{2\pi \epsilon} \mathrm{Li}_2(e^{\pi ik} x^N)+\sum_{l=0}^{N-1} \left(\frac{l+1/2}{N}+\frac{1}{2}\right)\log\big(1- e^{2\pi i k(l+1/2)/N} x\big)+O(\epsilon)
\end{split}\end{equation} 
The finite part of this expression is the discrete quantum dilogarithm at $\tau=k/N$,
\begin{equation}
 \Psi(x; k/N)_\mathrm{dis.}=\prod_{l=0}^{N-1}\big(1- e^{2\pi ik(l+1/2)/N}x\big)^{-(l+1/2)/N-1/2}.
\end{equation} 
Restricting to an irreducible (finite) representation of the quantum torus algebra at $q=\exp(2\pi i k/N)$, our discrete quantum dilogarithm function reduces, up to an irrelevant overall normalization\footnote{\ The relative normalization constant depends on the particular irreducible representation, that is, it depends on the numerical value of the central element $x^N$.}, to the discrete quantum dilogarithm defined in refs.\cite{baxter,Faddeev:1993rs}. To see this, it is enough to check that it satisfies the same difference equation
\begin{equation}
 \frac{\Psi(e^{2\pi i k/N}\,x; k/N)_\mathrm{dis.}}{\Psi(x; k/N)_\mathrm{dis.}}= \frac{(1-e^{\pi ik}x^N)^{1/N}}{(1- e^{\pi i k/N}x)}.
\end{equation}

In conclusion, at $q=\exp(2\pi i k/N)$, the adjoint action \eqref{adj2} corresponds to the combined effect of the classical sympectomorphism given by the Hamiltonian flow generated by $H_\gamma\equiv \mathrm{Li}_2( e^{\pi i l} e^{\pi i k s_\gamma} U_{N\gamma})$, which corresponds to the polar term in eqn.\eqref{1overeps}, together with the adjoint action of the finite matrix $\Psi(e^{2\pi k s_\alpha/N} \cu_\gamma; k/N)_\mathrm{dis.}$ acting on each irreducible representation\footnote{\ At $q$ a root of unity, the irreducible representations of the quantum torus algebras are universal up to a rescaling of the generators (which corresponds to changing the numerical values of the corresponding Frobenius central elements).} of the
quantum torus algebra, which is induced by the finite part of eqn.\eqref{1overeps}.

In this section we present an illustrative example, to show in concrete terms how the general principles work
for the $y^2=x^3$ Argyres-Douglas theory.  More examples are postponed to Appendix \ref{apprat}.

\subsection{The $(A_2,A_1)$ model}

For $(A_2,A_1)$ the quantum torus algebra is $XY=q\,YX$. If $q$ is a \textit{primitive} $N$--th root of unity, $X^N$ and $Y^N$ are \textit{central} elements in the torus algebra. Then, on an irreducible module, they act as $c$--numbers $X^N=\lambda$ and $Y^N=\mu$. The irreducible modules $V_{\lambda,\mu}$ are classified, up to unitary equivalence, by the complex numbers $\lambda$, $\mu$. In $V_{\lambda,\mu}$, $X$, $Y$ are represented by the explicit $N\times N$ matrices
\begin{align}
 X&= (\lambda)^{1/N} \mathrm{diag}(1, q^2, q^4,\cdots, q^{2(N-1)})\label{XXX}\\
 Y_{ij}&= (\mu)^{1/N}\, \delta_N(i-j-1)\label{YYY}
\end{align}  
where $\delta_N(x)=\tfrac{1}{N}\sum_k e^{2\pi i k x/N}$ is the Kronecker delta mod $N$. The choice of $N$--th roots in eqns.\eqref{XXX}\eqref{YYY} is irrelevant (up to unitary equivalence). Notice that
\begin{equation}
 \det(X)=(-1)^{N-1}\, \lambda,\qquad \det(Y)=(-1)^{N-1}\,\mu.
\end{equation} 

\subsubsection{The monodromy $M$ and the quantum Frobenius map}\label{sec:monqFrob}
The map
\begin{equation}
 M\colon \quad \begin{aligned}
                & X\mapsto Y^{-1}\\
& Y\mapsto (1-q^{1/2} Y)X
               \end{aligned}
\end{equation} 
is an automorpshim of the algebra $XY=qYX$ (for both signs of the square--root $q^{1/2}$), but not of the irreducible representation $V_{\lambda,\mu}$. Indeed
\begin{equation}\label{chaneg}
 M\colon \quad \begin{aligned}
                & \det(X)\mapsto \det(Y)^{-1} & &\Rightarrow& &\lambda\mapsto \mu^{-1}\\
& \det(Y)\mapsto \det(1-q^{1/2} Y)\, \det(X) & &\Rightarrow& &\mu\mapsto \lambda\,(1- q^{N/2}\mu),
               \end{aligned}
\end{equation}
that is the monodromy $M$ acts on\footnote{\ Note that from $q^N=1$ we have $q^{N/2}=\pm 1$.} $q^{N/2}X^N$, $q^{N/2}Y^N$ the same way as the classical monodromy acts on $X$, $Y$, in agreement with the quantum Frobenius theorem of section \ref{sec:quantFrob} (with $s_\gamma\equiv 0$).
The irreducible representation $V_{\lambda,\mu}$ is invariant under $M$ (leading
to non-vanishing expectation values in the traces) only if $\mu=\lambda^{-1}$ and $q^{N/2}\lambda$ satisfies the `golden ratio' equation
\begin{equation}\label{golden}
 (q^{N/2}\lambda)^2-(q^{N/2}\lambda)-1=0.
\end{equation}

Iterating the map \eqref{chaneg} five times we get back the original $\lambda$ and $\mu$. So, in general, to represent the adjoint action of $M$ we must consider the vector space
\begin{equation}\label{Bifrep}
 \bigoplus_{k=0}^4 V_{M^k(\lambda), M^k(\mu)}.
\end{equation}
of dimension $5N$. Only if $\lambda=\mu^{-1}$ and eqn.\eqref{golden} is satisfied we can represent its action in a shorter module of dimension $N$.
For generic $\lambda$, $\mu$ we have 
\begin{equation}
 \mathrm{Tr}(M^k)= 0 \ \text{if } k\neq 0\mod 5.
\end{equation}  
since $M$ permutes the summands in eqn.\eqref{Bifrep}.

On the contrary, on the short `golden' representations (for a given $q$, there are $4$ of them, corresponding to the choice of a root $q^{N/2}=\pm 1$ and a solution to the quadratic equation \eqref{golden}), 
the monodromy $M$ is represented by an $N\times N$ matrix of the form
\begin{equation}
 M_{mn}=D_m\, q^{-mn}\qquad\text{with } D_{m+N}=D_m. 
\end{equation}
where $D_m$ satisfies the difference equation
\begin{equation}
 D_{m-1}=D_m\,\Big(\lambda^{2/N}-q^{-m +1/2}\lambda^{1/N}\Big),
\end{equation} 
whose general solution is a constant $C$ times the inverse discrete quantum dilogarithm, (see \S.\,\ref{sec:quantFrob})
\begin{equation}
 D_m= \frac{C}{\lambda^{2(m+1)/N}\: (\lambda^{-1/N}q^{1/2}; q^{-1})_{m+1}}.
\end{equation}

The periodicity condition $D_{m+N}=D_m$ is satisfied given that
\begin{equation}
 D_{m+N}= \big((q^{N/2}\lambda)^2-(q^{N/2}\lambda\big)^{-1}\, D_m\equiv D_m,
\end{equation} 
as a consequence of eqn.\eqref{golden}.

$M^5$ commutes with $X$ and $Y$ and hence we must have $M^5=(\det\,M)^{5/N}\cdot\boldsymbol{1}$. 
We choose the overall factor $C$ so that $M^5=\boldsymbol{1}$. This fixes $C$ up to multiplication by a fifth--root of unity.
\medskip

Then the trace of $M(N)$ in a `golden' representation is
\begin{equation}
 \mathrm{tr}\,M(N) = C\,\sum_{n=0}^{N-1} \frac{q^{-n^2}\, (\lambda^{-2/N})^{n+1}}{(\lambda^{-1/N}q^{1/2}; q^{-1})_{n+1}} 
\end{equation}

Note that using \ref{golden}, if we let $\Phi=q^{N/2} X^N$ and as before define the expectation values as
$$\langle\cdots\rangle ={\Tr}\big[\cdots M\big]$$
Then we have
$$\langle \Phi^2 \cdots\rangle =\langle (\Phi +1)\cdots\rangle$$
where $...$ stands for any line operators.  In other words the line operator $\Phi$
is localized on a subspace which realizes the Verlinde algebra of the $(2,5)$ model:
$$\Phi \times \Phi=1+\Phi$$
exactly as we found in the $q\rightarrow 1$ limit of the irrational version of this model.

\subsubsection{Eigenvalue multiplicities}

Having normalized $M(N)$ such that $M(N)^5=\mathbf{1}$, its eigenvalues are fifth--roots of unity, and
\begin{equation}\label{traces}
 \mathrm{Tr}(M_N^\ell)= \sum_{k=0}^4 N_k(N)\, e^{2\pi i k\ell/5},
\end{equation} 
where $N_k(N)$ are non--negative integers, namely the multiplicities of the eigenvalue $\exp(2\pi i k/5)$ (and 
$\sum_k N_k(N)=N$).
$M(N)$ is well defined up to multiplication by a fifth--root of unity. The map $M(N)\rightarrow e^{2\pi i m/5} M(N)$ preserves 
\emph{cyclically} permutes the set $\{N_0(N),N_1(N),N_2(N),N_3(N),N_4(N)\}$. Thus we may speak of the eigqnvalue multiplicities only up to \textit{cyclic} permutations.

The sets $\{N_0(N),N_1(N),N_2(N),N_3(N),N_4(N)\}/(\text{cyclic})$ will depend on finitely many choices: we have
$\phi(N)$ choices for the primitive $N$--th root $q$, two choices for the root $q^{1/2}$, and two choices for the solution to the golden equation \eqref{golden}.

Taking a uniform choice of $q$ ($q=\exp(2\pi i/N)$, say), for each selection of $q^{1/2}$ and $\lambda$, the set
\begin{displaymath}
\{N_0(N),N_1(N),N_2(N),N_3(N),N_4(N)\}/(\text{cyclic}) 
\end{displaymath}
will have the following general properties:
\begin{enumerate}
 \item Up to cyclic permutations, one has
\begin{equation}
N_k(N)=[N/5]+a_k(N),
 \end{equation}
where the $a_k(N)$'s satisfy
\begin{enumerate}
 \item  $|a_i|\leq 1$,
\item $a_i<0$ only for $N=0\mod 5$;
\end{enumerate}
\item the small deviations from equidistribution, $a_k(N)$, are periodic in $N$ mod $10$ (mod $5$ for some choices).
\end{enumerate}
 
These properties have been checked explicitly using Mathematica for $N$ up to $50$.
The $a_i(N)$ for various choices can be worked out.  There is an interesting pattern
that emerges for certain choice of $q$ which reflects the R-charge of the chiral
operators associated to $x^m$ deformations of the Argyres-Douglas model. As noted
in section \ref{rtwist} these are non-trivial only for $m\not=2 $ mod 3, and lead
to R-charges
$$R_m={2m+4\over 5}$$
As we increase $N$ by 1, we add one extra eigenvalue to $M$ which ends up being
the ${\rm exp}(2\pi i R)$ for the next chiral field.  

For simplicity, we restrict ourselves to odd\footnote{\ The match works also for $N$ even with a suitable choice of the roots. } $N$'s. Consider the subset of the first $N$ chiral fields starting with $x^3$, and let $\varrho_k(N)$ ($k=0,1,2,3,4$) be the number of the fields in this 
subset with R--charge $\frac{k}{5}$ mod $1$. 
 Choosing $q^{1/2}= \exp(2\pi i/N)$ (this gives $q$ a primitive $N$--th root only for $N$ odd) and the positive root of the Golden ratio equation we get the results in table \ref{comparisonmult}.

\begin{table}
 \begin{center}\begin{footnotesize}                     
  \begin{tabular}{|c|c|c|}\hline
$N$ mod $10$ & $\{\varrho_0(N),\varrho_1(N),\varrho_2(N),\varrho_3(N),\varrho_4(N)\}$ & $\{N_0(N), N_1(N), N_2(N), N_3(N), N_4(N)\}$\\\hline  
 $10 m+1$ & $\{2m+1,2m,2m,2m,2m\}$ & $\{2m+1,2m,2m,2m,2m\}$\\\hline
$10 m+3$ & $\{2m+1,2m+1,2m+1,2m,2m\}$ & $\{2m+1,2m+1,2m+1,2m,2m\}$\\\hline
$10 m+5$ & $\{2m+1,2m+1,2m+2,2m+1,2m\}$ & $\{2m+1,2m+1,2m+2,2m+1,2m\}$\\\hline
$10 m+7$ & $\{2m+1,2m+1,2m+2,2m+2,2m+1\}$ & $\{2m+1,2m+1,2m+2,2m+2,2m+1\}$\\\hline
$10 m+9$ & $\{2m+2,2m+1,2m+2,2m+2,2m+2\}$ & $\{2m+2,2m+1,2m+2,2m+2,2m+2\}$\\\hline
  \end{tabular} \end{footnotesize}
 \end{center}
\caption{\label{comparisonmult} Comparison between the $R$--charges of the first $N$ chiral fields and the eigenvalues multiplicities for $M_N$. The set $\{N_0(N),N_1(N),N_2(N),N_3(N),N_4(N)\}$ is well defined only up to cyclic permutations.}
\end{table}

Although the match is perfect, we do not have a deep explanation of this
fact.

\section{Towards an explanation of RCFT models in $(G, G')$ theories} \label{speculate}

We have seen in the various examples considered in previous sections deep connections between
the BPS data of $\cn=2$ theories in $d=4$ with RCFT's in 2 dimensions. 
Here we offer an explanation of some of these results.  We  apply various string dualities to our setup which leads naturally to 
the corresponding CFT's.  

For $\cn=2$ theories given by pairs of ADE singularities, the trace of the monodromy or fractional monodromy,
lead to characters of RCFT's.  Furthermore we have seen that insertion of line operators
corresponds to changing the corresponding character.  In other words we have a structure
of the form
$${\rm Tr}(\prod X_i^{n_i} ) M(q)=\chi_{n_i}(q)$$
where ${n_i}$ form a (redundant) basis of labels for the characters of the conformal theory.
We would like to explain how such a result may come about from the perspective of the 
four dimensional theory.  

The basic idea for explaining the appearance of RCFT relies on string dualities, and uses many details of our construction.
For simplicity let us consider the case of $(A_{n-1},A_{m-1})$:
$$x^n=y^m+uv$$
  As discussed in \S.\,\ref{5bp}, 
this corresponds to considering
$m$ M5 branes in flat space, fibered over the $\C_x$ plane, where
the $M5$ branes are separated for $x\not=0$.
Now, according to our prescription, where we replace the 4d spacetime with
$$\mathbb{R}^4\rightsquigarrow  T^2\times C.$$
Let us parameterize the world volume of the $m$ M5 branes by 
$$T^2\times C\times \C_x,$$
  According to our construction the two cycles of $T^2$ are twisted:
For one of the circles, as we go around it we mod out by R-twist (or fraction thereof).  Let us call that
the R-circle.  For the other one we mod out by the action of rotation of $C$ represented
by $q$ (combined with an $SU(2)_R$ transformation).    Let us call that circle, the $q$-circle.
Viewing the R-circle as the 11-th circle, and taking
it to be small, we obtain an effective IIA description, where the $m$ M5 branes are replaced by $m$
D4 branes, whose positions depend on $x$, captured by the expectation
value of an adjoint scalar in the gauge multiplet.  Let $\Phi$  be an
adjoint scalar in the $SU(m)$ gauge theory on the brane.   Then we can read from the geometry that it has an expectation value which
depends on $x$:
$${\rm det}(\lambda -\Phi(x))=\lambda^m-x^n$$

According to our construction, in order to get the characters of the $SU(n)_m/U(1)^{n-1}$ theory we need to use a fractional
R-symmetry (see \S.\ref{sec:frakMon}), corresponding to modding out by $x\rightarrow {\rm exp}(2\pi i/n)\cdot x$, and at the same
time rotating $C$ by ${\rm exp}(-2\pi i/n)$.  Since we have taken the R-circle as the 11-th circle,
and this is invisible to the D4 brane, it implies that the D4 brane worldvolume is
$$S^1 \times (\C_x,C)/{\Z_n}$$
In other words the $SU(m)$ gauge theory of the D4 branes lives on an $A_{n-1}$ singularity.
Even though in the 11 dimensional sense there is no singularity (which is reflected by some
RR-fluxes being turned on in the 4d ALE space \cite{Schwarz:1995bj}) the path-integral of the M5 branes will
localize to configurations as if the space has a singularity, i.e. on fixed point of the R-symmetry action.
In other words, the $SU(m)$ gauge theory of the D4 brane living in 5 dimension lives on
 a four dimensional space with an $A_{n-1}$ singularity at $(x=0,p)$ (where $p$
is the tip of $C$). Sufficiently close to the origin where $\langle \Phi \rangle \sim 0$
we have an approximate  $\cn=4$
$SU(m)$ theory on a space with $A_{n-1}$ singularity.
 The partition function of this theory \cite{Vafa:1994tf} is captured by Euler
class of moduli space of $SU(m)$ instantons and according to Nakajima \cite{MR95i:53051}
these are in 1-1 correspondence with the elements of Hilbert space of
characters of $SU(n)$ at level $m$.   Moreover the choice of which representation of
$SU(n)_m$ one gets, according to Nakajima, depends on the choice of boundary
conditions, i.e. a flat connection at infinity,  which is in 1-1 correspondence with maps
$$\phi: {\Z_n\rightarrow}U(m)$$
The space of such $\phi$'s is isomorphic to the choice of characters of $SU(n)_m$.
Let us compare these with our predictions. 
For precisely this  theory  we had found that ${\rm Tr} K(q^{-1})$
gives the characters of $SU(n)_m/U(1)^{n-1}$. This is very close to the partition function
of instantons:  The states of $SU(n)_m$ can be decomposed to the representations of level $m$ parafermions of $SU(n)$
(which make up the states of $SU(n)_m/U(1)^{n-1}$) tensored with $n-1$ free bosons.

The choice of the characters of the Nakajima theory, is dictated by
the choice of boundary conditions for the gauge field, which matches what we had predicted
for our theory, namely the range of allowed boundary condition leading to non-vanishing partition function
is dictated by fixed points of R-symmetry action which
are in 1-1 correspondence with characters arising from monodromy trace.
But we have seen more is true in our context:
The insertion of suitable line operators
change the characters of CFT.  Thus, to match how characters arise in Nakajima's story, we would predict that 
insertion of such line
operators should change the boundary conditions of the
gauge theory, i.e. it should change the flat connection at infinity for the D4 brane.
We now show how this arises.  To do this, it is convenient to deform the theory
so that the geometry is given by $\Sigma: y^m=x^n-1$.  The line operator we inserted
correspond, in this setup to insertions of the $B$ field on M5 brane over
2-cycles consisting of a 1-cycle $\gamma \in H_1(\Sigma)$ times $S^1_q$, invariant
under the $\Z_n$ action:
$$X_\gamma ={\rm exp}\bigg(i \int_{\gamma \otimes S^1_q} B\bigg)$$
We insert this operator and ask if the connection on the D4 brane at infinity has
changed.  This is the same as asking if $\int_{\{x=0\} \times C/\Z_n} F^i$ has changed, where $i$ denotes
some element of Cartain of $SU(m)$ which can also be identified with the choice of the difference of the gauge
field between the  $i$-th and $i+1$-st sheet
 of $\Sigma$ as viewed as an $m$-fold cover over $\C_x$.   Lifting this to the M-theory, implies
that in the M5
brane setup this is the same as asking if
$$\int_{S^1_R \otimes (\{x=0\},{C/ \Z_n})}dB^i$$
has changed.  This has indeed changed by one quantum (for suitable cycles $\gamma$) because the insertion of
$X_\gamma$ in the M5 brane language correspond to having an M2 brane
ending on the cycle $ \gamma \otimes S^1_q$ and the flux this induces can be meausured
by $dB$ on the cycle surrounding it, which is $S^1_R \otimes (\{x=0\},{C/ \Z_n})$.  Thus
indeed the map of Nakajima matches what we have found in terms of insertion of line operators.

In the Nakajima approach, which we now see is the same as ours,
 the characters of 2d RCFT
arises in the Hilbert space of the quantum mechanical system, but the two dimensional space in which the CFT lives is
invisible, and can only be made visible
after applying suitable string dualities, as in \cite{Dijkgraaf:2007sw}.   This basically
involves an 11/9 flip in which direction we consider the extra dimension of type IIA.
So far we viewed the R-circle of the $m$ M5 branes, whose worldvolume is given by $T^2 \times C\times C_x$,
as the extra dimension. Following \cite{Dijkgraaf:2007sw} we now consider the circle
rotation along the anti-diagonal circle in $C\times C_x$ as the 11-th dimension, replacing $C\times C_x\rightsquigarrow \R^3$.  We thus end up with $m$ D4 branes
with worldvolume 
$$T^2\times \R^3$$
where we mod out by an extra $\Z_n$ action as we go around the R-circle.  Thus the origin of $\R^3$
has `effectively'  $n$ D6 branes  filling $T^2\times \R^5$, where $\R^5$
is the space transverse to the D4 brane.    In other words we have
$m$ D4 branes intersecting $n$ D6 branes at the origin of $\R^3 \times \R^5$.  This leads to open
strings stretched between D4 and D6 brane leading to
fermions in the bifundamental representaions of $U(n)\times U(m)$, i.e. a conformal system
realizing $U(nm)$ current algebra at level 1. On the other hand
the gauge theory on the D4 branes is dynamical, leading to gauging the level  $U(nm)_1$:
$$U(nm)_1/U(m)_n =SU(n)_m$$
living on $T^2$.  This is the same as the Nakajima system, but now the 2d space is visible
as our $T^2$. If we take the Cartan $U(1)^{n-1}\subset SU(n)$ to also
be dynamical this would lead precisely to the parafermionic partition function we have obtained.

There remains some points to clarify:  In particular the role of the parameter $q$.  In the context 
of Nakjima the parameter $q$ is related to the gauge
coupling constant $\tau$ by
 $q={\rm exp}(2\pi i\tau)$.  For us the $q$ parameterizes the action of the rotation on $C$
(as well as an $SU(2)_R$ rotation) as we go around $S^1$.   Also in the context of the
brane intersecting on $T^2$, the partition function of the chiral fermions would give
the corresponding characters only if we identify $q={\rm exp}(2\pi i\tau)$ where $\tau$
is the modulus of the $T^2$.  The connection of this with the $q$ rotation of the cigar
is a bit mysterious.

It would be interesting to connect the physics of 4d more directly to TBA systems. So far we have obtained only chiral characters for the CFT, and not the full partition function.
For this to appear we need the anti-chiral characters.  This naturally suggests using the
tt* geometry and replacing the cigar $C$ by an $S^2$ (somewhat reminiscent
of the proposal of \cite{Nekrasov:2010ka} as to how Liouville characters arise).  This would naturally
explain, also the appearance of the doubled commuting torus algebras that we have
noted in this paper,
in the context of irrational $q$.  We would thus conjecture
that compactifying the $\cn=2$ theory on an R-twisted circle times an $S^2\times T^2$
with a suitable R-twist on the circle, leads
to the quantum mechanics of the TBA system on $T^2$.   
Moreover, certain deformations of the 4d theory, should correspond to integrable deformations
of the CFT leading to TBA systems.
   It would also be interesting to see if there
is any relation between the way TBA arises here as compared to the setup of \cite{Nekrasov:2009rc}

Even though there are a number of open questions
 we feel that the basic explanation we have found for our findings is on the right track.

\section{4d/2d worldsheet correspondence conjecture}\label{4d2d}

In the previous sections we have seen that there is an interesting correspondence between
4d CFT's and  2d CFT's, roughly related by geometric reduction from 4 dimensions on the Melvin cigar.
As already notes, along the way we found another connection with 2d systems:  The quivers
of the 4d theory and their mutations matched that encountered for 2d quivers encoding
the solitons of $\cn=2$ systems.
In this section we first review how quivers arise naturally both in 4d and 2d $\cn=2$ theories.
We then
propose a correspondence which is roughly
a 2d worldsheet/4d target correspondence.   We show how this  can be used to import
the 2d $\cn=2$ classification program into the 4d one. 

Similar ideas relating worldsheet dynamics and 4d solitons have
also been considered
in \cite{Ritz:2006rt}.

\subsection{4d theories, 2d theories and quiver mutations}\label{subsec:quiver-mut}

As already mentioned the structure of the 4d quivers and their transformations
mirrors that in the quiver diagram summarizing the BPS data of $\cn=2$ theories in 2d.
 In the 2d context the quiver has one node for each {\it vacuum} of the theory,
and arrows representing solitons going from one vacuum to another.  More
precisely, $B_{ij}$ is identified with
the integral skew--symmetric matrix $\mu_{ij}$ which is defined by the IR asymptotics of the supersymmetric index $Q_{ij}$ \cite{Cecotti:1993rm,Cecotti:1992qh}.  $|\mu_{ij}|$ is the number of BPS kinks interpolating between the vacua $|i\rangle,\:|j\rangle$, and the sign of $\mu_{ij}$ is determined by the Fermi number fractionalization in the given kink sector as explained in \cite{Cecotti:1993rm}.   For example, if we consider a 2d Landau-Ginzburg theory 
with superpotential
$W(X) = X^{n+1} + \cdots$, with suitable choice of couplings (for example
$W=T_{n+1}(X)$ where $T_{n+1}(X)$ is the $(n+1)$-th Chebyshev polynomial), then
from the explicit solution of the $tt^*$ equations
\cite{Cecotti:1991me} we see that the BPS soliton quiver is exactly
the $A_n$ Dynkin diagram with alternating sinks and sources.

As in the 4d case, the quiver we attach to a 2d theory is not uniquely determined.
Apart 
from the trivial possibility of permuting the \textsc{susy} vacua, there are two natural options.  
One possibility is to change the sign of a vacuum $|i\rangle \rightarrow -|i\rangle$;
this has the effect of inverting all the arrows starting or ending at the $i$--th vertex of $Q$.
We may also move in the coupling space (without changing the UV CFT).  Sooner or later we cross some 
wall of marginal stability at which the BPS multiplicities, and hence the quiver exchange matrix $\mu_{ij}$, jump. We thus end up with a different quiver $Q^\prime$.  As in the 4d case, remarkably, this transformation
of quivers is exactly a quiver mutation --- in other words quiver mutation is identified with 2d wallcrossing!


\subsection{The 4d/2d classification correspondence}\label{subsec:4d2dcorr}

In this section we explain our conjectured 4d/2d correspondence.  Not surprisingly, the 
idea is that the 4d theory
attached to a given quiver should correspond to a 2d theory attached to the {\it same} quiver.
This is a very strong statement. 
Not only does this conjecture lead to a classification program for $\cn=2$ theories
in 4d based on the simpler 2d case, it also leads to a number of additional predictions
for the corresponding 4d theory.

\smallskip

We begin by considering Type IIB on a Calabi-Yau with an isolated singularity.
As a warmup, let us take 
a Calabi-Yau 2-fold (i.e. K3).  In this case it is known that the worldsheet theory of the Type II
string has a subsector with a Landau-Ginzburg potential corresponding to the
ADE singularity type \cite{Ooguri:1996wj}.  The full ${\cal N}=2$ theory
in this case is composed of the minimal model sector with ${\hat c}<1$ plus a Liouville
sector with ${\hat c}>1$, combining to the critical value 2. 
Once we deform the singularity the LG model is deformed to a massive one.  (The coefficients
of the superpotential involve Liouville fields to compensate for the deficit in dimension of the relevant
LG operators, so the full worldsheet theory remains conformal.)
We thus get a map associating to each Calabi-Yau a 2d theory on the string worldsheet, whose
non-universal part is a Landau-Ginzburg model.  As in \cite{Gukov:1999ya} one can 
argue on general grounds that any isolated singularity
would have to lead to a $W$ which has $\hat c < d-1$, i.e. $\hat c < 1$ for $d=2$.  
Thus we recover the classification of
ADE singularities of K3 from the classification of 2d ${\cal N}=2$ models with $\hat c<1$.
Similarly, for general Calabi-Yau $d$-folds, we would expect that the worldsheet theory
involves a universal Liouville sector (with $\hat c>1$), coupled to a non-universal part which distinguishes between
different Calabi-Yau's.  

In particular, in the case of 
Calabi-Yau 3-fold hypersurface singularities, we expect to find associated worldsheet theories with
${\hat c}< 2$.
This is our proposed explanation of the 4d/2d correspondence in these cases:  the 4d theory coming from IIB on 
a Calabi-Yau 3-fold corresponds to a 2d theory coming from the non-universal part of the 
string worldsheet theory.

In particular, we propose that if we consider any CY 3-fold developing a hypersurface singularity $f = 0$, 
the worldsheet theory will contain an LG model with the superpotential $f$.
At least for the $A$ type singularities this
has been studied, and indeed works as expected \cite{Eguchi:2001fm}.
Let us roughly explain why, assuming this proposal, the quivers appearing in 2d and 4d would be the same.
Consider $f(x_i)=0, i=1, \dots , 4$, as a hypersurface singularity deformed by
relevant fields, defining a local, non-singular CY.  Let us single out the constant term and
write in fact $f(x_i)-u=0$.  Let $W = f(x_i)$.   
To each critical point $(x_i)$ of $W$, there
is a unique associated choice $u$ for which a 3-cycle collapses:  just choose
$u$ so that $ f(x_i)-u =0$ at the critical point.  Generically this gives a conifold 
singularity, i.e. a collapsing $S^3$.  Thus, at this point, we have a 
massless BPS state.  If we change $u$ a little bit, this
BPS state picks up some mass; we call it a ``vanishing cycle.''
If we consider D3-branes wrapped on these vanishing cycles to be our basis of elementary BPS states,
then they will be the nodes of our 4d quiver.  On the other hand, the critical points of $W$ are exactly
the nodes of the 2d quiver!
Next we may ask what are the electric-magnetic pairings of the BPS charges, which would give the numbers
of arrows in the 4d quiver.  In the IIB realization this should be given by the 
intersection product of the vanishing cycles.  But this intersection product is a familiar object
in the study of LG models:  it is simply counting the solitons connecting 
vacua $i$ and $j$ \cite{Cecotti:1993rm}--- and this is exactly what determines the number of arrows of the 2d quiver.
So the quivers indeed seem to match.

Our conjecture is more general than this example:  we do not require that the worldsheet theory comes from a
Calabi-Yau 3-fold singularity --- it could be any conformal field theory with 
the appropriate central charge.  Moreover, the ``non-universal sector'' we consider need not be 
a Landau-Ginzburg model in general.

\subsection{Consequences of the 2d classification}

We have already noted that quiver mutation corresponds to wall-crossing in 2d, an operation which 
does not change the theory.
It is thus no surprise that the known results about the classification of quivers up to mutation are 
essentially the same as the results about $2d$ $\cn=2$ models. Just to mention a few such correspondences:  
both $2d$ $\cn=2$ minimal models and finite--type quiver--mutation classes (or representation--finite path 
algebras) are classified by finite $ADE$ Dynkin diagrams.  The classification for mutation equivalent quivers with $\leq 3$ vertices \cite{cctv,cluster-intro} corresponds to the classification of $2d$ \textsc{susy} models with at most $3$ vacua (see \S\S.6.2,6.3 of \,\cite{Cecotti:1993rm}).  In fact all of these correspond
to theories with $\hat c\leq 2$ and so would have to correspond to some $\cn=2$ system in $4d$.  We find
that this is indeed the case in \S.\ref{quiver-ident}.

Indeed, the $2d$ classification program of \cite{Cecotti:1993rm} may be rephrased in purely quiver--theoretic language as follows: \textit{find all mutation--classes of quivers (without 1-cycles or 2-cycles) whose Coxeter element has eigenvalues which are all roots of unity.}\footnote{The spectrum of the Coxeter element is a mutation--invariant of the quiver class.}  Indeed, the Coxeter element of a quiver \cite{MR2197389} is exactly the $2d$ monodromy of the $2d$ $\cn=2$ model with that BPS quiver, which was shown in \cite{Cecotti:1993rm} to have 
only roots of unity for eigenvalues.  (Although this monodromy
is analogous to the $M$ we have been considering for 4d theories, we emphasize that the two are
not the same and in fact have different orders, and are not identified by our 4d/2d correspondence!
$M$ is a target space monodromy as opposed to worldsheet monodromy.) In addition
for the $2d$ theory to correspond to $\cn=2$ theory in $4d$, its ${\hat c}<2$.

Thus, in the $2d$ context we have a somewhat non-obvious criterion for what is an acceptable quiver:
If the Coxeter element of the quiver has eigenvalues which are not pure phase (corresponding
to R-charges in the 2d theory not being real), it is not acceptable!
Assuming the validity of our 4d/2d correspondence, this translates into a 
rather nontrivial condition on the quivers which can arise
in 4d theories.  We will exploit this condition more fully in Section \ref{quiver-ident} below.

\section{Classification and identification of the quivers} \label{quiver-ident}

In this section we discuss the quivers corresponding to some specific $\cn=2$ theories.  
First we use the known classification of 2d models with at most 3 vacua to classify 
$\cn=2$ theories where the BPS spectrum is generated by at most 3 objects.
Next we use this correspondence to give examples of quivers associated to 
a larger class of known $\cn=2$ theories.

\subsection{Theories with 1, 2 and 3 generators}

{\bf 1 generator.}
In string theory we can get a theory with a single BPS state just by taking
Type IIB on the conifold:
$$W(x_i) = x_1^2+x_2^2+x_3^3+x_4^2 = \mu.$$
The corresponding 2d theory is a Landau-Ginzburg model with superpotential $W(x_i)$,
i.e. it has a vacuum with no solitons.

The trace of the quantum monodromy $M$ in this case is just the partition function
of a complex fermion:
$$\Tr\,M = \prod(1-q^{n+{1\over 2} }X)(1-q^{n+{1\over 2}}X^{-1})$$
That this does not correspond to a conformal fixed point can be seen by the fact
that $M$ has infinite order:  it acts by\footnote{This monodromy is of course closely related 
to the fact that the theory has a running coupling.}
$$X\rightarrow X, \qquad Y\rightarrow q^{1/2}\,Y X.$$

{\bf 2 generators.}
This corresponds to quiver diagrams with 2 nodes.  If the two nodes are disconnected we simply have
two non-interacting $U(1)$ theories, each with its own electron.
If the quiver diagram is connected, the 2d classification
in \cite{Cecotti:1993rm} would imply that there can be only one or two arrows connecting the two nodes.

\begin{equation*}
\begin{gathered}
 \xymatrix{1\ &&\ 2 \ar[ll]}\\
\text{The }A_1\ \text{quiver}
\end{gathered}\hskip 1.5cm
\begin{gathered}
 \xymatrix{1\  \ar@<0.7ex>[rr]\ar@<-0.7ex>[rr] &&\  2}\\
\text{The }\hat A_1\ \text{(Kronecker) quiver} 
\end{gathered}
\end{equation*} 

If there is only one arrow, then this BPS spectrum is that of
the first Argyres-Douglas CFT (in the strong coupling region):
$$W(x_i)=x_1^2+x_2^3+x_3^2+x_4^2=u.$$
The corresponding 2d theory is the first non-trivial minimal $\cn=2$ model, given by
the same $W$.  The 4d monodromy in this case has order 5, as we have already seen
(even though the 2d monodromy is order 3).

If there are two arrows, then the quiver is the affine ${\hat A_1}$ Dynkin diagram (also known as the Kronecker quiver \cite{auslander}).
This BPS spectrum corresponds to the pure $SU(2)$ Yang-Mills theory.   The corresponding
3-fold geometry is given by
$$W(x_i)=(e^{x_1}+x_2^2 +e^{-x_1})+x_3^2+x_4^2=u,$$
where we recognize the term in parentheses as the Seiberg-Witten curve.
The corresponding 2d Landau-Ginzburg is known to be the theory 
corresponding to the sigma model on $\C\PP^1$
(where the $W$ is the mirror description of it).
The monodromy $M$ in this case has infinite order,
consistent with the fact that the pure $SU(2)$ theory is not conformal.

It is interesting that the conjectured 2d-4d correspondence forbids the existence of a 4d
$\cn=2$
theory with two BPS generators whose charges have DSZ inner product $\ge 3$.
For example we cannot have an ${\cal N}=2$ theory in 4d which has only two solitons
with (electric, magnetic) charges $(1,0)$ and $(0,3)$.
It is reassuring that no such theories are known.

\begin{figure}
 \begin{align*}
 &\begin{gathered}
  \xymatrix{ &\ 3 & \\ \ 2\ \ar[ur] \ar[rr] &&\ 1} \\
\text{canonical }A_3
  \end{gathered}
&\begin{gathered}
 \xymatrix{ &\ 3 &\\
\ 2\ \ar[ur] && \ 1\ar[ll]}\\
\text{linear }A_3
\end{gathered}\\
\\
&\begin{gathered}
 \xymatrix{& \ 3\ \ar@<1.3ex>[dr]\ar@<0.9ex>[dr]\ar@<0.5ex>[dr]\ar@{}[dr]|\vdots\ar@<-1.1ex>[dr]\ar@<-1.5ex>[dr]\ar@<-1.9ex>[dr] &\\
\ 2\  \ar@<1.3ex>[ur]\ar@<0.9ex>[ur]\ar@<0.5ex>[ur]\ar@{}[ur]|\vdots\ar@<-1.1ex>[ur]\ar@<-1.5ex>[ur]\ar@<-1.9ex>[ur] &&\ 1 \ar@<1.3ex>[ll] \ar@<2ex>[ll] } \\
\phantom{\bigg|}(2,y,y)\ \text{quiver}
 \end{gathered}
& \begin{gathered}
 \xymatrix{& \ 3\ \ar@<0.66ex>[dr]\ar[dr]\ar@<-0.66ex>[dr] &\\
\ 2\ \ar@<0.66ex>[ur]\ar[ur]\ar@<-0.66ex>[ur] &&\ 1 \ar@<0.9ex>[ll] \ar@<1.56ex>[ll]\ar@<2.22ex>[ll] } \\
\phantom{\bigg|}(3,3,3)\ \text{quiver}
 \end{gathered}
 \end{align*}
\caption{\label{3quivers} The allowed quivers with three nodes. (The two $A_3$'s are equivalent).}
\end{figure}
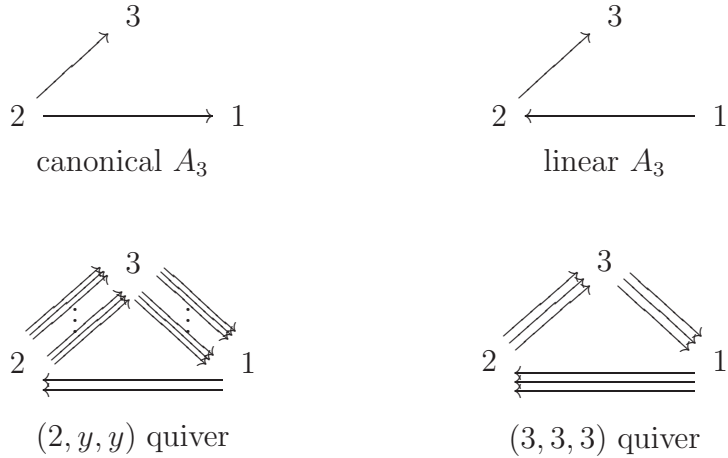

{\bf 3 generators.}
Last, let us look at quivers with 3 nodes.
The disconnected ones correspond to combinations of the
cases already discussed.  The allowed connected ones in the 2d setting have been
classified in \cite{Cecotti:1993rm}:  namely, up to mutations, the number of
arrows (oriented, say, clockwise going around the quiver) can be
$$(1,1,0), (2,y,y), (3,3,3),$$
see figure \ref{3quivers}.

The case $(1, 1, 0)$ is known to correspond to the LG model with $W=X^4$;
we immediately infer that this corresponds to the next Argyres-Douglas theory,
corresponding to the Calabi-Yau threefold with
$$W=x_1^2+x_2^4+x_3^2+x_4^2=u.$$
As noted before, this theory has monodromy $M$ of order 3, and should thus
correspond to a CFT where all R-charges have denominator 3; it indeed does.

The case $(3,3,3)$ is known to correspond to the $2d$
sigma model into $\C\PP^2$, for which the mirror superpotential is
$$W=e^{x_1}+e^{x_2}+e^{-x_1-x_2}e^{-t}.$$
We would expect this to correspond to Type IIB on the local 3-fold given by
$$W=e^{x_1}+e^{x_2}+e^{-x_1-x_2}e^{-t}+x_3^2+x_4^2=0,$$
which is indeed the mirror of $Y = \co(-3) \rightarrow \C\PP^2$
\cite{Hori:2000kt,Hori:2000ck}.  (Note that the quiver we are discussing is also
that for $\C^3 / \Z_3$, which is the singular limit of $Y$.)

Finally, what about the case $(2, y, y)$?  We now argue that this case 
corresponds to $SU(2)$
gauge theory coupled to one matter field in the spin-$j = y/2$ representation of $SU(2)$.
The case $y = 0$ we already discussed above when we considered two-node quivers:  it indeed corresponds to
the pure $SU(2)$ theory in four dimensions.
The case $y=1$ corresponds, as discussed in \cite{Cecotti:1993rm}, to the 2d Bullough-Dodd LG model, 
given by
$$W=e^{x_1}+e^{-2x_1}.$$
According to our correspondence, it should then correspond to the 4d theory given by the local
Calabi-Yau
$$W=e^{x_1}+x_2^2+e^{-2x_1}+x_3^2+x_4^2=u.$$
To see the relation to the $SU(2)$ theory with one fundamental hypermultiplet, let us make
the change of variables
$${\tilde x}_2= (x_2+e^{-x_1}).$$
This gives
$$W=e^{x_1}+{\tilde x}_2^2 -2x_2 e^{-x_1}+x_3^2+x_4^2=u,$$
which is indeed a special case of the SW curve for $N_f$ fundamental hypermultiplets (with $N_f=1$):
$$W=e^{x_1}+{\tilde x}_2^2 +P_{N_f}(x_2) e^{-x_1}+x_3^2+x_4^2=u.$$
The monodromy in this case is again of infinite order.

Next let us consider $y=2$.  This example was shown in \cite{Cecotti:1993rm} to correspond to a 2d 
LG theory with
$$W={\cal P}(X),$$
where $X$ is a periodic coordinate on the torus, and ${\cal P}(X)$ is the Weierstrass function.
The 2d-4d relation would map this to a IIB local geometry given by
$$W(x_i)=({\cal P}(X_1)+x_2^2)+x_3^2+x_4^2=u.$$
We recognize this as the Seiberg-Witten geometry for the $SU(2)$ ${\cal N}=2^*$ theory, i.e.
$SU(2)$ with a massive adjoint field, as expected.  The methods of
\cite{Gaiotto:2009hg} imply that the quantum monodromy
in this case is just $1$,\footnote{This follows from the fact that the theory can be realized
in terms of the $(2,0)$ SCFT on the torus with only ``regular'' punctures; nontrivial monodromy arises
only in cases where there are irregular punctures.} 
reflecting the fact that the theory is a mass deformation of
a conformal fixed point with R-charges all integral.

The cases with $y>2$ were not fully explored in \cite{Cecotti:1993rm}, except
to suggest that they should correspond to 2d theories which are not
asymptotically free, and thus are partially incomplete.  This matches
our conjectured equivalence with $SU(2)$ theories with one matter
field of spin $j=y/2>1$:  these theories are similarly 
not asymptotically free, and thus partially incomplete.

It is quite remarkable that all the theories covered by the 2d classification do map to some
known ${\cal N}=2$ theories in 4d.  Not only this is a very nice check of the 2d-4d
correspondence, it strongly suggests that we have a complete list of all
4d ${\cal N}=2$ systems in which the BPS spectrum is generated by three objects!

\subsection{Quivers for $SU(N)$ Yang-Mills}

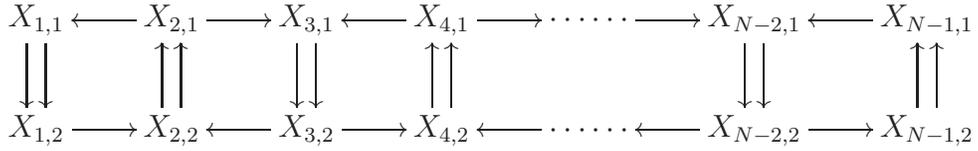
\begin{figure}
\begin{equation*}
 \xymatrix{X_{1,1}\ar@<0.7ex>[d]\ar@<-0.7ex>[d] & X_{2,1}\ar[l] \ar[r] & X_{3,1}
\ar@<0.7ex>[d]\ar@<-0.7ex>[d] & X_{4,1}\ar[l] \ar[r] &\cdots\cdots \ar[r] & X_{N-2,1} \ar@<0.7ex>[d]\ar@<-0.7ex>[d]
& X_{N-1,1}\ar[l]\\
X_{1,2} \ar[r] & X_{2,2} \ar@<0.7ex>[u]\ar@<-0.7ex>[u] & X_{3,2}\ar[l] \ar[r] &
X_{4,2}  \ar@<0.7ex>[u]\ar@<-0.7ex>[u] &\cdots\cdots \ar[l] & X_{N-2,2} \ar[l] \ar[r] & X_{N-1,2}  \ar@<0.7ex>[u]\ar@<-0.7ex>[u] }
\end{equation*}
\caption{\label{AN1Ahatquiver} The quiver $A_{N-1}\, \square\, \widehat{A}_1$}
\end{figure}

Exploiting the 2d-4d correspondence and the fact that quivers corresponding to many LG models
are known, we can now propose quivers corresponding to many $\cn=2$ gauge theories. 
For example, if we consider pure $SU(N)$ Yang-Mills theory, the Seiberg-Witten curve is
known to be
$$W=e^{x_1} + P_N(x_2) + e^{-x_1}.$$
From the perspective of the corresponding 2d LG model, separating this into two summands,
we find that the corresponding quiver
is the product of the ${\hat A}_1$ Dynkin diagram and 
a quiver corresponding to the $(LG)_N$ model.
(So it has $2 \times (N-1)$ nodes, and link structure inherited from the two separate quivers, see figure \ref{AN1Ahatquiver}.)

It should be straight-forward to extend these constructions to obtain quiver digrams for
various $\cn=2$ theories.

%
%
%
%

\section*{Acknowledgements}

We thank B. Keller for many valuable correspondences on his work and the related
literature on the cluster algebras.  We would also like to thank A. Goncharov, K. Intriligator,
N. Nekrasov and S. Shatashvili for valuable discussions.

The research of AN was supported in part by NSF grant PHY-0804450.
The research of CV was supported in part by NSF grant PHY-0244821.

\newpage
\appendix

\section{$2d$ BPS spectra and $tt^*$ in the presence of (many) collinear vacua} \label{appaligned}

In this appendix we extend the arguments already sketched in the appendices of refs.\cite{Cecotti:1993rm,Cecotti:2010qn} about the subtleties of the quantum $2d$ amplitudes when many vacua are aligned in the $Z$--plane, that is, when many (possibly infinitely many) BPS states are exactly on their walls of marginal stability. As we shall see in an explicit $2d$ example such $2d$ theories are directly relevant for the $4d$ gauge theories and, in particular, the functions expressing the solution to the $2d$ $tt^*$ equations are the same entering in the solution of the corresponding GMN equations \cite{Gaiotto:2008cd} for the quantum corrected hyperK\"ahler metric of the $\cn=4$ $\sigma$--model obtained by compactification down to $3d$.

We begin by recalling some points discussed in \cite{Cecotti:2010qn}.   \medskip

\subsection{The case of collinear vacua with a central charge lattice}

In section \ref{pure} we reduced the derivation of KS monodromy to the 2d case discussed
in \cite{Cecotti:2010qn}, using the
Melvin cigar geometry.  However in order to apply the results of that paper we need to extend the discussion
there to the case where we have infinitely many collinear vacua, as will be the case for us:  As discussed
in section \ref{pure} the space of chiral operators is spanned by $\cu_i$ which form a quantum
torus algebra.  Suppose we have a $U(1)^n$ theory.  Let us choose a canonical
basis for electric and magnetic charge line operators denoted by $\cu_i,\cv_i$.
Then we can choose an $n$ dimensional
lattice to label the electric charges ${\bf Q}$ where $\cu_i$ act by
$$\cu_i |{\bf  Q}\rangle =q^{Q_i}  |{\bf Q}\rangle$$
$$\cv_i  |{\bf  Q}\rangle =  |{\bf Q}+e_i\rangle$$

On  the other hand when we have a given 4d soliton, say with unit magnetic $i$-charge, it
contributes to the 2d monodromy by infinitely many instantons, indexed by $n$ leading to
$\prod_{n=0}^{\infty} (1-q^{n+s+{1\over 2}}\cv_i)$.  For simplicity of notation 
let us absorb the $s+{1\over 2}$ shift in a redefinition of $\cv_i$.
 In particular let us take one soliton contribution
at a time:  For $n=0$ we get a contribution
to the 2d monodromy operator $S$ given by
$$S=(1-\cv_i)$$
This operator is given by $1+T$ where $T$ counts the electric vacua differing by 1 units of electric charge.
Indeed this is what was shown to be the case when we have one soliton connecting aligned vacua.
See in particular the discussion in the refs.\cite{Cecotti:1993rm,Cecotti:2010qn}.  There are two
additional effects: only subtlety
here is that we have infinitely many vacua, instead of finite number of them.  Below we confirm
through a concrete example that no subtletly arises due to this additional ingredient.  The other
one is that we do not have only 1 soliton in 2d.  Namely one 4d soliton gives infinitely many
2d solitons (of higher spins indexed by $q^n$).  Thus adding the effect of next soliton is
$$S=(1-\cv_i)(1-q \cv_i)$$
this is indeed the effect one would expect from the arguments in \cite{Cecotti:1993rm,Cecotti:2010qn}
for two solitons.  Here the extra factors of $q$ refer to taking into account shifting the spin of the vacua
as well.  In other words, now we have two solitons connecting the adjacent electric charge states,
one of which shifts the spin, and the other does not.  Continuing in this way, we obtain the
full quantum dilogarithm.\footnote{Here we have assumed a fermionic instantonic structure.  In the
bosonic case we would simply obtain the bosonic partition function instead.}

In the next section we check that having infinitely many aligned vacua works
as in the finite dimensional case without leading to additional subtleties.

\subsection{A check in a solvable model}\smallskip

We confirm the above expectactions by an explicit computation in a LG model, having a lattice $\mathbb{Z}$ of classical vacua, for which the tt* equations may be integrated in closed form. This is the LG model with superpotential 
\begin{equation}
	W(X)=\lambda(e^{gX}-\mu X)+\mathrm{const}
\end{equation}
which has an infinite number of physically equivalent (massive) vacua and a spectrum of the BPS central charges $\{Z_{ab}\}=\{2(W(X_a)-W(X_b))\}$  equal to $\mathbb{Z}$, up to an overall complex factor.

\subsubsection{The Chiral Ring} 
The critical points are at
\begin{equation}
X_{k}=\frac{1}{g}\log\frac{\mu}{g}+\frac{2\pi i}{g}k\qquad \mathrm{with}\ \ k\in\mathbb{Z}. 
\end{equation} 
The corresponding critical values are
\begin{equation}
W_k:=W(X_k)=-\frac{2\pi i\,\lambda\mu}{g}\,k,
\end{equation} 
where we adjusted the additive constant in $W(X)$ in a convenient way. The Hessian at the critical points is $H(X_k)=g\lambda\mu$, which is independent of $k$. In the sector $\mathcal{H}_{kh}$ the central charge is ${Z}_{hk}=4\pi i \lambda\mu(h-k)/g$, and the BPS mass
\begin{equation}
m_{hk}=4\pi\, \frac{|\mu\lambda|}{|g|}\, |h-k|.\label{eq:solitionmasses}
\end{equation}

The model has a lattice--translation symmetry $T$
\begin{equation}
T\colon\;\ X\rightarrow X+2\pi i/g.
\end{equation} 
Let $l_k$ be the chiral primary associated with the $k$--th critical point; using the above symmetry, we can consider the Bloch--wave chiral primary operators
\begin{equation}
\mathcal{O}(\theta)=\sum_{k\in\mathbb{Z}} e^{i k\,\theta}\, l_k,\qquad 0\leq \theta\leq 2\pi.
\end{equation} 
As $\theta\rightarrow 0^+$, $\mathcal{O}(\theta)\rightarrow 1$, the identity operator, while as $\theta\rightarrow 2\pi^-$ we have $\mathcal{O}(\theta)\rightarrow H(X)/(g\mu\lambda)$, where $H(X)=W^{\prime\prime}(X)$ is the Hessian of the super--potential (the chiral primay of largest charge). Although everything should be periodic in the angle $\theta$, the above  identification of the operators implies that most quantities will be discontinuous at the point $\theta=2\pi$. Under the symmetry $T$ one has $\mathcal{O}(\theta)\rightarrow e^{i\theta}\,\mathcal{O}(\theta)$, and so the RG flow do not mix the Bloch--wave operators. The chiral ring $\mathcal{R}$ is simply
\begin{equation}
\mathcal{O}(\theta)\:\mathcal{O}(\theta^\prime)=\mathcal{O}(\theta+\theta^\prime).\label{eq:ringchiral}
\end{equation}

\subsubsection{$tt^*$ Equations}  

Let $C_\lambda$ be the matrix representing multiplication in $\mathcal{R}$ by
\begin{displaymath}
 \lambda^{-1}W(X)=-2\pi i(\mu/g)\sum_k k\,l_k.
\end{displaymath}
In the Bloch--wave basis it reads
\begin{equation}
C_\lambda\rightarrow -2\pi\,\frac{\mu}{g}\,\frac{\partial}{\partial\theta}.
\end{equation} 
Consider now the ground--state (tt*) metric $g_{k\bar h}$. Because of the symmetry $T$, it depends only on the difference $k-h$, and hence it has a representation
\begin{equation}
g_{k\bar h}=\int\limits^{2\pi}_{0}\frac{d\theta}{2\pi}\, e^{i(h-k)\theta}\, G(\theta),
\end{equation} 
where $G(\theta)=\langle \overline{\mathcal{O}(\theta)}|\mathcal{O}(\theta)\rangle$ (vacua with different $\theta$'s are orthogonal, of course). Hence the function $G(\theta)$ is \emph{real}, \emph{positive}, and \emph{periodic}
\begin{equation}
G(\theta+2\pi)=G(\theta).
\end{equation} Since the topological metric is proportional to $1$, the \emph{reality constraint} \cite{Cecotti:1991me} is very easy, and it reads
\begin{equation}
|g\lambda\mu|^2\, G(\theta)\, G(-\theta)=1.
\end{equation}
We write
\begin{equation}
G(\theta)=\frac{1}{|g\lambda\mu|}\exp[L(\theta)],
\end{equation} where
\begin{align}
&L(\theta)=L(\theta+2\pi)\\
&L(\theta)=-L(-\theta),
\end{align} and we expect $L(\theta)$ to be discontinuous (or otherwise non--analytic) at $\theta=2\pi$.

The dependence on $\lambda$ is governed by the $tt^*$ 
equation
\begin{equation}
\partial_{\bar\lambda}\partial_\lambda L(\theta)+\frac{4\pi^2|\mu|^2}{|g|^2}
\partial_\theta^2 L(\theta)=0,
\end{equation}
which, for this peculiar model, happens to be \emph{linear}.
Since $L(\theta)$ is real, periodic and odd,
\begin{equation}
L(\theta)=\sum^{\infty}_{m=1}L_m(\lambda)\, \sin(m\theta).
\end{equation} 
$G(\theta)$ depends on $\lambda$ only trough $|\lambda|=\rho$, since its phase can be cancelled by a redefinition of the Fermi fields. Then the $tt^*$ equation reduces to
\begin{equation}
\left(\frac{\partial^2}{\partial\rho^2}+\frac{1}{\rho}\frac{\partial}{\partial\rho}\right)
L_m(\rho)-16\pi^2\left\vert \frac{\mu}{g} \right\vert^2\, m^2\,L_m(\rho)=0.\label{eq:ttstarbessel}
\end{equation}
Set $M=4\pi \rho |\mu/g|$ ($M\equiv m_{k,k+1}$, the mass of the basic soliton, eqn.\eqref{eq:solitionmasses}). Then the general solution of eqn.\eqref{eq:ttstarbessel} can be written in terms of modified Bessel functions
\begin{displaymath}
L_{m}(M)=\gamma_m\, K_0(mM)+ \tilde\gamma_m I_0(mM)
\end{displaymath} 
where $\gamma_m$ and $\tilde\gamma_m$ are real constants to be determined. In the IR limit, $M\rightarrow\infty$, $I_0(mM)$ blows up exponentially, which is unphysical, and hence we must set $\tilde\gamma_m=0$. Thus
\begin{equation}
L(\theta,M)=\sum^{\infty}_{m=1}\gamma_m\, \sin(m\,\theta)\, K_0(mM),
\end{equation} and the $Q$--index is\begin{equation}
Q(\theta, M)\equiv-\frac{M}{2}\frac{\partial L(\theta,M)}{\partial M}=\frac{1}{2}\sum^{\infty}_{m=1}\gamma_m\, \sin(m\,\theta)\, (mM)\, K_1(mM).\label{eq:qindexthetaM}
\end{equation}

\subsubsection{IR Limit}  In the point basis the $Q$ index corresponds to the infinite \emph{Hermitean} matrix
\begin{equation}
Q_{kh}(M)=\int\limits_0^{2\pi}\frac{d\theta}{2\pi} e^{i(h-k)\theta} \,Q(\theta,M).
\end{equation} 
From the explicit form, eqn.\eqref{eq:qindexthetaM}, we see that
\begin{equation}
Q_{kh}=-\frac{\mathrm{sign}(k-h)}{4i}\, \gamma_{|k-h|}\, (|k-h|M)\, K_1(|k-h|M).\label{eq:qkhexact}
\end{equation}

On the other hand, from the  theory of the $Q$--index we know that in a \emph{general} $\mathcal{N}=2$ model, in the IR limit, \begin{equation}
Q_{kh}\simeq -\,i N_{kh}\,\exp(i\pi f_{kh})\,\frac{1}{2\pi}\, (m_{kh}\,\beta)\, K_1(m_{kh}\, \beta),\label{eq:irlimitqindex}
\end{equation}
 where \emph{generically}\footnote{That is, in absence of vacuum alignement, which is not the case of the present model.} $N_{kh}$ is an integer such that $|N_{kh}|$ is the number of BPS solitons in the sector $\mathcal{H}_{kh}$, having mass $m_{kh}$, and $f_{kh}$ is related to the fractional nature of the Fermi number in the given non--trivial sector (in the model at hand,
$f_{kh}=0$  since $H(Z)$ is proportional to $1$). $\beta$ is the inverse temperature which we have set to $1$ in the above computations. There is, however, an important \textit{caveat}: eqn.\eqref{eq:irlimitqindex} is valid under the assumption that there is no (non--trivial) sequence of indices $h_1,h_2,\dots,h_l\equiv h$ such that\begin{displaymath}\label{eq:chainofsolitons}
m_{kh}=m_{kh_1}+m_{h_1h_{2}}+m_{h_2h_3}+\cdots+m_{h_{l-1}h}
\end{displaymath} because otherwise in the RHS of eqn.\eqref{eq:irlimitqindex} we have also the contribution of the configuration of the $l$ solitons having a total mass degenerate with that of the basic soliton. Hence, for the model at hand, the formula \eqref{eq:irlimitqindex} can be trusted only for $k-h=\pm 1$.  

Comparing eqns.\eqref{eq:qkhexact} and \eqref{eq:irlimitqindex} for $|k-h|=1$, we get for the first coefficient
\begin{equation}\label{eq:gammaoneenne}
{\;\gamma_{1}=\pm\frac{2N_{k,k\pm 1}}{\pi}\;}
\end{equation} 

\subsubsection{UV Regime} One has $\lim_{z\rightarrow 0}(z\, K_1(z))=1$, and hence
\begin{equation}
q(\theta):=\lim_{M\rightarrow 0}Q(\theta,M)=
\frac{1}{2}\sum_{m=1}^\infty\gamma_m\sin(m\theta).
\end{equation} 
The function $\hat q(\theta):=q(\theta)-q(0)$, should correspond to the chiral charge of the chiral primary operator $\mathcal{O}(\theta)$ at the UV fixed point (this operator should have a definite charge, since it is chiral primary and does not mix with the others). Hence, from the ring structure, eqn.\eqref{eq:ringchiral}, we get the constraint
\begin{equation}
 \hat q(\theta+\theta^\prime)=\hat q(\theta)+\hat q(\theta^\prime),
\end{equation} from which we infer
\begin{equation}
q(\theta)=\alpha(\theta-\pi)\qquad\ \mathrm{for}\ 0\leq\theta\leq 2\pi,
\end{equation} 
for some \emph{positive} constant $\alpha$. In particular, $q(\theta)$ is discontinuous at $\theta=2\pi$, as expected.
 Now 
\begin{equation}\label{eq:dentedisega}
\theta-\pi=-2\sum_{m=1}^\infty\frac{\sin(m\theta)}{m}\qquad \ \mathrm{for}\ 0<\theta<2\pi,
\end{equation} 
that is,
\begin{equation}\label{eq:coefficients}
\gamma_m=-\frac{4\alpha}{m},
\end{equation}
and, comparing with eqn.\eqref{eq:gammaoneenne}, we get 
$\alpha=N/(2\pi)$, where $N>0$ is the number of {basic} solitons of mass $M$. Then
\begin{equation}
q(\theta)=N\,\frac{\theta-\pi}{2\pi}.
\end{equation} 

The WKB method gives $N=1$. Indeed, {the discontinuity} $q(2\pi)-q(0)=1$ {should} {be equal to the central charge} $\hat c$ {at the} UV {super--conformal limit} ($\lambda\rightarrow 0$). Hence $\hat c=1$, the right result for a \emph{free} $\mathcal{N}=2$ model with \emph{one} chiral superfield. 

Eqn.\eqref{eq:coefficients} has a natural interpretation. There is no soliton of mass $mM$ for $m\neq 1$, and the term in $Q(\theta,M)$ proportional to $(mM)K_1(mM)$ arises from a chain of $m$ basic solitons as in eqn.\eqref{eq:chainofsolitons}. It appears that such a $m$ soliton contribution has a relative factor $1/m$ with respect to the one we would get from a genuine BPS soliton of mass $(mM)$ soliton. 

\medskip

\subsubsection{Relation with $4d$ $\cn=2$ QED} 
\smallskip

We note the identity
\begin{equation}
\frac{\partial}{\partial\theta}\log G(\theta,M)=
-\frac{2}{\pi}\,\sum^{\infty}_{m=1}\cos(m\phi)\, K_0(mM),
\end{equation}
where the function in the \textsc{rhs} is precisely the one appearing in the wall--crossing analysis for $\cn=2$ QED
in four dimensions \cite{Gaiotto:2008cd}. This is not at all an accident. It is further evidence of the fact that the same structures and geometries control the wall--crossing in $4d$ and in $2d$. 

This is yet another detailed confirmation of the basic $2d$/$4d$ correspondence, in the target sense.

\section{A generalized twistor construction} \label{app:hk-ambient}

In this section we describe a slight generalization of the construction of Section \ref{deforming}, where
the ambient space $\C^2$ is replaced by a more general hyperk\"ahler manifold.

\subsection{A simple Calabi-Yau 3-fold}

Let $Q$ be a hyperk\"ahler 4-manifold, with complex structures $J^\zeta$, K\"ahler forms
$\omega^\zeta$, and normalized holomorphic symplectic forms $\Omega^\zeta$
($\zeta \in \C\PP^1$).
Let 
\begin{equation}
X = Q \times \C^\times. 
\end{equation}
The structures $(J^{\zeta = 1}, \omega^{\zeta = 1}, \Omega^{\zeta = 1})$
make $Q$ into a Calabi-Yau manifold.  The cylinder $\C^\times$ is also Calabi-Yau in a standard way.
So $X$ is a Calabi-Yau threefold with the obvious product structure.

\subsection{A conformal brane}

Let $\Sigma \subset Q$ be a fixed 1-dimensional complex submanifold in complex structure $J^{\zeta = 0}$.
Note that since $\Sigma$ is complex inside $(Q, J^{\zeta = 0})$ it is special Lagrangian inside $(Q, \omega^{\zeta = 1},
\Omega^{\zeta = 1})$.  Then define
\begin{equation}
 L_0 = (\Sigma \times S^1) \subset X.
\end{equation}
This is the product of two special Lagrangian submanifolds and hence it is again special Lagrangian.

\subsection{A non-conformal brane}

More generally, let $R: U(1) \times Q \to Q$ be an action of $U(1)$ on $Q$.
Write $R(\vartheta): Q \to Q$ for the action of $e^{i \vartheta} \in U(1)$.
Suppose that $R(\vartheta)$ is an isometry and $R(\vartheta)^* J^\zeta = J^{e^{-i \vartheta} \zeta}$,
i.e. $R(\vartheta)$ rotates the twistor sphere by $e^{i \vartheta}$.
Then define 
\begin{equation}
L = \bigcup_{\vartheta} \left( R(\vartheta)\Sigma \times \{e^{i \vartheta}\} \right) \subset X.
\end{equation}
$L$ is ``fiberwise special Lagrangian'', i.e. its restriction to the fiber of $X$ over any
$\zeta = e^{i \vartheta} \in \C^\times$ is special Lagrangian.
The whole $L$ however is not Lagrangian inside the whole $X$, unless $\Sigma$ happens to be fixed 
by the $U(1)$ action, 
in which case we reduce to $L = L_0$ above.

We can consider holomorphic discs $D \subset X$ which have boundary along $L$.  Suppose $D$ is contained
at a single point $e^{i \vartheta}$ of the $\C^\times$ fiber of $X$.  Then $R(-\vartheta) D$ 
is a holomorphic disc on $(Q, J^{\zeta = e^{i \vartheta}})$ with boundary on $\Sigma$.

%

\subsection{Examples}

\begin{itemize}

 \item Take $C$ to be a compact Riemann surface,
$Q$ a sufficiently small $U(1)$-invariant 
neighborhood of the zero section in $T^*C$.  $Q$ then admits a unique $U(1)$-invariant hyperk\"ahler metric
\cite{MR1848662}, with the desired properties.  Take $\Sigma$ any branched $n$-fold cover
of $C$ in $T^* C$.  This gives a non-conformal brane, which should 
become conformal in the limit where
$\Sigma$ degenerates to $n$ copies of the zero section.

\item Take $Q = \R^4$ with its flat metric.  Identify it with $\C^2$ with coordinates $(x,y)$.  
An hyperk\"ahler structure is determined by giving the
K\"ahler and holomorphic symplectic forms in complex structure $\zeta = 0$:  they are
\begin{equation}
 \omega_3 = dx \wedge d\bar{x} + dy \wedge d\bar{y}, \qquad \omega_+ = dx \wedge dy.
\end{equation}
As usual, if we define $\omega_- = \overline{\omega_+}$ the 
general holomorphic symplectic form is
\begin{equation} \label{eq:Omegazeta}
\varpi(\zeta) = - \frac{i}{2\zeta} \omega_+ + \omega_3 - \frac{i}{2} \zeta \omega_-.
\end{equation}

Then let $R(\vartheta)$ act by $(x, y) \to (e^{m i \vartheta / (m+n)} x, e^{n i \vartheta / (m+n)} y)$.
Note this takes $\varpi(\zeta) \to \varpi(e^{-i \vartheta} \zeta)$, so it acts in the 
desired way on the twistor sphere.

Take $\Sigma = \{y^m = x^n\} \subset Q$.  This is fixed by all $R(\vartheta)$ and hence 
gives a conformal brane.  This recovers the construction of the main text.

\end{itemize}

\section{Diagram folding and non--simply laced $Y$--systems}\label{sec:diagramfol}

Many $ADE$ models have a simpler formulation in terms of non--simply laced Dynkin diagrams of smaller rank.
Here we limit to two simple examples just to illustrate the technique. In this way, also the $Y$--system associated to
non--simply laced Dynkin diagrams will play a r\^ole for the quantum monodromy theory.

\subsection{The $A_3\rightarrow B_2$ folding}

The $A_3$ quantum torus algebra, associated to the quiver $\begin{diagram} \node{1}\node{2} \arrow{w}\arrow{e}\node{3}\end{diagram}$, is generated by three \emph{invertible} operators $X_i$ ($i=1,2,3$)
satisfying the quantum relations
\begin{equation}
X_2X_1=qX_1X_2,\quad X_2X_3=qX_3X_2,\quad X_1X_3=X_3X_1.
\end{equation} 
In particular, the operator $Z=X_1X_3^{-1}$ commutes with all generators and hence belongs to the center of the algebra.
This also implies that $Z$ is invariant under monodromy: $Z\rightarrow M^{-1}ZM\equiv Z$.

On any irreducible representation of the quantum torus algebra, $Z$ is a $c$--number. Then we remain with a reduced algebra with just two generators, $X_2$ and $X_3$, which then must be associated to a rank $2$ Lie algebra. The Dynkin diagram for the reduced algebra is obtained by folding the original one:
\begin{displaymath}
\Big( \begin{diagram}\node{1}\node{2}\arrow{w}\arrow{e}\node{3}\end{diagram}\Big)\ \longrightarrow\
\Big(\ \xymatrix{2\ \ar@2[rr]&&\ 3,1}\ \Big)
\end{displaymath}
Assume, for simplicity, that $Z$ has the special value $-1$. Then the quantum monodromy reads
\begin{equation}
 \begin{split}
  M
&=\Psi(-X_3;q)\, \Psi(X_3;q)\, \Psi(X_2;q)\,\Psi(-X^{-1}_3;q)\, \Psi(X^{-1}_3;q)\, \Psi(X^{-1}_2;q)\\
&\sim \Psi(X_2;q)\, \Psi(X_3^{-2};q^2)\, \Psi(X_2^{-1};q)\,\Psi(X_3^2;q^2).
 \end{split}
\end{equation}
where $\sim$ means equivalence up to conjugacy.

Consider the sequence of operators $Y_k$, $k\in\mathbb{Z}$ with
$Y_0=X_2$, $Y_{-1}=X_3^{-2}$, satisfying the quantum relations
\begin{equation}
 Y_{k+1}Y_{k}= q^2Y_{k}Y_{k+1}
\end{equation} 
and the recursion relation
\begin{equation}\label{recB2}
 \Psi(Y_{k+1}; q_{k+1})\, Y_k\, \Psi(Y_{k+1};q_{k+1})^{-1}=Y^{-1}_{k+2},
\end{equation} 
where
\begin{equation}
 q_k=\begin{cases}
      q & k\ \text{even}\\
q^2 & k\ \text{odd}.
     \end{cases}
\end{equation} 
Then
\begin{equation}
\begin{cases}
 Y_{k+2}Y_{k}=(1-q\, Y_{k+1})& k\ \text{even}\\
Y_{k+2}Y_{k}=(1-q^{1/2}\, Y_{k+1})(1-q^{3/2}Y_{k+1})& k\ \text{odd},
\end{cases}
\end{equation} 
which are eqns.(60) of ref. \cite{MR2567745}
for the mutations in the $B_2$ cluster algebra\footnote{\ Up to the redefinition $Y_k\rightarrow -Y_k$.}. In particular, the sequence is periodic mod $6$: $Y_{k+6}=Y_k$. Notice that $6$ is $h+2$ for $B_2$.

The recursion relation implies the following remarkable expression for $M$
\begin{equation}
 M=\Psi(Y_k;q_k)\, \Psi(Y_{k-1};q_{k-1})\,\Psi(Y_{k-2};q_{k-2})\, \Psi(Y_{k-3};q_{k-3})\qquad \forall\, k\in \mathbb{Z}
\end{equation}
which, together with eqn.\eqref{recB2} gives
\begin{equation}
 Y_k M\equiv M Y_{k-4}\equiv M Y_{k+2},
\end{equation} 
so the quantum monodromy $M$ of the $A_3$ model implements $Y_k\rightarrow Y_{k+2}$ on the reduced $B_2$ quantum algebra, and hence must have order $3$. Of course, this coincides with the order in table \ref{Table} for $A_3$.

The discussion for a generic value of the central element $Z$ is similar.

\subsection{The $D_4\rightarrow G_2$ folding}

The $D_4$ quantum torus algebra has four invertible generators $Y,X_i$ ($i=1,2,3$) with relations
\begin{equation}
 YX_i=qX_iY,\qquad X_iX_j=X_jX_i,\qquad i,j=1,2,3.
\end{equation} 
Again, the elements $X_iX_j^{-1}$ are central, hence monodromy invariants. Therefore we have a reduced quantum algebra with only two generators, which should correspond to the rank $2$ Dynkin diagram obtained by folding the $D_4$ one, which is the $G_2$ diagram:
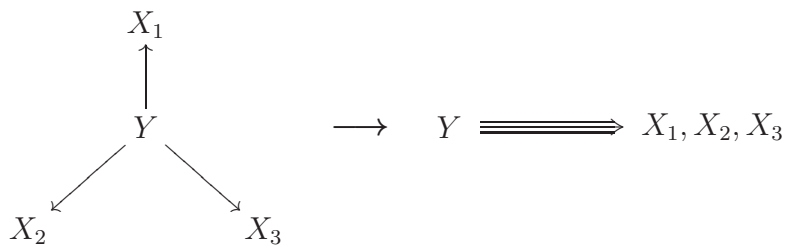
\vglue 12pt

\begin{equation*}
\begin{gathered}\xymatrix{ & X_1 & \\
& Y\ar[u]\ar[dl]\ar[dr] & \\
X_2 & & X_3}\end{gathered} \quad \boldsymbol{\longrightarrow}\quad \xymatrix{Y\ \ar@3[rr]& &\ X_1,X_2,X_3}
\end{equation*} 
\vglue 12pt

For simplicity, assume the numerical values of such central elements is such that $X_k=e^{2\pi i k/3}X_3$ ($k=1,2,3$), and set
\begin{equation}
 q_k=\begin{cases}
      q & k\ \text{even}\\
q^3 & k\ \text{odd}.
     \end{cases}
\end{equation} 
Again, we have a sequence $\{Y_k\}_{k\in\mathbb{Z}}$ of operators satisying $Y_{k+1}Y_k=q^3Y_kY_{k+1}$ and
the recursion relation
 \begin{equation}
 \Psi(Y_{k+1}; q_{k+1})\, Y_k\, \Psi(Y_{k+1};q_{k+1})^{-1}=Y^{-1}_{k+2},
\end{equation}
which implies
\begin{equation}
 \begin{aligned}
         &Y_{k+2}\, Y_k=(1-q^{3/2}Y_{k+1}) & & k\ \text{even}\\
& Y_{k+2}\, Y_k=(1-q^{1/2}Y_{k+1})(1-q^{3/2}Y_{k+1})(1-q^{5/2}Y_{k+1}) & & k\ \text{odd}
        \end{aligned} 
\end{equation}
which are eqns.(60) of \cite{MR2567745} for $G_2$ (up to $Y_k\rightarrow - Y_k$).
The sequence is periodic mod $h_{G_2}+2=8$, while the physical monodromy of the $D_4$ model reads
\begin{equation}
 M=\Psi(Y_{k+3};q_{k+3})\, \Psi(Y_{k+2};q_{k+2})\, \Psi(Y_{k+1};q_{k+1})\, \Psi(Y_k;q_k),\qquad \forall\, k\in\mathbb{Z},
\end{equation}  
so that $M^{-1}Y_kM\equiv Y_{k+4}$, and hence the $D_4$ quantum monodromy has order $2$, in agreement with table \ref{Table}.

\section{The $(A_n,A_1)$ theory in the linear BPS chamber}\label{app:linearchamber}

\subsection{Quiver mutation--equivalence}

In this section we briefly consider the monodromy operator for the $(A_n, A_1)$ theories,
in the alternative ``linear'' chamber corresponding to the quiver
\begin{equation}
 \begin{diagram}\label{Anlinquiver}
  \node{X_n}\node{X_{n-1}}\arrow{w}\node{X_{n-2}}\arrow{w}
 \node{\cdots}\arrow{w}
 \node{X_{2}}\arrow{w}\node{X_{1}}\arrow{w}
\end{diagram}
\end{equation} 
where we write at each vertex the corresponding quantum algebra generator. This quiver is \emph{mutation--equivalent} to the canonical $A_{n}$ one with only sink and source nodes
\begin{equation}\label{Ancanquiver}
 \begin{diagram}
  \node{n}\node{n-1}\arrow{w}\arrow{e}\node{n-3}
 \node{\cdots}\arrow{w}\arrow{e}
 \node{2}\node{1}\arrow{w}
\end{diagram}
\end{equation}  
in fact, one can mutate the quiver \eqref{Ancanquiver} into the \eqref{Anlinquiver} by a sequence of mutations at sink/source nodes distint of the $n$--th one. This can be seen by induction on $n$. For $n=2$ there is nothing to show. Assume the claim is true for the $A_n$ quiver and consider the $A_{n+1}$ one. By mutations at the sinks $n-1, n-3, n-5,\dots$ we put the full subquiver with the $(n+1)$--th node omitted in the canonical $A_n$ form. By the induction hypothesis, we can put the subquiver into the linear form by a sequence of mutations at vertices distinct from the $n$--th one. Thus, rearranging the $A_n$ subquiver we do not modify the direction of the $n+1\ \leftrightarrow\ n$ arrow, which has already the right orientation. The resulting quiver is the linear $A_{n+1}$ one. 

Therefor the two quivers correspond to the same class of $2d$ $\cn=2$ models. By the $2d/4d$ correspondence discussed in this paper, the corresponding two $4d$ $\cn=2$ gauge theories should also belong to the same class of the proposed classification.
Indeed, in section \ref{sec:a2nlincham} we saw that, thanks to a new magical $q$--series identity, the traces of the monodromies of the two model agree.
\medskip

The phase--ordering of the BPS states is, of course, different in the linear chamber with respect to the one in the canonical chamber, and apriori their could have been extra BPS states.   However, we will see
that with the assumption of the same number of BPS states, we get consistent results
validating the existence of this chamber\footnote{This can in principle be established using the methods of
\cite{Shapere:1999xr}.}.  Furthermore,
in general we do not have a simple rule for the phase--order. However, in the Dynkin diagram models a practical way of finding an ordering which leads to a consistent monodromy was
to select a sequence of elementary cluster--mutations $\prod cq_k$ such that the corresponding sequence of quiver--mutations $\prod_k\mu_k$ implements the inversion $I$ of the quantum algebra, as needed to revover the correct PCT structure. 

Let us consider the linear $A_n$ quiver \eqref{Anlinquiver}. For graphical convenience we rewrite (a segment of) the linear quiver in the form
\begin{equation}\label{quiverXlin}
 \begin{diagram}
 \node{}\node{X_{k+5}} \node{} \node{X_{k+3}}\arrow{sw} \node{} \node{X_{k+1}}\arrow{sw}\node{}\node{}\arrow{sw,..,-}\\
\node{}\arrow{ne,..,-}\node{}\node{\boldsymbol{X}_{k+4}}\arrow{nw}\node{}\node{\boldsymbol{X}_{k+2}}\arrow{nw}\node{}\node{\boldsymbol{X}_{k}}\arrow{nw}\node{}
 \end{diagram} 
\end{equation} 
Let us try first, \textit{e.g.}, the even/odd order as in the canonical chamber. Then we perform first the mutations at the even (or odd) nodes which we draw in the lower position in \eqref{quiverXlin} (the ones written in bold face). One would get the quiver
\begin{equation}\label{quiverXXlin}
 \begin{diagram}
 \node{}\arrow{e,..,-}\node{X_{k+5}X_{k+6}}\arrow{se} \node{} \node{X_{k+3}X_{k+4}}\arrow{se}\arrow[2]{w} \node{} \node{X_{k+1}X_{k+2}}\arrow{se}\arrow[2]{w}\node{}\arrow{w,..,-}\\
\node{}\arrow{ne,..,-}\node{}\node{{X}_{k+4}^{-1}}\arrow{ne}\node{}\node{{X}_{k+2}^{-1}}\arrow{ne}\node{}\node{{X}_{k}^{-1}}
 \end{diagram} 
\end{equation} 

Now the result of the mutation at the upper nodes would depend on their order since they do not longer commute. However, no choice of order would eliminate the orizontal arrows in the quiver \eqref{quiverXXlin} reproducing the original one \eqref{quiverXlin}.

On the contrary let us perform first $\mu_1$, then $\mu_2$, after $\mu_3$, $\cdots$ starting from the quiver \eqref{Anlinquiver}:
\begin{align*}
&\mu_1\ \Rightarrow\  
\begin{diagram}
\node{} \arrow{e,..,-}\node{X_5}\node{X_4}\arrow{w}\node{X_3}\arrow{w}\node{X_2}\arrow{w}\arrow{e}\node{X_1^{-1}} 
\end{diagram} \\
&\mu_2\ \Rightarrow\  
\begin{diagram}
\node{} \arrow{e,..,-}\node{X_5}\node{X_4}\arrow{w}\node{X_3}\arrow{w}\arrow{e}\node{X_2^{-1}}\node{X_1^{-1}}\arrow{w} 
\end{diagram} \\
&\mu_3\ \Rightarrow\  
\begin{diagram}
\node{} \arrow{e,..,-}\node{X_5}\node{X_4}\arrow{w}\arrow{e}\node{X_3^{-1}}\node{X_2^{-1}}\arrow{w}\node{X_1^{-1}}\arrow{w} 
\end{diagram} \\
&\cdots\hskip 1cm \cdots\hskip 4cm \cdots\hskip 4cm \cdots\\
&\mu_{n-1}\ \Rightarrow\  
\begin{diagram}
\node{X_n}\arrow{e}\node{X^{-1}_{n-1}}\node{X_{n-2}^{-1}}\arrow{w}\arrow{e,..,-}\node{X_2^{-1}}
\node{X_1^{-1}}\arrow{w} 
\end{diagram}\\
&\mu_{n}\ \Rightarrow\  
\begin{diagram}
\node{X_n^{-1}}\node{X^{-1}_{n-1}}\arrow{w}\arrow{e,..,-}\node{X_3^{-1}}\node{X_2^{-1}}\arrow{w}
\node{X_1^{-1}}\arrow{w} 
\end{diagram},
\end{align*}
so the quiver--mutation
$$\prod_\mathrm{linear\ order}\mu_k$$
maps the linear quiver back to itself up to the inversion automorphism $I$ of the quantum torus algebra. Then the expression
\begin{equation}\label{monlincaabst}
M(q)= \bigg(\prod_\mathrm{linear\ order}\cq_k\bigg)^2,
\end{equation} 
has the right PCT structure to be a quantum monodromy. The fact that it has the right order, traces, and fixed points confirms the identification.

\subsection{Quantum monodromy}

From eqn.\eqref{monlincaabst} we see that in the linear chamber the quantum monodromy $M(q)$ has the explicit form
\begin{multline}
	M(q)= \Psi(X_1;q)\, \Psi(X_2;q)\, \Psi(X_3;q)\cdots \Psi(X_{n};q)\times\\
	\times\Psi(X_1^{-1};q)\, \Psi(X_2^{-1};q)\, \Psi(X_3^{-1};q)\cdots \Psi(X_{n}^{-1};q).\hskip 1.6cm
\end{multline}
Since the linear quiver is simply--laced, the adjoint action of this operator can be computed using the techniques of section \eqref{simplyla}, and turns out to be:
\begin{align}
&\begin{aligned}
 \text{for }&  1 \le k \le n-2: \\
& X_k \to N\left[\frac{X_{k+2} (1 - X_2 + X_2 X_3 - X_2 X_3 X_4 + \cdots +(-1)^{k-1} \prod_{i=2}^k X_i)}{1 - X_2 + X_2 X_3 - X_2 X_3 X_4 + \cdots +(-1)^{k-1} \prod_{i=2}^{k+2} X_i}\right]\label{MMo1}
\end{aligned}\\
\intertext{and}
&\begin{aligned}
 X_{n-1} &\to (-1)^{n}\,N\left[\frac{1 - X_2 + X_2 X_3 - X_2 X_3 X_4 + \cdots + (-1)^{n}\prod_{i=2}^{n-1} X_i}{\prod_{i=2}^{n} X_i}\right], \\
X_{n} &\to N\Big[X_1 - X_1 X_2 - X_1 X_2 X_3 + \cdots +(-1)^{n} \prod_{i=1}^{n} X_i\Big],\label{MMo3}
\end{aligned}
\end{align}
where $N[\cdots]$ stands for the normal order.\smallskip

Here we just record the simplest nontrivial example:  in case $n=2$ this action is simply
\begin{align}
 X_1 &\to \frac{1}{X_2}, \\
 X_2 &\to X_1(1 - X_2),
\end{align}
and fixed points are at
\begin{align}
 (X_1, X_2) &= \left( \frac{1}{x}, x \right)
\end{align}
where $x$ is one of the 2 roots of the golden ratio equation
\begin{equation}
 x^2 + x - 1 = 0.
\end{equation}

\section{Explicit construction of $\mu_\square$ for $A_m\,\square\, A_n$ quivers}\label{ape:agymnastics}

In this appendix we explicitly construct the rational map $\mu_\square$ for $A_m\,\square\,A_n$ quivers, establishing eqns.\eqref{ksquare1}--\eqref{ksquaren} of the main text.  The quiver $A_m\,\square\,A_n$ is represented in figure \ref{squaremnquiver}.

We consider an inner plaquete of the quiver $A_m\,\square\,A_n$:
\begin{equation}
 \begin{diagram}
  \node{}\node{\vdots}\node{\vdots}\arrow{s}\node{}\\
\node{\cdots}\arrow{e}\node{X_{2l+1,2k+1}}\arrow{n}\arrow{s}\node{X_{2l+2,2k+1}}\arrow{w}\arrow{e} \node{\cdots}\\
\node{\cdots} \node{X_{2l+1,2k+2}}\arrow{w}\arrow{e} \node{X_{2l+2,2k+2}}\arrow{n}\arrow{s}\node{\cdots}\arrow{w}\\
\node{} \node{\vdots}\arrow{n}\node{\vdots}\node{}
 \end{diagram}  
 \end{equation} 
where at each node we write the corresponding cluster variable. 

We
perform the sequence of elementary quiver--mutations which defines $\mu_\square$. For plaquetes on the boundary of the quiver $A_m\, \square\,A_n$ the following expressions must be modified in the obvious way. 

The mutation $\mu_{+1,-1}$ produces the following quiver/basis of the algebra:
\begin{equation}
 \begin{diagram}
  \node{}\arrow{se}\node{\vdots}\node{\vdots}\arrow{sw}\arrow{se}\node{}\\
\node{\cdots}\node{X_{2l+1,2k+1}}\arrow{w}\arrow{e}\arrow{n}\arrow{s}\node{X_{2l+2,2k+1}^{-1}}\arrow{n}\arrow{s} \node{\cdots}\arrow{w}\\
\node{\cdots}\arrow{ne}\arrow{se} \node{X_{2l+1,2k+2}}\arrow{w}\arrow{e} \node{\prod_{s=1}^3X_{2l+2,2k+s}}\arrow{ne}\arrow{nw}\arrow{se}\arrow{sw}\node{\cdots}\arrow{w}\\
\node{} \node{\vdots}\arrow{n}\node{\vdots}\arrow{n}\node{}
 \end{diagram}  
 \end{equation} 
then the application of $\mu_{-1,+1}$ gives the quiver
\begin{equation}
 \begin{diagram}
  \node{}\node{\vdots}\arrow{s}\node{\vdots}\node{}\\
\node{\cdots}\node{\prod_{s=0}^2X_{2l+1,2k+s}}\arrow{e}\arrow{w}\node{X_{2l+2,2k+1}^{-1}}\arrow{n}\arrow{s} \node{\cdots}\arrow{w}\\
\node{\cdots}\arrow{e} \node{X_{2l+1,2k+2}^{-1}}\arrow{n}\arrow{s} \node{\prod_{s=1}^3X_{2l+2,2k+s}}\arrow{w}\arrow{e}\node{\cdots}\\
\node{} \node{\vdots}\node{\vdots}\arrow{n}\node{}
 \end{diagram}  
 \end{equation} 
and $\mu_{+1,+1}$
\begin{equation}
 \begin{diagram}
  \node{}\node{\vdots}\arrow{s}\node{\vdots}\arrow{s}\node{}\\
\node{\cdots}\arrow{ne}\arrow{se}\node{\prod_{s=0}^2X_{2l+1,2k+s}}\arrow{e}\arrow{w}\node{\prod_{s=-1}^3X_{2l+2,2k+s}}
\arrow{ne}\arrow{se}\arrow{nw}\arrow{sw} \node{\cdots}\arrow{w}\\
\node{\cdots} \node{X_{2l+1,2k+2}^{-1}}\arrow{n}\arrow{s}\arrow{w}\arrow{e} \node{\prod_{s=1}^3X_{2l+2,2k+s}^{-1}}\arrow{n}\arrow{s}\node{\cdots}\arrow{w}\\
\node{}\arrow{ne} \node{\vdots}\node{\vdots}\arrow{ne}\arrow{nw}\node{}
 \end{diagram}  
 \end{equation} 
Finally, $\mu_{-1,-1}$ gives back the original quiver with a mutated basis
\begin{equation}
 \begin{diagram}
  \node{}\node{\vdots}\node{\vdots}\arrow{s}\node{}\\
\node{\cdots}\arrow{e}\node{\prod_{s=0}^2X_{2l+1,2k+s}^{-1}}\arrow{n}\arrow{s}\node{\prod_{s=-1}^3X_{2l+2,2k+s}}\arrow{w}\arrow{e} \node{\cdots}\\
\node{\cdots} \node{\prod_{s=0}^4 X_{2l+1,2k+s}}\arrow{w}\arrow{e} \node{\prod_{s=1}^3X_{2l+2,2k+s}^{-1}}\arrow{n}\arrow{s}\node{\cdots}\arrow{w}\\
\node{} \node{\vdots}\arrow{n}\node{\vdots}\node{}
 \end{diagram}  
 \end{equation} 

In conclusion, $\mu_\square$ implements the classical rational map:
\begin{align}
& X_{2l,2k+1} \mapsto X_{2l,2k-1}X_{2l,2k}X_{2l,2k+1}X_{2l,2k+2}X_{2l,2k+3}\\
&X_{2l,2k+2}\mapsto X^{-1}_{2l,2k+1}X^{-1}_{2l,2k+2}X^{-1}_{2l,2k+3}\\
&X_{2l+1,2k+1}\mapsto X^{-1}_{2l+1,2k}X^{-1}_{2l+1,2k+1}X^{-1}_{2l+1,2k+2}\\
&X_{2l+1,2k+2}\mapsto X_{2l+1,2k}X_{2l+1,2k+1}X_{2l+1,2k+2}X_{2l+1,2k+3}X_{2l+1,2k+4}\\
\intertext{with the \textit{convention}}
&X_{l,k}=1\ \text{for } k=0\ \text{or } k>n,\\
\intertext{and the \textit{exceptions}}
&\bullet \quad X_{2l,1}\mapsto X_{2l,2}X_{2l,3}\\
&\bullet\quad \begin{cases}
               X_{2l,n}\mapsto X_{2l,n-2}X_{2l,n-1} & n\ \text{odd}\\
X_{2l+1,n}\mapsto X_{2l+1,n-2}X_{2l+1,n-1} & n\ \text{even}
              \end{cases}
\end{align}
where the conventions and the exceptions follow from the consideration of the boundary plaquetes.\smallskip
 
A remarkable property of $\mu_\square$ is that it maps $X_{l,k}$ into a rational function of $X_{l,k^\prime}$ with a \emph{fixed} $l$. Thus, to show that $(\mu_\square)^{m+1}=\mathbf{1}$ we may work at fixed $l$. In other words, we may effectively replace $G$ by the trivial quiver $A_1$.

\section{Rational cases for $(A_3,A_1)$ and $(A_4,A_1)$}\label{apprat}

In this appendix we discuss two additional examples for rational values of $q$, which
shows very similar features to the $(A_2,A_1)$ case already discussed in the main
body of the paper.

\subsection{The $(A_3,A_1)$ model}\label{apprat2}

The $(A_3,A_1)$ model has a quantum algebra $\ca_3$ which is generated by three \textit{invertible} elements, $X$, $Y$ and $Z$ with
\begin{equation}
 XY=q YX,\qquad YZ=q ZY,\qquad XZ=ZX.
\end{equation}
In particular $XZ$ commutes with everything; therefore $XZ$ is a non--zero complex number in any given irreducible representation of $\ca_3$. We call this number $\rho$,
\begin{equation}
 XZ=\rho,\qquad \rho\in\mathbb{C}^*.
\end{equation} 
The central element $XZ$ is invariant under the monodromy $M$. Thus we may consider the monodromy action restricted to $Y,Z$:
\begin{equation}
 M\colon \quad\begin{aligned}
               &Y \mapsto Z^{-1}(1-q^{1/2}Y^{-1})\\
&Z\mapsto \rho\, \big(1-q^{1/2}Y+YZ\big)Z^{-1}.\label{quantummon}
              \end{aligned}
\end{equation} 
Iterating three times the quantum monodromy \eqref{quantummon} one gets the identity map.

Again, we set $q$ equal to a primitive $N$--th root of unity.
$Y^N$ and $Z^N$ become central, and act as scalars in any given irreducible representation of the quantum $\ca_3$ algebra. An irreducible representation is labelled by the values of $\rho$ and these two central elements that we call, respectively,
$y$ and $z$. Then, up to isomorphism, there is a unique representation of the $\ca_3$ algebra (where $X\equiv \rho Z^{-1}$)
\begin{align}
 Y&=y^{1/N}\mathrm{diag}(q^k)_{k=0,\cdots, N-1}\label{rep1}\\
 Z_{ij}&=z^{1/N}\,\delta_N(i-j-1)\equiv \frac{z^{1/N}}{N}\sum_{k=0}^{N-1}q^{(i-j-1)k},\label{rep2}
\end{align}
the choices of the $N$--roots in the overall factors being irrelevant up to isomorphism. Again, one has
$\det[Y]=(-1)^{N-1}y$, $\det[Z]=(-1)^{N-1}z$,
and the action of the monodromy $M$ on the central elements is determined by
\begin{equation*}
\begin{aligned}
               \det[Y] \mapsto &\det[Z]^{-1}\det[1-q^{1/2}Y^{-1}]
=(-1)^{N-1}\:\frac{1-(q^{N/2}y)^{-1}}{z}\\
\det[Z]\mapsto &\: (-1)^{N-1}\,\frac{\rho^N}{z}\, \det[1-q^{1/2}Y+YZ)]=(-1)^{N-1}\rho^{N}\,\frac{1-q^{N/2}y+yz}{z},
              \end{aligned}
\end{equation*}
which just says that the central elements $q^{N/2}X^N$, $q^{N/2}Y^N$ and $q^{N/2}Z^N$ transform under the quantum monodromy as the classical variables $X$, $Y$, and $Z$, in agreement with the general result of sect.\,\ref{sec:quantFrob}. 

The (non--singular) fixed points $(q^{N/2}y,q^{N/2}z)$ of the monodromy are
\begin{align}
 &q^{N/2}z=\frac{q^{N/2}y-1}{(q^{N/2}y)^2}\label{fix1}\\
 &\text{where } (q^{N/2}y)^3+\rho^{-N}=0.\label{fix2}
\end{align}
\smallskip

The monodromy $M$ is defined by the properties
\begin{gather}
 YM=M Z^{-1}(1-q^{-1/2}Y^{-1})\label{mon1}\\
ZM= \rho M(1-q^{1/2}Y+YZ)Z^{-1}.\label{mon2}
\end{gather}
 In the representation \eqref{rep1}\eqref{rep2}, these equations become the recursion relations 
\begin{gather}
 \frac{M_{k,l}}{M_{k,l-1}}=(\mu\lambda)^{-1}\, q^{-k}\, (1-\lambda^{-1}q^{1/2-l})\label{ee3}\\
\frac{M_{k,l}}{M_{k-1,l}} =\frac{\mu}{\rho\lambda}\,\frac{q^{-l}}{1-\lambda\, q^{k-1/2}}.\label{ee4}
\end{gather}
To solve the recursions \eqref{ee3}\eqref{ee4}, we write
 $M=AFB$,
with $A,B$ diagonal matrices and $F_{mn}=N^{-1/2}\,q^{-mn}$ the {Fourier transform},
\begin{align}
 M_{k,l}&=\frac{1}{\sqrt{N}}\: A_k\, q^{-kl}\,B_l\label{moma1}\\
 A_k&= \left(\frac{\mu}{\rho\lambda}\right)^{\!\!k}\: \prod_{s=1}^{k} (1-\lambda\, q^{s-1/2})^{-1}\label{moma2}\\
 B_l&=(\mu\lambda)^{-l}\: \prod_{r=1}^l(1-\lambda^{-1}q^{1/2-r}).\label{moma3}
\end{align}
The diagonal matrices, $A,B$, are just discrete quantum dilogarithms.
Again, the periodicity property $A_{k+N}=A_k$ and $B_{l+N}=B_l$ holds if and only if the classical fixed point conditions
\eqref{fix1}\eqref{fix2} are satisfied by $(q^{1/2}\lambda)^N\equiv q^{N/2}y$ and $(q^{1/2}\mu)^N\equiv q^{N/2}z$.
\medskip

$M^3$ is a central element of the quantum algebra. We normalize it by the condition $M^3=\mathbf{1}$, which fixes it up to multiplication by a thrid root of unity.
Its eigenvalues are $\exp(2\pi i k/3)$ and
\begin{equation}
 \mathrm{Tr}_N\, M^\ell= \sum_{k=0}^2 N_k(N)\, \exp(2\pi i k\ell/3),
\end{equation}
with the set of eigenvalue multiplicities
 $\{N_0(N),N_1(N), N_2(N)\}$
well--defined only {up to cyclic permutations.}

As in the $(A_2,A_1)$ case, we have (finitely) many different realizations of the cluster algebra at a given $N$, corresponding to
the choices of the different root choices. The eigenvalue multiplicities, being integers, are independent of the free paramater $\rho=XZ$.

Numerical experiments show that the structure of the sets  $\{N_0(N),N_1(N), N_2(N)\}$ for the $(A_3,A_1)$ model is very similar to that of the $(A_2,A_1)$ model: one has
\begin{align}
 N_k(N)&= [N/3]+a_k(N), && |a_k|\leq 1, && a_k(N)=a_k(N+6),
\end{align} 
(the periodicity being valid for coherent choices of the roots for the different $N$'s).

\subsection{The $(A_4,A_1)$ model}
The $\ca_4$ quantum algebra is generated by four (invertible) generators $X_k$ satisfying
\begin{equation}
 X_kX_{k+1}=q\,X_{k+1}X_k,\qquad X_jX_k=X_kX_j\ \ \text{if }|j-k|\geq 2.
\end{equation} 
An irreducible representation of the $\ca_4$ quantum algebra with $q$ a (primitive)
$N$--root of unity has dimension $N^2$, and it is specified (up to isomorphism) by four (non--zero) complex numbers $x_1,x_2,x_3,x_4$.
The four generators are explicitly
\begin{align}
X_1&= x_1^{1/N}\, S^{-1}_N\otimes \mathbf{1}
&X_2&=x_2^{1/N}\, \mathfrak{Z}_N\otimes \mathbf{1}\label{m42}\\
X_3&=x_3^{1/N}\, S_N\otimes S_N^{-1}
&X_4&=x_4^{1/N}\, \mathbf{1}\otimes \mathfrak{Z}_N,\label{m44}
\end{align}
where  $\mathfrak{Z}_N$ and $S_N$ are the $N\times N$ matrices
\begin{align}
 &\mathfrak{Z}_N=\mathrm{diag}(\zeta^\ell)_{\ell\in \mathbb{Z}/N\mathbb{Z}}, &&
(S_N)_{kl}= \delta_N(k-l- 1)\equiv \frac{1}{N}\sum_{r=0}^{N-1} q^{(k-l- 1)}.
\end{align} 

The quantum monodromy $M$ acts on $\ca_4$ as\footnote{\ This formula was deduced in the linear BPS chamber, see appendix \ref{app:linearchamber}.}
\begin{align}
 X_1^{-1}M& = M\Big((1-q^{1/2}X_2)X_3^{-1}+X_2\Big)\label{mond1}\\
X_3M&= M\Big(X_4^{-1}-q^{-1/2}X_3^{-1}X_4^{-1}+q^{-1}X_2^{-1}X_3^{-1}X_4^{-1}\Big)\label{mond2}\\
X_4M&= M\Big(X_1-q^{-1/2}X_1X_2+q^{-1}X_1X_2X_3-q^{-3/2}X_1X_2X_3X_4\Big)\label{mond3}\\
X_2X_4M&= M\Big( (1-q^{1/2}X_2)X_1X_4\Big).\label{mond4} 
\end{align}
One looks for an $N^2\times N^2$ matrix, $M_{(k,l)(m,n)}$ (here $k,l,m,n=1,2,\dots, N$) with the properties
\eqref{mond1}--\eqref{mond4} in the given irreducible representation of $\ca_4$. As in the previous examples,
these conditions give recursion relations for the entries of the monodromy matrix. The general solution of the recursion is
\begin{equation}\label{Mklmn}
 \begin{aligned}
         M_{(k,l)(m,n)}=C\cdot &(x_1/x_2)^{m/N}(x_1x_2x_3/x_4)^{n/N}(x_1x_2)^{-k/N}(x_1x_2x_3x_4)^{-l/N}\times\\
&\times q^{m(n-k-l)-nl}\, \prod_{r=0}^{m-1}\big(1-x_2^{1/N}q^{r+1/2}\big)\times\\
&\times \prod_{s=0}^{n-1}\frac{1-x_4^{1/N}q^{s+1/2}}{1-(x_2/x_4)^{1/N}q^{k-s-1}}\, \prod_{t=0}^{k-1}\frac{1-(x_2/x_4)^{1/N}q^t}{1-x_2^{1/N}q^{t+1/2}},
        \end{aligned}
\end{equation} 
which is, of course, a combination of discrete quantum dilogarithms.

The condition of periodicity mod $N$ in the four indices of $M_{(k,l)(m,n)}$
 fixes the allowed values of the central elements $x_k$ for a consistent representation of the $(A_4,A_1)$ quantum cluster algebra:
\begin{align*}
 M_{(k,l+N)(m,n)}=M_{(k,l)(m,n)}\quad &\Rightarrow\ (q^{N/2}x_1)(q^{N/2}x_2)(q^{N/2}x_3)(q^{N/2}x_4)=1\\
M_{(k+N,l)(m,n)}=M_{(k,l)(m,n)}\quad &\Rightarrow\ \left|\begin{aligned}&(q^{N/2}x_1)(q^{N/2}x_2)\big(1-q^{N/2}x_2\big)=\\ &\: =\big(1-(q^{N/2}x_2)/(q^{N/2}x_4)\big)
\end{aligned}\right.\\
M_{(k,l)(m+N,n)}=M_{(k,l)(m,n)}\quad &\Rightarrow\ (q^{N/2}x_1)(1-q^{N/2}x_2)=q^{N/2}x_2\\
M_{(k,l)(m,n+N)}=M_{(k,l)(m,n)}\quad &\Rightarrow\ \left|\begin{aligned}&(q^{N/2}x_1)(q^{N/2}x_2)(q^{N/2}x_3)\big(1-(q^{N/2}x_2)/(q^{N/2}x_4)\big)=\\ &\: =\big(1-q^{N/2}x_4)\big)
(q^{N/2}x_4),\end{aligned}\right.
\end{align*} 
which gives
\begin{equation}\label{fix41}
 \big(q^{N/2}x_1,q^{N/2}x_2,q^{N/2}x_3,q^{N/2}x_4)=\left(\lambda-\frac{1}{\lambda}, \frac{1}{\lambda},\lambda,\lambda+1\right)
\end{equation} 
where $\lambda$ is a solution to the cubic equation
\begin{equation}\begin{split}\label{fix42}
 &\lambda^3+\lambda^2-2\lambda-1=0\\
& \Rightarrow\quad \lambda \in \left\{ 2\, \cos\,\frac{2\pi}{7}, 2\, \cos\,\frac{4\pi}{7},
2\, \cos\,\frac{6\pi}{7}\right\}.\end{split}
\end{equation}  
Needless to say, eqns.\eqref{fix41}\eqref{fix42} just state that the central quantity $q^{N/2}X_k^N$ should be equal to
the classical variable $X_k$ at a fixed point of the (classical) monodromy, in agreement with the quantum Frobenius theorem.
\medskip

The $N^2\times N^2$ matrix $M$, eqn.\eqref{Mklmn}, satisfies $M^7=\mathbf{1}$ (for a suitable overall constant $C$).
Again, one can study the multiplicities $N_k(N)$ of the seven possible eigenvalues $\exp(2\pi i\ell/7)$ (well--defined up to cyclic permutations).

The numerical experiments show again a strong tendency towards equidistribution. In the range we explored ($N\leq 13$) one has
\begin{equation}
 N_k(N)=[N^2/7]+a_k(N),\qquad \text{with } a_k(N)=0,1.
\end{equation}

\newpage

\bibliography{4dmon}

\end{document}